\documentclass[aps, prx, twocolumn, superscriptaddress]{revtex4-2}

\usepackage{graphicx}
\usepackage{amsmath}
\usepackage{amsthm}
\usepackage{amssymb}

\usepackage[dvipsnames]{xcolor}
\usepackage{braket}
\usepackage{siunitx}
\usepackage{tikz}
\usepackage{quantikz}
\usepackage[
 hidelinks,
pdfauthor={Colin Scarato, Johannes Kn{\"o}rzer, Markus K. Hoffmann, Leon C. Sander, Luca Hofele, Shengpu Wang, Kilian Hanke, Ashay Sathe, Dominic Hagmann, Alexander Flasby, Michael J. Hartmann, Petr Zapletal, Andreas Wallraff and Christoph Hellings},
pdftitle={Hybrid Quantum-Classical Neural Networks for Recognizing Quantum Phases},
]{hyperref}
\usepackage[capitalize]{cleveref}
\usepackage[all]{hypcap}
\usepackage{ifplatform}
\usepackage{placeins}
\usepackage[T1]{fontenc}
\usepackage{microtype}

\newcommand{\affiliationethzphys}{\affiliation{Department of Physics, ETH Zurich, CH-8093 Zurich, Switzerland}}
\newcommand{\affiliationethzqc}{\affiliation{Quantum Center, ETH Zurich, CH-8093 Zurich, Switzerland}}
\newcommand{\affiliationpsiqchub}{\affiliation{ETH Zurich - PSI Quantum Computing Hub, Paul Scherrer Institute, CH-5232 Villigen, Switzerland}}
\newcommand{\affiliationerlangen}{\affiliation{Department of Physics, Friedrich-Alexander Universit{\"a}t Erlangen-N{\"u}rnberg (FAU), Erlangen, Germany}}
\newcommand{\affiliationerlangenmpi}{\affiliation{Max Planck Institute for the Science of Light, Staudtstra\ss{}e 2, 91058 Erlangen, Germany}}
\newcommand{\affiliationerlangenquint}{\affiliation{Quint Computing GmbH, Erlangen, Germany}}

\newcommand{\ptopo}{y}
\newcommand{\stopo}{S_{\mathrm{topo}}}
\newcommand{\yt}{y_{\mathrm{t}}}
\newcommand{\ytEER}{y_{\mathrm{t,EER}}}
\newcommand{\nb}{N_x}
\newcommand{\cz}{{\mathrm{C}}Z}
\newcommand{\figpanel}[1]{(#1)}

\begin{document}

\title{Hybrid Quantum-Classical Neural Networks for Recognizing Quantum Phases}

\author{Colin~Scarato}\affiliationethzphys\affiliationethzqc
\author{Johannes~Kn{\"o}rzer}\affiliationethzphys\affiliationethzqc\affiliationpsiqchub
\author{Markus~K.~Hoffmann}\affiliationerlangen
\author{Leon~C.~Sander}\affiliationerlangen
\author{Luca~Hofele}\affiliationethzphys\affiliationethzqc
\author{Shengpu~Wang}\affiliationethzphys\affiliationethzqc
\author{Kilian~Hanke}\affiliationethzphys\affiliationethzqc
\author{Ashay~Sathe}\affiliationerlangen
\author{Dominic~Hagmann}\affiliationethzphys\affiliationethzqc\affiliationpsiqchub
\author{Alexander~Flasby}\affiliationethzphys\affiliationethzqc\affiliationpsiqchub
\author{Michael~J.~Hartmann}\affiliationerlangen\affiliationerlangenmpi\affiliationerlangenquint
\author{Petr~Zapletal}\affiliationerlangen
\author{Andreas~Wallraff}\affiliationethzphys\affiliationethzqc\affiliationpsiqchub
\author{Christoph~Hellings}\affiliationethzphys\affiliationethzqc

\date{June 26, 2026}

\begin{abstract}
Identifying quantum phases of matter is key to understanding strongly correlated materials,
but remains a challenging task for both conventional computers and current quantum processors.
Here, we introduce and implement a hybrid quantum-classical neural network for quantum phase recognition by
combining a hardware-efficient parameterized quantum circuit and a feedforward neural network.
We jointly train both components with superconducting quantum hardware in the optimization loop,
to experimentally demonstrate
a classifier for the quantum phases
of surface code lattices with up to $4 \times 4$ sites in a magnetic field.
To learn nonlocal features of the topological phase,
we train the hybrid neural network
to distinguish topological ground states of the surface code
from a featureless ensemble of product states.
This allows the trained classifier to distinguish topological ground states
from randomly chosen product states,
even when subjected to any single-qubit Pauli error.
The classifier reaches accuracies above $\SI{85}{\%}$ in single-shot measurements,
and above $\SI{99}{\%}$ when averaging over ten measurements.
We expect hybrid neural networks such as the one presented here
to be a promising approach for characterizing quantum states
in scenarios where classical methods exhibit an unfavorable scaling of sample complexity.
\end{abstract}

\maketitle

\section{Introduction}\label{sec:introduction}

Machine learning encompasses model-agnostic methods
applicable to high-dimensional problems common in quantum many-body physics~\cite{Carleo2019}.
Motivated by the success of classical machine learning,
quantum machine learning~\cite{Wittek2014, Biamonte2017, Cerezo2022a}
explores the use of quantum processors for learning tasks and data analysis.
Applications span quantum chemistry~\cite{Xia2018a},
high-energy physics~\cite{Guan2020},
and condensed-matter physics~\cite{Cong2019, Bravo-Prieto2020, Wu2023, Acampora2025}.
By directly processing quantum data,
quantum machine learning avoids
the measurement overhead of converting quantum states to classical datasets.
This can enable empirically efficient training procedures~\cite{Abbas2021, Wiersema2021, Rudolph2024}
and, in some cases,
provable efficiency gains over classical machine learning~\cite{Huang2021j, Huang2021f}.

The ability to efficiently process quantum data
is crucial for identifying quantum phases of matter,
which provide insight into the low-energy physics of many-body systems.
While conventional quantum phases can usually be characterized by local order parameters,
topologically nontrivial phases evade such descriptions.
A paradigmatic example is the two-dimensional topological phase
realized by the surface code~\cite{Bravyi1998, Kitaev2003}.

A common strategy for identifying quantum phases from data
is to use classical neural networks~\cite{Carrasquilla2017, Nieuwenburg2017}
or classical shadows~\cite{Huang2021g},
which enable the extraction of local observables.
These approaches are effective for phases characterized by local order parameters,
but they face challenges for topological phases, whose defining features are nonlocal.
These challenges have motivated the development of methods for characterizing topological quantum states, including local error correction~\cite{Cong2024}, approaches based on tensor networks and quantum simulation~\cite{Liu2024s,Wahl2025}, and learning from randomized measurements~\cite{Teng2025}.
As an alternative,
quantum machine learning provides a framework for quantum phase recognition
that operates directly on quantum many-body states.
Existing approaches range from quantum kernel methods~\cite{SanchoLorente2022}
and quantum hypothesis testing~\cite{Tanji2026}
to parameterized quantum circuits~\cite{Cong2019, Chen2025e}.

Parameterized quantum circuits can implement
nonlocal transformations in large Hilbert spaces
to process classical~\cite{Havlicek2019, Henderson2020, Liu2021s, Peters2021, Senokosov2024}
and quantum data~\cite{Romero2017b, Beckey2022, Gellerc2021, Gong2023}.
They have been shown to generalize from few states used during training to unseen states~\cite{Caro2022, Caro2023}
and are, in this context,
often referred to as quantum neural networks~\cite{Schuld2018c, Farhi2018, Beer2020, Broughton2020}.
To identify topological quantum phases,
parameterized circuits have been optimized analytically
and in numerical simulations~\cite{Cong2019, Uvarov2020, Herrmann2022, Liu2022h, Sander2025, Aktar2025}.
For one-dimensional systems,
including spin chains~\cite{Chen2025e} and quasiperiodic lattice models~\cite{Ren2022a},
parameterized circuits to distinguish quantum phases have been trained directly on quantum hardware.

Here, we introduce
a hybrid quantum-classical neural network architecture for quantum phase recognition,
and train it to recognize
the two-dimensional ground states of the surface code in a magnetic field~\cite{Trebst2007}.
The input data are quantum states
generated on the quantum processor using a numerically optimized ground-state preparation circuit.

The first stage of the hybrid quantum-classical neural network is a parameterized quantum circuit,
which we construct from continuous sets of two-qubit controlled-phase ($\cz_\theta$) gates
and of single-qubit $Y_\theta$ gates.
We use hardware-native implementations of these continuously parameterized gates
as this approach enables low-depth quantum circuits with high fidelities~\cite{Lacroix2020, Cirstoiu2020}.
For the $\cz_\theta$ gate set,
we choose the implementation from~\cite{Scarato2025},
which provides robustness to control pulse distortions
and parameterizes the conditional phase as a linear function of a single pulse parameter.

\begin{figure*}
	\centering
	\includegraphics[trim={0 0.3cm 0 0.2cm},clip]{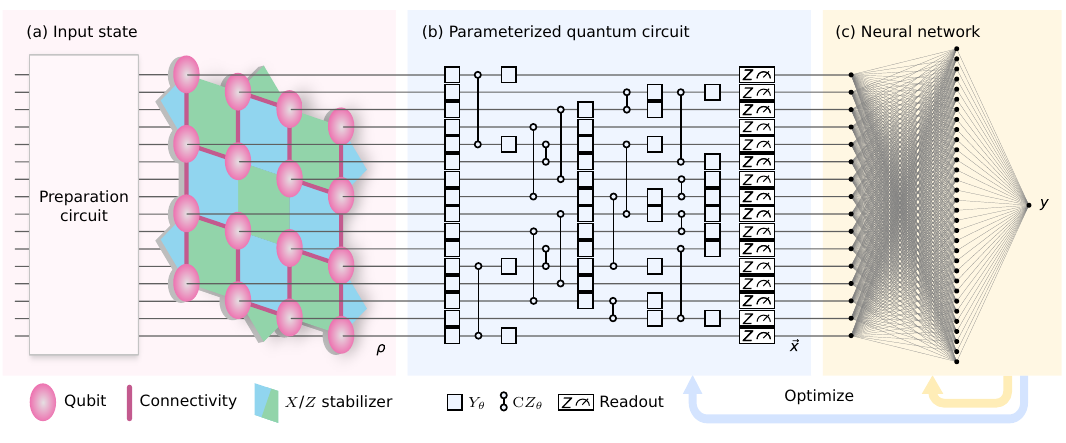}
	\caption{
		Architecture of the hybrid neural network.
		\figpanel{a}~The input state $\rho=\ket{\psi}\!\bra{\psi}$ to a hybrid neural network is
		prepared on a lattice of 16 qubits (pink circles).
		The connectivity used in the experiment is indicated as pink lines.
		The blue and green polygons indicate the $X$ and $Z$ stabilizers
		of the $4 \times 4$ surface code, whose ground state is
		used as an example of a topological input state.
		\figpanel{b}~Parameterized quantum circuit used in the experiment,
		consisting of arbitrary-angle $Y_\theta$ gates and controlled arbitrary-phase $\cz_\theta$ gates,
		followed by measurements in the computational basis.
		\figpanel{c}~Classical neural network
		mapping the measured bit string $\vec{x}=x_1x_2...x_N$ to the output~$\ptopo$.
		The nested loops of the joint optimization of the quantum circuit and the classical neural network
		are indicated by the two arrows below the blue and yellow boxes.
		The architecture is outlined in \cref{sec:introduction},
		the state preparation~\figpanel{a} is discussed in \cref{sec:sc},
		and the hybrid neural network~\figpanel{b}-\figpanel{c}
		in \cref{sec:implementation}.
	}
	\label{fig:concept}
\end{figure*}

Each bit string~$\vec{x}$ measured on the quantum hardware
is then processed by a classical feedforward neural network.
The output $\ptopo\in[0,1]$ provides a probability estimate
that represents the confidence of having recognized the topological phase.
The high-level architecture is illustrated by the colored boxes in \cref{fig:concept}:
the input state (red),
the parameterized quantum circuit (blue), and the classical neural network (yellow box),
which are explained in more detail in the following sections.

We train the quantum circuit and the classical neural network with a joint optimization method
by updating the parameterized quantum circuit
in an outer optimization loop with access to the quantum hardware
and repeatedly training the classical neural network in an inner loop.
In this procedure,
the parameterized quantum circuit learns a nonlocal basis transformation of the input state.
This ensures that the bit strings corresponding to the two classes of states
are sufficiently separable by the classical neural network.

In previous work~\cite{Huang2021g, Cong2019, Liu2022h, Aktar2025, Chen2025e},
training procedures have relied
on training data obtained from both sides of a phase transition.
The training data then typically include a
``locally-easy'' phase~\cite{Bermejo2026}, which can be classified based on local observables.
In contrast, we present a training method
that primarily relies on measurements of topological ground states.
We demonstrate that the trained hybrid neural network
recognizes the features of topological states
and distinguishes them from a broad class of topologically trivial states.

We implement the hybrid quantum-classical neural network
on a superconducting quantum processor
hosting a grid of 17 flux-tunable transmon qubits.
An analysis of the experimental results shows that
the trained classifier generalizes to states not included in the training set.
We perform the classification of topological states
in single-shot measurements
and demonstrate that the classification remains robust
when subjecting the states to any randomly chosen Pauli error
and when comparing them to random tensor products of Pauli eigenstates.
The accuracy in inference increases above $\SI{99}{\%}$
when using $10$ measurement shots,
showing that the trained hybrid neural network can efficiently recognize
the topological phase from only a few measurements.
These experimental results are consistent with
the numerical study in the companion paper~\cite{Hoffmann2026},
where we show that the hybrid neural network reduces the sample complexity
of both training and inference
compared to a classical neural network
trained directly on randomized measurements in local Pauli bases.

\section{Surface code in a magnetic field}\label{sec:sc}

\newcommand{\stabset}[1]{{#1}}
\newcommand{\stabsetX}{\stabset{A}}
\newcommand{\stabsetZ}{\stabset{B}}
We experimentally study a quantum many-body system
featuring both a topologically ordered and a topologically trivial phase:
the surface code~\cite{Bravyi1998} in a magnetic field.
The Hamiltonian
\begin{align}\label{eq:surface-code-in-field}
	H =
    -\sum_{\stabsetX}^{} X_\stabsetX - \sum_{\stabsetZ}^{} Z_\stabsetZ
    \; \underbrace{-  \,h\,\sum_{i=1}^{N} Z_i}_{H_{\mathrm{Field}}}
\end{align}
is defined on a square lattice of $N$ qubits.
The $X$ stabilizers
$X_\stabsetX=\prod_{i\in \stabsetX}^{}X_i$
and $Z$ stabilizers $Z_\stabsetZ=\prod_{i\in \stabsetZ}^{}Z_i$
are indexed by the sets of qubits $\stabsetX$ and $\stabsetZ$,
as indicated for $N=16$ qubits by the blue and green polygons, respectively,
in \cref{fig:concept}\figpanel{a}.
The longitudinal $Z$ field is modeled by $H_{\mathrm{Field}}$
and parameterized by the field strength $h$.
The ground states of \cref{eq:surface-code-in-field}
undergo a phase transition between a topologically ordered state~\cite{Hamma2005} for $h=0$
and a paramagnetic state $|0 \dots 0\rangle$ aligned with the field for $h\to\infty$.
To lift the twofold degeneracy at $h=0$,
we select the ground state obtained by approaching $h=0$
from positive field strength, $h\rightarrow 0^+$.

We map the surface code lattices of sizes $3 \times 3$ and $4 \times 4$
onto the quantum processor,
in a way which is compatible with the hardware connectivity,
see \cref{app:device}.
The connectivity we use in the experiment with $N=16$ qubits
is shown in \cref{fig:concept}\figpanel{a}.

The ground state of the surface code in a magnetic field
has previously been prepared experimentally using auxiliary qubits~\cite{Bluvstein2022, Iqbal2024a, Cochran2024},
and for $h=0$ without auxiliary qubits~\cite{Satzinger2021}.
An extension to $h\geq 0$ without auxiliary qubits
has been proposed using continuously parameterized single-qubit rotations~\cite{Sun2023f}.
While the circuits proposed for $h=0$ prepare the ground state exactly~\cite{Bluvstein2022,Iqbal2024a,Satzinger2021},
the constructions for $h>0$ from~\cite{Cochran2024,Sun2023f} are approximate variational ansätze
whose good performance relies on the finite size of the system
and whose corresponding variational states become topologically trivial in the thermodynamic limit.
Here, we experimentally prepare
ground states for $h \in \{0, 0.05,\dots,1\}$
using $\cz_\pi$ gates and continuously parameterized $Y_\theta$ gates
in quantum circuits similar to those in~\cite{Satzinger2021, Sun2023f}
by optimizing the gate parameters in numerical simulations,
see \cref{app:preparation_circuits}.

\begin{figure*}
	\centering
	\includegraphics{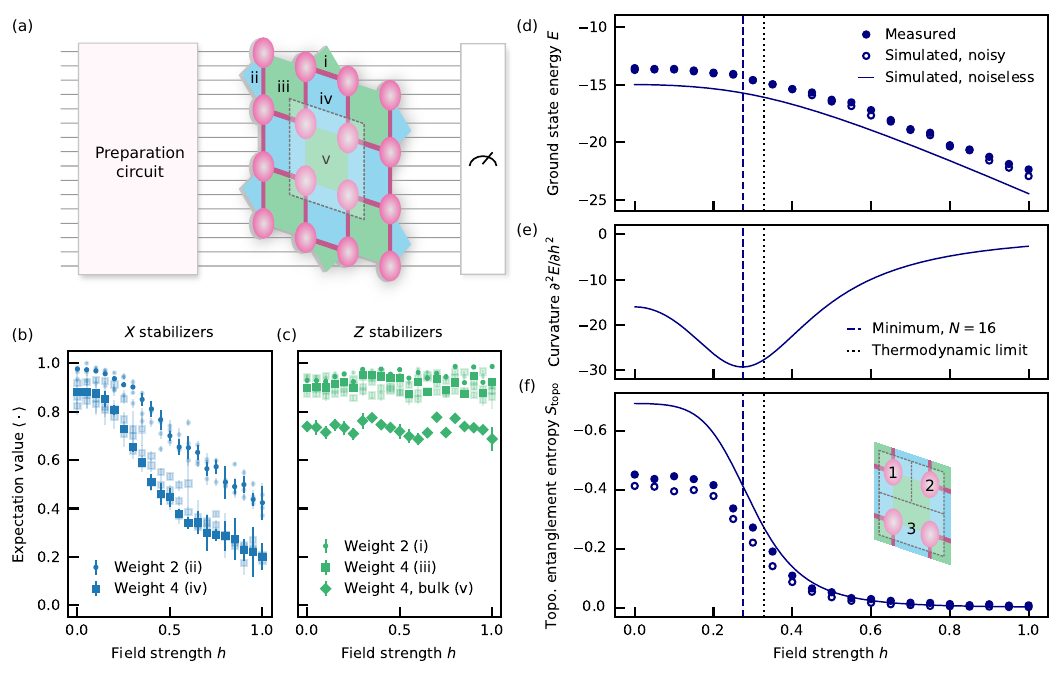}
	\caption{
		Ground state characterization in a $4 \times 4$ surface code.
		\figpanel{a}~Qubit lattice, illustrating the ground state prepared
		and directly measured in multiple bases (see main text).
		Roman numerals indicate a selection of representative stabilizers;
		the bulk stabilizer (v) is highlighted with a dashed square.
		\figpanel{b},~\figpanel{c}~Measured expectation values
		of the stabilizers labeled in \figpanel{a} (opaque markers)
		and the other stabilizers (transparent),
		as a function of the field strength $h$.
		\figpanel{d},~\figpanel{e}~Ground state energy $E$
		and its curvature $\partial^2 E/\partial h^2$:
		extracted from the stabilizers measured in \figpanel{b}~and \figpanel{c} (filled circles),
		from simulations considering gate and idling errors (empty circles) and from noiseless simulations (solid line).
		The dashed blue line indicates the minimum of $\partial^2 E/\partial h^2$
		and the dotted black line the phase boundary in the thermodynamic limit,
		see main text for details.
		\figpanel{f}~Topological entanglement entropy $\stopo$
		of the bulk stabilizer as a function of the field strength $h$,
		from \cref{eq:stopo}.
		The legend is as in panel~\figpanel{d}, and the results are averaged over the four possible tripartitions,
		one example of which is shown in the inset.
	}
	\label{fig:preparation}
\end{figure*}

To verify that the target ground states
at selected field strengths
are prepared faithfully,
we measure the system in multiple bases
after executing the preparation circuit as sketched in \cref{fig:preparation}\figpanel{a}.
First, we perform $8192$ simultaneous single-shot measurements
of all qubits either in the $X$ or $Z$ basis.
We then multiply and average the measurement results
to obtain the expectation values of the stabilizers.
From ten experimental runs,
we extract the mean stabilizer values and the standard errors
shown in \cref{fig:preparation}\figpanel{b}-\figpanel{c}.
The expectation values of the $X$ stabilizers
decrease as a function of field strength,
whereas those of the $Z$ stabilizers do not
since the $Z$ stabilizers commute with $H_{\mathrm{Field}}$ in \cref{eq:surface-code-in-field}.
The weight-four stabilizers,
and in particular the bulk $Z$ stabilizer,
deviate the most from their ideal values, e.g., from $1.0$ at $h=0$.
These stabilizers are most sensitive to
gate and idling errors in the state preparation,
as confirmed by Kraus operator simulations described in \cref{app:kraus},
which account for qubit relaxation, dephasing, and residual $ZZ$ coupling.

From the measured stabilizers and Pauli-$Z$ expectation values,
we compute the dimensionless ground state energy,
see the filled circular markers in \cref{fig:preparation}\figpanel{d}.
Standard errors from the ten experimental runs are below $0.3$ and not visible at the scale of the plot.
Noiseless simulations of the unitary evolution (solid line)
and the Kraus operator simulations from \cref{app:kraus} (empty circles)
are in good agreement with the experimental results.
In case of a vanishing field, $h=0$,
the $15$ stabilizers each contribute $-1$
to the total ground state energy $E=-15$, see \cref{eq:surface-code-in-field}.
For large $h$, the magnetic field is dominant,
yielding a linear dependence of the energy on $h$.
As an indicator for the crossover between the topological and the trivial phases,
we determine the minimum of the second derivative
of the energy obtained in noiseless simulations,
see the dashed blue line in \cref{fig:preparation}\figpanel{e}.
The phase transition in the thermodynamic limit~\cite{Dusuel2011}
is marked by a dotted black line.

Finally, we verify the presence of topological order in the prepared states.
The entanglement entropy~$S$ of a subsystem of a qubit lattice
is defined as the entropy of the reduced density matrix of the subsystem.
In the presence of two-dimensional topological order,
the entanglement entropy scales linearly with the boundary length of the subsystem,
but also includes a constant term known as the topological entanglement entropy.
This term can be expressed as~\cite{Kitaev2006, Satzinger2021}
\begin{align}\label{eq:stopo}
	\stopo =& S_{\{\mathrm{1}\}} + S_{\{\mathrm{2}\}} + S_{\{\mathrm{3}\}}\\
	& - S_{\{\mathrm{1,2}\}} - S_{\{\mathrm{2,3}\}} - S_{\{\mathrm{1,3}\}} + S_{\{\mathrm{1,2,3}\}}\nonumber,
\end{align}
where $S_{\{\mathrm{i,\dots}\}}=-\ln\mathrm{Tr}(\rho_{\{\mathrm{i,\dots}\}}^2)$
are 2-R\'enyi entropies,
and the indices label the density matrices $\rho_{\{\mathrm{i,\dots}\}}$
formed from three regions partitioning the subsystem, see the example of regions labeled $1$, $2$, and $3$ in the inset of \cref{fig:preparation}\figpanel{f}.

From a full state tomography of the bulk stabilizer (v),
we calculate the R\'enyi entropy using the method from~\cite{Brydges2019}
for each of the possible tripartitions
and average the resulting topological entanglement entropy over the four tripartitions,
see the filled circular markers in \cref{fig:preparation}\figpanel{f}.
The standard errors extracted from bootstrapping are up to $0.01$,
and are therefore not visible at the scale of the plot.
We observe that $\stopo$ drops to zero close to the expected phase crossover,
which is marked by the dashed line.
This demonstrates the experimental preparation of
topologically non-trivial states in two spatial dimensions,
with a topological entanglement entropy decreasing in the field strength $h$.
The reduced contrast compared to noiseless simulations (solid line) is caused by gate and idling errors, as confirmed by the Kraus operator simulations from \cref{app:kraus} (empty markers).

\section{Training the hybrid neural network}\label{sec:implementation}

To implement the hybrid quantum-classical neural network for quantum phase recognition,
we construct the parameterized quantum circuit
from the structure of the inverse of the ground-state preparation circuit.
This choice of ansatz,
shown in \cref{fig:concept}\figpanel{b} for $N=16$ qubits,
ensures that there is a set of gate parameters for which
the quantum circuit implements a unitary that maps the ground state for $h=0$
to a product state $\ket{0 \dots 0}$.
By disentangling the state,
this nonlocal basis change facilitates classification of quantum states based on measurements of individual qubits.
Quantum circuits constructed from
inverse preparation circuits were previously applied in experiments
for quantum phase recognition in one-dimensional spin chains~\cite{Herrmann2022, Chen2025e}
and in a numerical study of two-dimensional topological phases~\cite{Sander2025}.

We implement the quantum circuit using a hardware-efficient gate set,
consisting of continuously parameterized $\cz_\theta$ gates~\cite{Scarato2025}
and $Y_\theta$ gates,
see \cref{app:device} for implementation details.
Unconditional qubit reset based on the method from~\cite{Zhou2021}
enables us to use a short repetition period
of $\SI{6.0}{\micro\second}$,
comprising $\SI{2.5}{\micro\second}$ of qubit reset,
$\SI{0.6}{\micro\second}$ each for executing
the gate sequences of the state preparation and the parameterized circuit,
$\SI{0.4}{\micro\second}$ for qubit readout,
and an idle time of $\SI{1.9}{\micro\second}$.

The classical part of the hybrid machine learning architecture
consists of a fully-connected feedforward neural network with three layers,
which processes the measured bit strings.
For $N=16$ qubits, we construct the input layer
with 16 nodes and the hidden layer with 32 nodes,
followed by a single-node output layer,
see \cref{fig:concept}\figpanel{c}.
As activation functions,
we employ a rectified linear unit for the hidden layer and a sigmoid for the output.

We construct a training set consisting of the ground state for $h=0$
as representative state for the topological phase (labeled as 1),
and a maximally mixed state (labeled as 0).
The latter is equivalent to letting the trivial phase be represented
by a featureless ensemble of product states and is emulated in classical processing,
see \cref{app:training}.

Given sets of labeled training states $\mathcal{T}_L$ with label $L\in\{0,1\}$,
we use the binary cross-entropy cost function~\cite{Dawid2022}
\begin{align}
	c = \frac{1}{2}\sum_{L=0}^{1}\frac{1}{|\mathcal{T}_L|}\sum_{\rho\in\mathcal{T}_L}\frac{1}{N_{\rho}}\sum_{i=1}^{N_{\rho}}\left[ L\ln\frac{1}{\ptopo_i}+(1-L)\ln\frac{1}{1-\ptopo_i} \right]
	\label{eq:bxe}
\end{align}
where $N_{\rho}$ is the number of measurements $i$
available for the input state $\rho$ after leakage rejection,
see \cref{app:device},
and $\ptopo_i$ is the output of the classical neural network
for the $i^{\mathrm{th}}$ measured bit string.
The expression in square brackets in \cref{eq:bxe}
is the cross-entropy between label and output,
and the cost function $c$ averages this quantity over all measurements,
see \cref{app:training} for details.

We implement the joint training in two nested loops,
as summarized here and detailed in \cref{app:training}.
At each iteration of the outer optimization loop,
we randomly sample eight sets of gate parameters around the current parameters.
Based on 8192 repetitions of measurements for these eight sets of parameters,
we train the classical neural network using all bit strings available after leakage rejection.
This inner optimization loop is implemented
with standard methods including backpropagation and the ADAM optimizer~\cite{Kingma2017}.
From the eight values of the cost function $c$ produced by the trained classical neural network,
we compute an approximate gradient as in~\cite{Salimans2017}
to update the parameters of the quantum circuit.

\begin{figure*}
	\centering
	\includegraphics{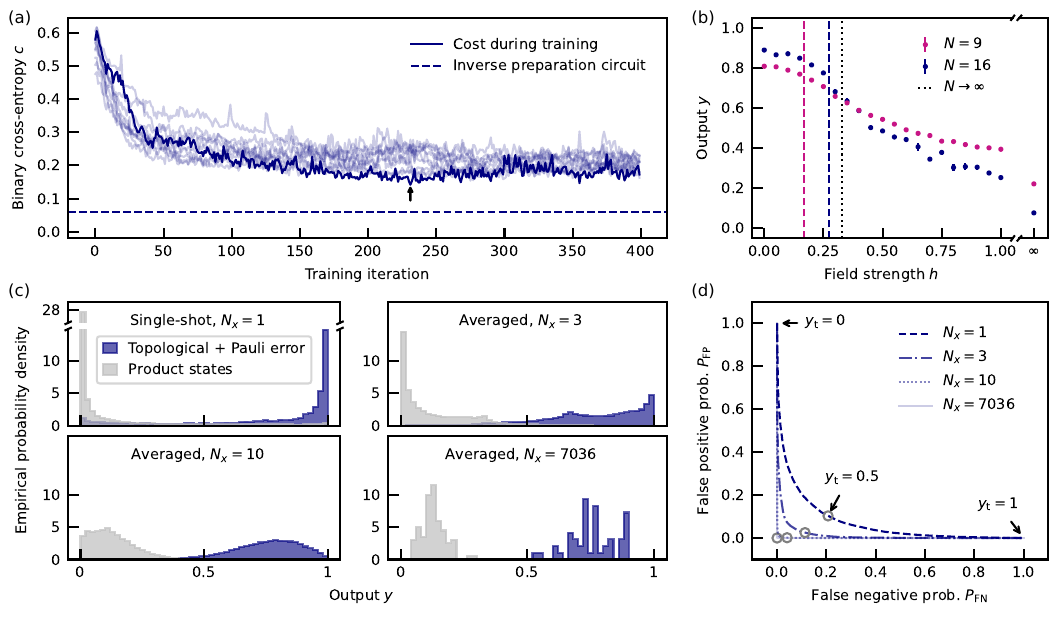}
	\caption{
		Training and benchmarking of the hybrid neural network.
		\figpanel{a}~Binary cross-entropy cost during training,
		for the $4 \times 4$ surface code, starting from ten random initializations.
		The lowest achieved cost is marked with an arrow.
		The corresponding training trace is drawn as an opaque curve,
		the nine others as transparent curves.
		Dashed blue line: binary cross-entropy achieved
		with the inverse preparation circuit when training only the classical neural network.
		\figpanel{b}~Averaged output of the trained hybrid neural networks
		when sweeping the field strength $h$ that parameterizes the ground states provided at the input,
		plotted for $N=16$ (blue) and $N=9$ (pink).
		The vertical lines mark the expected phase crossover, see details in the main text.
		\figpanel{c}~State classification results for $N=16$ qubits.
		Input states are the topological ground state for $h=0$
		with a single-qubit Pauli error (blue) and random product states (gray);
		the output~$\ptopo$ is histogrammed
		after averaging by groups of $\nb\in\{1, 3, 10, 7036\}$ experimental runs.
		\figpanel{d}~Classification errors when providing the hybrid neural network output~$\ptopo$
		to a binary classifier with decision threshold $\yt$,
		plotted for the four cases from panel~\figpanel{c}.
		The circles correspond to choosing a threshold of $\yt=0.5$.
	}
	\label{fig:results}
\end{figure*}

We perform the joint training for ten random initializations of all parameters.
The binary cross-entropy after training the classical neural network
is shown in \cref{fig:results}\figpanel{a}
as a function of training iteration of the quantum circuit.
The high initial cost reflects the fact that
a random quantum circuit does not extract features
that allow the trained classical neural network
to separate the topological states from the trivial states.
The decrease in binary cross-entropy indicates
that training makes the two classes of states increasingly distinguishable
by reducing the overlap between the distributions of bit strings measured for the two classes,
as confirmed by supplemental data in \cref{app:training}.
For our choice of training set,
the optimum is achieved for a quantum circuit that implements
the inverse of the ground-state preparation for $h=0$, see \cref{app:training}.
This yields a cross-entropy of $0.06$ in the experiment,
see the dashed line,
and $4\times 10^{-4}$ in noiseless simulations.

We select the gate parameters corresponding
to the lowest cross-entropy achieved during the joint training,
see the black arrow in \cref{fig:results}\figpanel{a}.
For the classical neural network, we then perform a final training iteration
in which we extend the training set by including measured data
for the ground state at $h \to \infty$
as an additional representative member of the trivial phase.
This final training step
can be interpreted as transfer learning~\cite{Mari2020},
see the discussion of variations of the final training step in \cref{app:variants}.

As the mixed state can be seen as a surrogate for a featureless ensemble of product states,
see \cref{app:training},
the training procedure promotes learning the specific features of the topological state.
This lets the trained hybrid neural network distinguish the topological state
from a broad class of trivial states as demonstrated in the next section.
When instead training with a particular class of topologically trivial states,
the hybrid neural network could instead focus on the simpler task of learning
locally-easy~\cite{Bermejo2026} features of the trivial states
such as the magnetization~\cite{Trebst2007},
see supplementary experimental results in \cref{app:variants}.

\section{Recognizing quantum phases}\label{sec:results}

We first benchmark the trained hybrid neural network
on the task of recognizing the quantum phases of the surface code
in a magnetic field while varying the field strength $h$.
For $N=16$ qubits, the output $\ptopo$ of the hybrid neural network
transitions smoothly from $\ptopo\approx 0.9$ for the topological state
to $\ptopo\approx 0.1$ for the paramagnetic state,
see the blue markers in \cref{fig:results}\figpanel{b}.
This indicates that the hybrid neural network
can generalize to states between $h=0$ and $h\to\infty$,
which have not been seen during training.
Similar behavior with slightly reduced contrast is obtained
for an implementation with $N=9$ qubits
presented in \cref{app:circuits},
as indicated by the pink markers.
In analogy to \cref{fig:preparation}\figpanel{d}-\figpanel{f},
we mark the expected phase crossovers for both lattice sizes by dashed lines
and the phase transition in the thermodynamic limit by a dotted line.

To probe the robustness of our method,
we generate a test set by separately applying each of the 48 possible single-qubit Pauli errors to the topological ground state for $h=0$,
and by randomly choosing 100 tensor products of Pauli eigenstates.
We prepare $8192$ copies of each state,
yielding $7036$ single-shot values of the output $\ptopo$ per state after leakage rejection,
see \cref{app:device}.
The distributions of the output values are shown as histograms
in the upper left panel of \cref{fig:results}\figpanel{c}.
We observe that the main peaks of these distributions are well separated, with overlapping tails.
We then average the single-shot values by groups of size $\nb$,
see the other three panels of \cref{fig:results}\figpanel{c}.
The averaging reduces the overlap between the distributions,
and averaging over all available measurements ($\nb=7036$)
yields non-overlapping distributions for this test set.

We provide the output $\ptopo$ of the hybrid neural network
to a binary classifier with decision threshold $\yt\in [0, 1]$.
The state is classified as topological for $\ptopo\geq\yt$ and as trivial for $\ptopo<\yt$.
We estimate the error probabilities,
which correspond to either classifying
the trivial states as topological (false positive $P_{\mathrm{FP}}$)
or the topological states as trivial (false negative $P_{\mathrm{FN}}$).
For each $\nb$, the values of $(P_{\mathrm{FN}}, P_{\mathrm{FP}})$ form a parametric curve of $\yt$.
These so-called receiver operating characteristic curves~\cite{HosmerJr2000}
are shown in \cref{fig:results}\figpanel{d}.
At the threshold $\ytEER$ that yields equal error rates,
$P_{\mathrm{FN}} = P_{\mathrm{FP}}=:P$,
we evaluate the classification accuracy $1-P$.
For single-shot measurements ($\nb=1$), the classifier achieves
a classification accuracy above $\SI{85}{\%}$ at $\ytEER=0.31$.
The classification accuracy increases when increasing $\nb$.
Averaging over all available measurements achieves
ideal classification of this test set
because the measured output distributions do not overlap,
so that the curve for $\nb=7036$ consists of two straight lines.
The curve for $\nb=10$ is almost indistinguishable from this ideal curve,
and averaging over $\nb=10$ measurements achieves
a classification accuracy of $\SI{99}{\%}$ at $\ytEER=0.40$.

\section{Conclusion}\label{sec:outlook}

We have introduced and implemented
a hybrid quantum-classical machine learning architecture,
and we have applied it
to recognize the quantum phases of the $4 \times 4$ surface code
in a longitudinal magnetic field:
a topological phase at weak fields
and a topologically trivial phase at sufficiently large field strength.
After joint training of the quantum and classical components
with access to superconducting quantum hardware in the optimization loop,
the trained hybrid neural network can both
(i) recognize quantum phases across the ground-state phase diagram
and (ii) distinguish topological ground states with a local Pauli error
from randomly chosen trivial states.

The broad applicability of the method
is enabled by a training set that primarily relies on
the features of the topological phase
instead of rewarding learning of locally-easy features of a trivial phase.
This experimental implementation of quantum phase recognition
goes beyond previous studies of two-dimensional topological systems~\cite{Huang2021g, Sander2025, Aktar2025},
which have been limited to numerical simulations
and have addressed only one of the two classification tasks labeled (i) and (ii) above in isolation.
The hybrid neural network enables efficient recognition
of the topological phase from only a few measurements,
with a classification accuracy exceeding $\SI{85}{\%}$ for single-shot inference
and $\SI{99}{\%}$ when averaging over $10$ measurements.

As the number of gates in our approach
scales linearly with the number of qubits $N$
while the circuit depth scales as $\sqrt{N}$,
see \cref{app:circuits},
the method could be readily implemented on a larger qubit grid.
This would enable probing the physics of larger surface code lattices,
aiming at better estimates of the behavior in the thermodynamic limit.
For example,
a parameterized quantum circuit for a surface code lattice of size $15 \times 15$
would require $308$ two-qubit gates and $841$ single-qubit gates,
yielding a gate sequence duration of $\SI{1.1}{\micro\second}$ on an architecture similar to ours.

The hybrid quantum-classical neural network introduced here
is a promising framework for scenarios
where classical methods exhibit an unfavorable scaling of sample complexity,
such as in systems near criticality and with large correlation length.
In terms of the number of samples required in training and inference
to achieve a targeted classification accuracy,
our approach outperforms a classical neural network,
as demonstrated numerically for surface code models up to size $5 \times 5$ in the companion paper~\cite{Hoffmann2026}.
It is yet to be shown that the demonstrated trainability and performance persist in even larger systems.
Future work may explore the surface code in transverse fields,
and systems hosting two topological phases~\cite{Haller2023}
as studied in the companion paper~\cite{Hoffmann2026}.
In such applications,
we expect hybrid neural networks to be particularly useful
for learning tasks requiring robust processing of quantum data.
This includes robustness against gate and readout errors,
as well as against imperfectly prepared input states,
as we have studied by injecting Pauli errors.

Future deployment of hybrid neural networks
would benefit from studies of the trade-off between
trainability and expressivity~\cite{Holmes2021, Larocca2021, Ragone2024},
to go beyond our current approach of a physics-inspired ansatz
towards a fully data-driven method based on a model-agnostic circuit structure.
Hybrid quantum-classical neural networks could then be applied to
\emph{a priori} unknown quantum states observed in the laboratory.

\section*{Acknowledgments}

We thank Eugene Demler and
Richard Kueng for insightful discussions,
Adri\'an P\'erez-Salinas
and Frank Pollmann
for valuable feedback on the manuscript,
François Swiadek
for contributions to the quantum processor design software,
and Chrysander D. Hagen for contributions to the device design.
The team in Zurich acknowledges financial support
by IARPA and the Army Research Office,
under the Entangled Logical Qubits program,
and under Cooperative Agreement Number W911NF-23-2-0212,
by the Baugarten foundation,
by the Swiss National Science Foundation, R'equip grant 206021-170731,
and by ETH Zurich.
The views and conclusions contained in this document are those of the authors
and should not be interpreted as representing the official policies,
either expressed or implied, of IARPA,
the Army Research Office, or the U.S. Government.
The U.S. Government is authorized to reproduce and distribute reprints
for Government purposes notwithstanding any copyright notation herein.
The work in Erlangen is part of the Munich Quantum Valley,
which is supported by the Bavarian state government with funds from the Hightech Agenda Bayern Plus,
and it was supported by the EU program HORIZON-MSCA-2022-PF Project No. 101108476 HyNNet NISQ.

C.S., M.K.H., P.Z., M.J.H., and C.H.\ conceived the experiments.
C.S.\ performed the measurements
and analyzed the data.
J.K.\ provided guidance on analysis and interpretation of the experimental results.
S.W.\ and K.H.\ contributed to preliminary versions of the experiment.
C.S., K.H., and S.W.\ developed the experimental software framework.
C.S.\ and L.H.\ maintained the measurement setup and calibrated the device.
L.H.\ designed the device,
and D.H.\ and A.F.\ fabricated it.
M.K.H., L.S., A.S., P.Z., and C.S.\ performed numerical simulations.
C.S., J.K., and C.H.\ wrote the manuscript with input from all co-authors.
C.H., A.W., P.Z., and M.J.H.\ supervised the work.

\appendix
\crefalias{section}{appendix}

\section{Measurement setup and quantum device}\label{app:device}

\begin{figure*}
	\centering
	\includegraphics{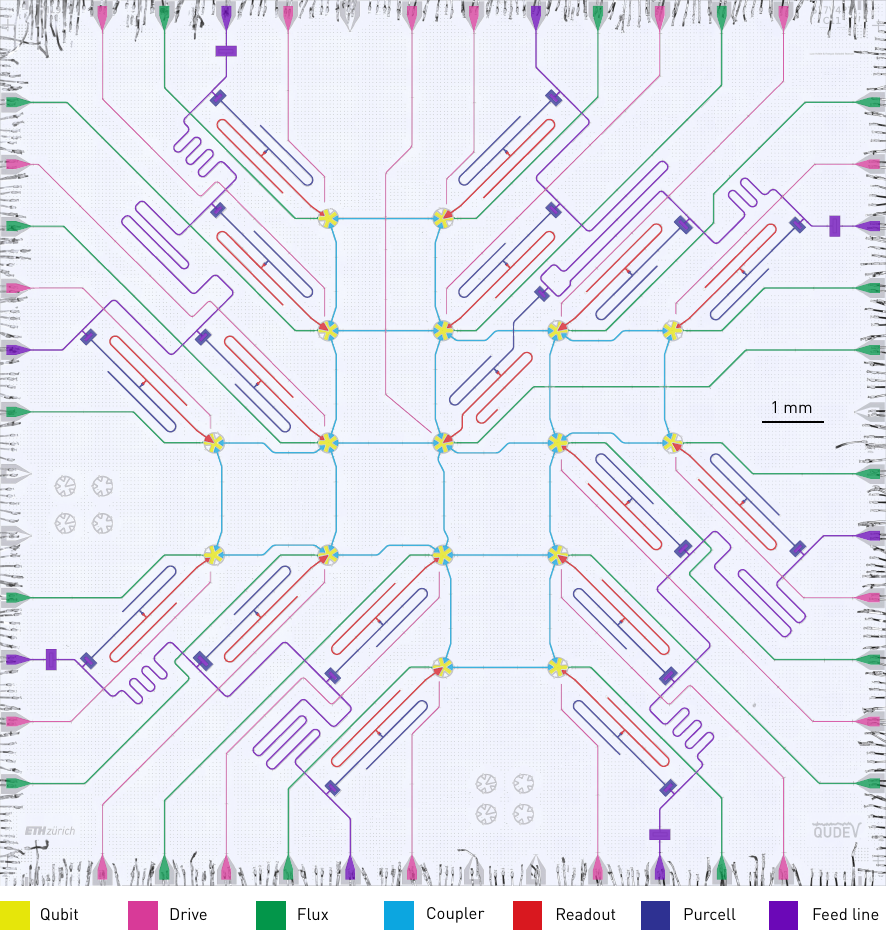}
	\caption{
		False-color micrograph of the superconducting quantum processor used in the experiment.
		Each transmon qubit consists of a capacitive island (yellow),
		which is capacitively coupled to a drive line (pink).
		The island is connected to the ground plane via a superconducting quantum interference device (SQUID),
		which is inductively coupled to a flux line (green).
		The static qubit-qubit coupling is implemented with fixed-frequency resonators (light blue).
		Each transmon is capacitively coupled to a readout resonator (red),
		which is coupled via a Purcell filter (dark blue) to a coplanar feed line (purple).
	}
	\label{fig:app_device}
\end{figure*}

We have realized the experiment on a quantum processor hosting 17 transmon qubits,
see the micrograph in \cref{fig:app_device}.
The device layout is similar to that presented in~\cite{Krinner2022},
with slightly adjusted qubit and resonator frequencies
to facilitate qubit reset, see below, while avoiding readout-induced leakage~\cite{Lacroix2025}.

We have defined the circuit elements using photolithography and reactive-ion etching
in a niobium film with a thickness of $\SI{125}{\nano\meter}$
deposited onto a high-resistivity intrinsic silicon substrate.
Aluminum/titanium/aluminum trilayer airbridges connect
the ground plane and establish cross-overs for coplanar waveguides.
We have fabricated the aluminum/aluminum-oxide/aluminum Josephson junctions
of the transmon qubits using electron-beam lithography and double-angle shadow evaporation.
Aluminum bandages~\cite{Bilmes2021} ensure the contact
between the leads of the Josephson junctions and the niobium base layer.
The device is installed in a dilution refrigerator
and operated at a temperature of $\SI{9}{\milli\kelvin}$.
We use the measurement setup described in~\cite{Krinner2022}
and compensate for flux pulse distortions
as well as for crosstalk of drive and flux control signals as described in Refs.~\cite{Hellings2025, Krinner2022}.
Coherence times and further device properties
are summarized in \cref{tab:device_params} and \cref{tab:device_params_pairs}.

\begin{table*}[]
	\begin{tabular}{|l|l|l|l|l|l|l|l|l|l|l|l|l|l|l|l|l|l|}
		\hline
		\textbf{Qubit index}														& 1 & 2 & 3 & 4 & 5 & 6 & 7 & 8 & 9 & 10 & 11 & 12 & 13 & 14 & 15 & 16 \\ \hline
		Qubit idle frequency $\omega_{\mathrm{ge}}/2\pi$ $(\SI{}{\giga\hertz})$					& 3.66 & 6.51 & 3.68 & 6.44 & 6.51 & 3.68 & 6.37 & 4.09 & 4.06 & 6.37 & 4.08 & 6.52 & 6.49 & 4.05 & 6.32 & 3.91 \\ \hline
		Anharmonicity $\alpha/2\pi$ $(\SI{}{\mega\hertz})$					& -172 & -151 & -172 & -155 & -156 & -173 & -160 & -170 & -170 & -158 & -169 & -151 & -156 & -169 & -159 & -170 \\ \hline
		Readout frequency $\omega_{\mathrm{RO}}/2\pi$ $(\SI{}{\giga\hertz})$		& 6.83 & 7.14 & 6.78 & 7.24 & 7.37 & 6.95 & 7.42 & 6.90 & 6.99 & 7.24 & 6.95 & 7.14 & 7.44 & 6.86 & 7.29 & 6.84 \\ \hline
		Three-state readout error $\epsilon_{\mathrm{RO}}$ $(\%)$					& 2.5 & 4.3 & 2.4 & 4.9 & 3.9 & 3.5 & 3.5 & 2.6 & 4.0 & 4.1 & 3.6 & 4.0 & 4.2 & 1.7 & 2.3 & 2.6 \\ \hline
		Residual population $1-P_{\ket{g},\mathrm{reset}}$ $(\%)$								& 0.53 & 0.07 & 0.40 & 0.47 & 0.13 & 0.08 & 0.10 & 0.19 & 0.15 & 0.06 & 0.35 & 0.15 & 0.03 & 0.05 & 0.23 & 0.19 \\ \hline
		Single-qubit RB error $\epsilon_{\mathrm{1Q}}$ $(\%)$						& 0.06 & 0.08 & 0.20 & 0.18 & 0.15 & 0.10 & 0.20 & 0.03 & 0.06 & 0.07 & 0.28 & 0.11 & 0.08 & 0.13 & 0.12 & 0.05 \\ \hline
		Lifetime $T_{\mathrm{1}}$ $(\SI{}{\micro\second})$ 						& 98.1 & 17.5 & 39.0 & 16.9 & 30.0 & 71.7 & 29.0 & 64.7 & 68.6 & 39.5 & 40.8 & 21.1 & 31.6 & 70.6 & 31.6 & 48.5 \\ \hline
		Ramsey decay time $T^*_{\mathrm{2}}$ $(\SI{}{\micro\second})$				& 74.0 & 17.3 & 42.2 & 16.5 & 10.6 & 46.0 & 9.4 & 45.4 & 34.2 & 8.5 & 13.7 & 11.6 & 10.0 & 57.2 & 30.6 & 44.8 \\ \hline
		Echo decay time $T^{\mathrm{E}}_{\mathrm{2}}$ $(\SI{}{\micro\second})$		& 101.3 & 14.7 & 50.6 & 27.9 & 24.9 & 63.0 & 14.9 & 50.3 & 47.2 & 18.1 & 26.3 & 15.8 & 41.5 & 51.8 & 31.4 & 46.6 \\ \hline
	\end{tabular}
	\caption{
		Qubit parameters and performance metrics.
		The qubits are indexed as indicated in \cref{fig:app_mapping}\figpanel{a}.
		The indicated ground state population $P_{\ket{g},\mathrm{reset}}$
		is the average over resetting from one of the lowest three energy levels;
		the largest quoted values of $1-P_{\ket{g},\mathrm{reset}}$ are limited by readout errors in the ground state.
	}
	\label{tab:device_params}
\end{table*}

\begin{table*}[]
	\begin{tabular}{|l|l|l|l|l|l|l|l|l|l|l|l|l|l|l|l|l|}
		\hline
		\textbf{Qubit pair}	& 2,6 & 7,8 & 10,9 & 15,11 & 2,3 & 5,9 & 12,8 & 15,14 & 7,3 & 12,11 & 13,9 & 4,8 & 5,6 & 10,14 & 5,1 & 12,16 \\ \hline
		Interaction frequency $\omega_{\mathrm{int}}/2\pi$ $(\SI{}{\giga\hertz})$		& 4.56 & 5.74 & 5.85 & 4.60 & 5.15 & 4.72 & 6.01 & 4.93 & 5.37 & 6.00 & 5.59 & 5.51 & 5.81 & 5.66 & 4.72 & 6.00 \\ \hline
		\begin{tabular}{@{}c@{}}Echo decay time at interaction freq. \\ (high-frequency qubit) $T^{\mathrm{E}}_{\mathrm{2,H}}$ $(\SI{}{\micro\second})$\end{tabular}	& 3.5 & 4.4 & 4.2 & 3.7 & 3.3 & 2.9 & 4.6 & 4.0 & 4.1 & 4.4 & 4.6 & 2.5 & 3.8 & 3.4 & 2.9 & 4.5 \\ \hline
		\begin{tabular}{@{}c@{}}Echo decay time at interaction freq. \\ (low-frequency qubit) $T^{\mathrm{E}}_{\mathrm{2,L}}$ $(\SI{}{\micro\second})$\end{tabular}	& 3.8 & 3.1 & 4.5 & 2.8 & 4.5 & 2.4 & 5.5 & 3.0 & 4.8 & 5.5 & 2.8 & 3.1 & 9.7 & 4.0 & 2.4 & 5.6 \\ \hline
		Interaction time $t_{\mathrm{int}}$ $(\SI{}{\nano\second})$ 						& 62.3 & 49.5 & 48.8 & 62.4 & 53.1 & 58.1 & 46.6 & 55.8 & 51.3 & 47.0 & 53.5 & 54.5 & 47.6 & 49.1 & 60.0 & 49.1 \\ \hline
		Total duration $t_{\mathrm{tot}}$ $(\SI{}{\nano\second})$ 						& 89.6 & 76.2 & 75.8 & 90.0 & 80.0 & 85.8 & 74.2 & 82.9 & 78.3 & 74.6 & 81.2 & 81.2 & 74.6 & 76.2 & 87.1 & 76.2 \\ \hline
	\end{tabular}
	\caption{
		Parameters and performance metrics of two-qubit gates.
		The qubits are indexed as indicated in \cref{fig:app_mapping}\figpanel{a}.
		In each pair, the first index indicates the high-frequency qubit
		involved in the gate and the second index the low-frequency qubit.
		The interaction frequency $\omega_{\mathrm{int}}/2\pi$ indicates
		the first-excited-state frequency of the low-frequency qubit during the gate.
		The interaction time $t_{\mathrm{int}}$ indicates the sum of the durations of the two frequency excursions.
		The total gate duration $t_{\mathrm{tot}}$
		additionally includes the idle time between the frequency excursions
		as well as the buffer times before and after the gates~\cite{Scarato2025}.
	}
	\label{tab:device_params_pairs}
\end{table*}

We implement single-qubit gates
using the derivative removal by adiabatic gates (DRAG) approach~\cite{Motzoi2009},
using a single-period cosine-squared pulse envelope with a duration of $\SI{40}{\nano\second}$.
After calibrating $Y$-type $\pi$ and $\pi/2$ rotations as in Ref.~\cite{Lazar2023a}
and benchmarking them with randomized benchmarking~\cite{Magesan2012a, Magesan2012},
we implement the continuous set of single-qubit gates
by linearly scaling the amplitudes of the $Y_\pi$ pulses for every qubit.
This approach neglects nonlinearities
of electrical components in the control lines~\cite{Lazar2023}
and can lead to
over- and under-rotation errors that impair the fidelity of the state preparation.
However, the training procedure of the hybrid neural network
includes measurements on the quantum hardware
and can, thus, implicitly adjust for potential over- and under-rotations.

We implement continuous sets of two-qubit gates
using a resonant interaction between the flux-tunable transmons,
which are capacitively coupled via far-detuned, fixed-frequency resonators.
Following the procedure in~\cite{Scarato2025},
the gates are activated by net-zero baseband flux pulses,
which create two time intervals of resonant population exchange.
This allows us to continuously parameterize
the conditional phase of each gate
by the idle time between the two population exchanges.
The net-zero pulses provide robustness
against pulse distortions on time scales longer than the gate duration,
and therefore facilitate the implementation of deep quantum circuits.

We perform frequency-multiplexed three-level single-shot readout~\cite{Magnard2018}
using a dedicated readout resonator and Purcell filter for each qubit.
The readout circuitry is grouped on four feed lines, see \cref{fig:app_device}.
The readout signal from each feedline is amplified
by a traveling-wave parametric amplifier and a cryogenic high-electron-mobility transistor amplifier,
followed by two low-noise amplifiers at room temperature.
We perform leakage rejection by excluding runs
in which leakage to the second excited state has been detected in the measurement.
For the state classification study reported in \cref{fig:results}\figpanel{c}-\figpanel{d},
we truncate all datasets to the size of
the smallest dataset after leakage rejection, which is $7036$ data points.

We reset each transmon to its ground state
using a modulated flux pulse to transfer population from the transmon into the readout resonator,
where the population decays to the readout feed line via the Purcell filter~\cite{Zhou2021}.
With this method,
we reset the population in the first or second excited state ($\ket{e}$, $\ket{f}$)
to the ground state ($\ket{g}$).
For qubits
that use the $\ket{f}$ state as an auxiliary level during two-qubit gates,
we also reset the population in the third excited state ($\ket{h}$)~\cite{Lacroix2025}.

\begin{figure}
	\centering
	\includegraphics{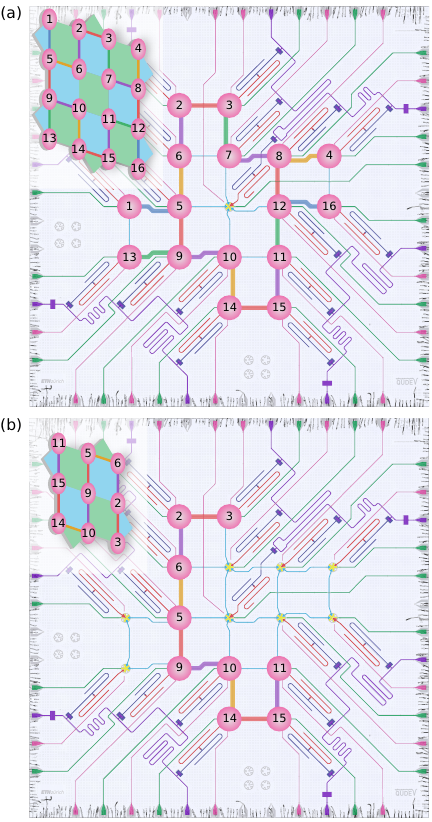}
	\caption{
		Mapping of the (a) $4 \times 4$ and (b) $3 \times 3$ surface code lattices
		to the quantum device displayed in \cref{fig:app_device}.
		The pink circles indicate the qubits, and
		the thick solid lines represent the connectivity used to perform two-qubit gates.
		The colors of these lines indicate groups of gates calibrated and executed in parallel.
	}
	\label{fig:app_mapping}
\end{figure}

Our experiments require data qubit lattices with direct connectivity.
Even though the quantum device does not include a regular $4 \times 4$ grid,
the connectivity required for experiments with the $16$-qubit surface code
can be implemented on the deformed lattice shown in \cref{fig:app_mapping}\figpanel{a}.
For $N=9$, the connectivity of a one-dimensional qubit chain is sufficient,
see \cref{fig:app_mapping}\figpanel{b}.

\section{Ground state preparation circuits}\label{app:preparation_circuits}

We prepare approximate ground states $U(\pmb{\theta})\ket{0\ldots0}$
of the Hamiltonian~\labelcref{eq:surface-code-in-field}
using parameterized quantum circuits implementing $U(\pmb{\theta})$.
Exploiting that these preparation circuits are not unique,
we choose constructions that match the hardware connectivity.
Reference~\cite{Sun2023f} proposed a variational ansatz for $U(\pmb{\theta})$
in the case of a boundary consisting of $Z$-type stabilizers of weight 2 or 3.
In contrast, we consider boundary conditions
in which the lattice of \mbox{weight-4} stabilizers is surrounded
by both $X$-type and $Z$-type \mbox{weight-2} stabilizers,
see \cref{fig:concept}\figpanel{a}.

\begin{figure}
	\centering
	\includegraphics{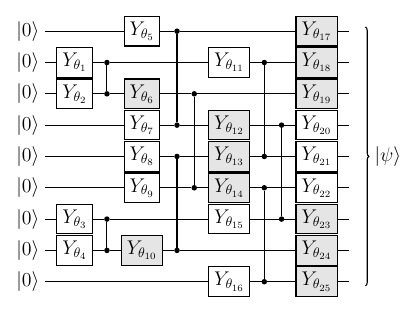}
	\caption{
		Preparation circuit for the ground state of a $3 \times 3$ surface code in a longitudinal field.
		Gates added to the circuit from~\cite{Satzinger2021} are indicated in gray.
		Qubits are ordered as in \cref{fig:concept_d3}.
	}
	\label{fig:app_prep_circuit_9_cz}
\end{figure}

As a starting point for the $3 \times 3$ surface code,
we take the quantum circuit introduced in Ref.~\cite{Satzinger2021}
to prepare the ground state of a surface code with an odd number of qubits $N$ for $h=0$.
We express the alternating layers of
controlled-NOT and Hadamard gates in this circuit
in terms of hardware-native $\mathrm{C}Z_\pi$ and $Y_{\pi/2}$ gates.
To obtain a variational circuit
for preparing ground states for field strengths $h \in \{0, 0.05,\dots,1\}$,
we replace the $Y_{\pi/2}$ gates with $Y_{\theta}$ gates
that are continuously parameterized by the angles $\pmb{\theta} = \{\theta_1,\theta_2,\ldots\}$,
and we insert additional $Y_{\theta}$ gates
such that each $\mathrm{C}Z_\pi$ gate is followed by $Y_{\theta}$ gates on both qubits,
see \cref{fig:app_prep_circuit_9_cz}.

\begin{figure}
	\centering
	\includegraphics{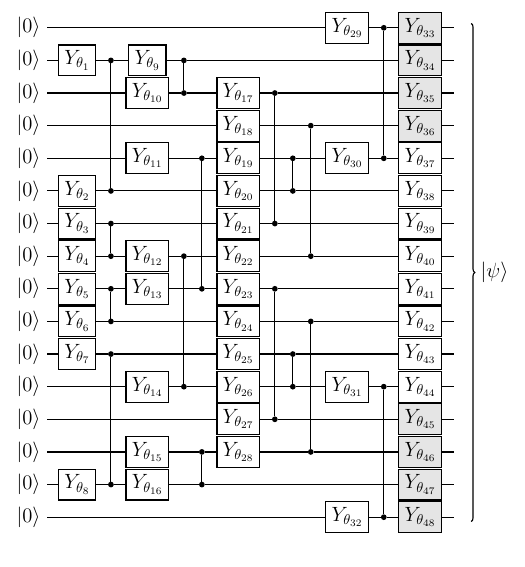}
	\caption{
		Preparation circuit for the ground state of a $4 \times 4$ surface code in a longitudinal field.
		Gates added to the exact preparation circuit for $h=0$ are indicated in gray.
		Qubits are ordered as in \cref{fig:concept}.
	}
	\label{fig:app_prep_circuit_16_cz}
\end{figure}

Extending the above construction to the even number of $N=16$ qubits
yields the circuit shown in \cref{fig:app_prep_circuit_16_cz}.
By design, the variational circuits can implement
an exact ground state preparation for $h=0$ for appropriately chosen gate parameters.

To numerically optimize the parameterized circuits $U(\pmb{\theta})$,
we minimize the energy $\langle H\rangle$ of the output state
with respect to the parameters $\pmb{\theta}$
using state-vector simulations with the L-BFGS optimizer,
as in Ref.~\cite{Herrmann2022}.
Starting this optimization from 100 random initializations for $N=16$ (50 times for $N=9$),
we select the best result $\pmb{\theta}$ in terms of the fidelity
$F = |\langle \psi_0|U(\pmb{\theta})\ket{0\ldots0}|^2$ with
the exact ground state $\ket{\psi_0}$ obtained via direct diagonalization.

\section{Kraus operator simulations}\label{app:kraus}

To model gate and idling errors,
we decompose all quantum circuits into
sequences of elementary operations represented by quantum channels of the form
\begin{align}\label{eq:kraus-decomp}
\mathcal{E}(\rho(t))= \sum_K K \rho(t) K^{\dagger},
\end{align}
where the Kraus operators $K$ acting on the quantum state $\rho(t)$
satisfy $\sum_K K^{\dagger}K = I$, with the identity matrix $I$.
Depending on the operation being modeled,
the channels below act either on single qubits or pairs of qubits.
In each case, $\rho$ denotes the density matrix of the corresponding subsystem,
while identity operations on unaffected qubits are left implicit.
The noise model described below is similar to that used in Ref.~\cite{Herrmann2022}.

We model qubit idling during a time $\Delta t$ as the concatenated channel
\begin{align}\label{eq:app_kraus_idle}
	(\mathcal{E}^{\mathrm{r}}_{\Delta t} \circ \mathcal{E}^{ZZ}_{\Delta t})(\rho).
\end{align}
where $\mathcal{E}^{\mathrm{r}}_{\Delta t}(\rho)$
models relaxation and dephasing of a single qubit
and $\mathcal{E}_{\Delta t} ^{ZZ}(\rho)$
accounts for the residual $ZZ$ couplings with its neighbors.
To decompose $\mathcal{E}^{\mathrm{r}}_{\Delta t}(\rho)$
in the form of \cref{eq:kraus-decomp} we use three Kraus operators,
\begin{align}
	K_1 &= \sqrt{\gamma_1}\sigma^{-},\nonumber\\
	K_2 &= \sqrt{\gamma_2}Z,\nonumber\\
	K_3 &= \sqrt{1-\gamma_2} \ket{0}\!\bra{0} + \sqrt{1 - \gamma_1 - \gamma_2}\ket{1}\!\bra{1},
\end{align}
where $\sigma^{-}$ is the qubit lowering operator,
$\gamma_1 = \Delta t/T_1$, and $\gamma_2 = \Delta t[1/(2T_2) - 1/(4T_1)]$,
with the coherence times $T_1$ and $T_2$, see \cref{tab:device_params}.
We express the channel accounting for the residual $ZZ$ couplings as
\begin{align}
	\mathcal{E}_{\Delta t} ^{ZZ}(\rho)= e^{-i \Delta t H_{ZZ}} \rho e^{i \Delta t H_{ZZ}},
\end{align}
with the Hamiltonian
\begin{align}\label{eq:h-zz}
	H_{ZZ}=\frac{1}{4}\sum_{\langle j, k\rangle}\alpha^{(j,k)}_{ZZ}(I_j-Z_j)(I_k-Z_k)
\end{align}
where the sum runs over
the considered pairs of neighboring qubits labeled by $j$ and $k$.
The residual coupling rates $\alpha^{(j,k)}_{ZZ}$ are estimated
from an analytical model taking into account the transmon frequencies and couplings~\cite{Krinner2020}.

Similarly, we simulate each $Y_{\theta}$ gate as the concatenated channel
\begin{align}\label{eq:app_kraus_y}
	(\mathcal{E}^{\mathrm{r}}_{\Delta t/2} \circ \mathcal{E}^{ZZ}_{\Delta t/2} \circ \mathcal{E}_Y \circ \mathcal{E}^{\mathrm{r}}_{\Delta t/2}  \circ \mathcal{E}^{ZZ}_{\Delta t/2} )(\rho),
\end{align}
which includes the ideal unitary
\begin{align}
	\mathcal{E}_Y(\rho) = Y_{\theta} \rho Y^{\dagger}_{\theta}.
\end{align}

We model a two-qubit $\mathrm{C}Z_\theta$ gate as the concatenated channel
\begin{align}
	(\mathcal{E}^{\mathrm{r}}_{\Delta t/2} \circ \mathcal{E}_{\theta} \circ \mathcal{E}^{\mathrm{r}}_{\Delta t/2}  )(\rho).
\end{align}
Here, the relaxation and dephasing channel $\mathcal{E}^{\mathrm{r}}_{\Delta t/2}$
is applied independently to each of the two qubits involved in the $\mathrm{C}Z_\theta$ gate.
To account for the reduced phase coherence during the gate~\cite{Scarato2025},
we use the echo decay times $T^{\mathrm{E}}_{\mathrm{2}}$ at the interaction frequency,
as summarized in \cref{tab:device_params_pairs}.
The unitary evolution
\begin{align}\label{eq:cz-channel-with-residual-zz}
	\mathcal{E}_{\theta}(\rho) = e^{-i(\theta|11\rangle\langle11| + \Delta t H_{ZZ})} \rho e^{i (\theta|11\rangle\langle11| + \Delta t H_{ZZ})}
\end{align}
includes the ideal $\mathrm{C}Z_\theta$ gate and residual $ZZ$ interactions.
We simulate each layer of simultaneous $\mathrm{C}Z_\theta$ gates
by applying the corresponding two-qubit channels in parallel.
The frequency excursions of each qubit involved in a $\mathrm{C}Z_\theta$ gate
cause increased $ZZ$ couplings with all neighboring qubits not involved in the gate,
which we take into account
by appropriately choosing the contributing terms from \cref{eq:h-zz} in \cref{eq:cz-channel-with-residual-zz}.

To reduce the memory requirements of the simulations for $N=16$ qubits,
we approximate the state evolution as an ensemble of $n$ stochastic quantum trajectories
\begin{align}
	\rho(t) = \frac{1}{n}\sum_{j=1}^n\ket{\psi_j(t)}\!\bra{\psi_j(t)}.
\end{align}
Each trajectory is propagated according to
\begin{align}
	\ket{\psi_j(t+\Delta t)} = K\ket{\psi_j(t)}/\sqrt{P^{j}_{K}},
\end{align}
where the Kraus operator $K$ is randomly drawn with probability
\begin{align}
	P^{j}_{K} = \bra{\psi_j(t)}K^{\dagger}K\ket{\psi_j(t)}.
\end{align}
We estimate a linear function of the quantum state,
such as the expectation value of an observable, as an ensemble average over all trajectories.
We use $n=10^4$ trajectories in all Kraus operator simulations.

\begin{figure}
	\centering
	\includegraphics{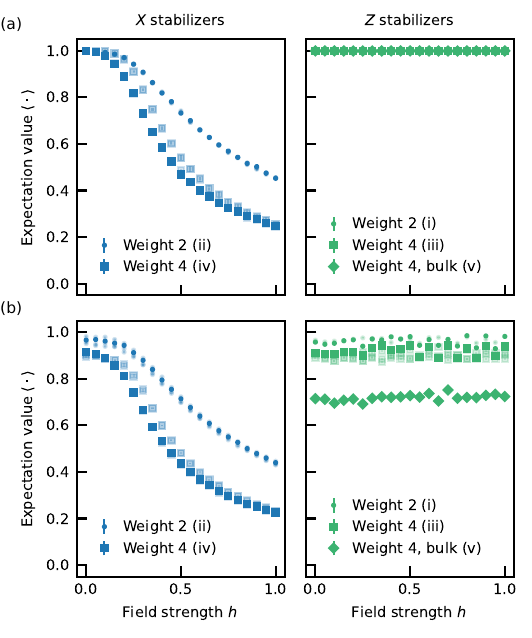}
	\caption{
		Simulated expectation values of the stabilizers
		labeled in \cref{fig:preparation}\figpanel{a} (opaque markers)
		and the other stabilizers (transparent),
		as a function of the field strength $h$.
		\figpanel{a}~Noiseless simulations.
		\figpanel{b}~Kraus operator simulations.
	}
	\label{fig:app_preparation_sim_stab}
\end{figure}

Based on the noiseless simulations and the Kraus operator simulations,
we calculate the expectation values of the stabilizers of the prepared states for $N=16$ qubits,
see \cref{fig:app_preparation_sim_stab}\figpanel{a} and \figpanel{b}, respectively.
The simulated effect of gate and idling errors
is consistent with the reduced contrast observed in the experimental results
in \cref{fig:preparation}\figpanel{b}-\figpanel{c}.
The weight-four stabilizers, and in particular the bulk $Z$ stabilizer
labeled (v) in \cref{fig:preparation}\figpanel{a},
are most affected by errors that occur during state preparation.

\begin{figure}
	\centering
	\includegraphics{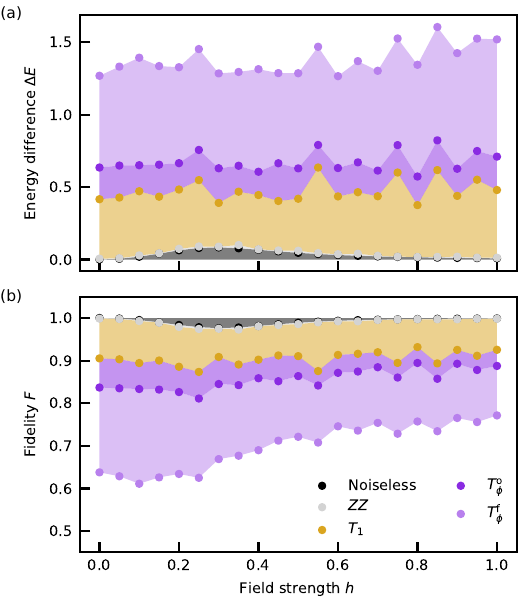}
	\caption{
		Energy and fidelity of the states prepared as a function of the field strength $h$,
		estimated from simulations that cumulatively include
		residual $ZZ$ couplings, relaxation ($T_1$),
		dephasing due to other sources than flux noise ($T_{\phi}^\mathrm{o}$),
		and dephasing due to flux noise ($T_{\phi}^\mathrm{f}$).
		\figpanel{a}~Difference between the dimensionless energy of the prepared state and of the exact ground state.
		\figpanel{b}~Fidelity of the prepared state with the exact ground state.
	}
	\label{fig:app_sim_breakdown}
\end{figure}

To quantify the contribution of different error channels,
we simulate the state preparation for $N=16$ qubits
while modeling only subsets of the error channels.
We then compute the difference $\Delta E = \langle H \rangle - E_0$
from the exact ground state energy $E_0$
as well as the fidelity $F$ with the exact ground state,
see the plots as a function of field strength $h$
in \cref{fig:app_sim_breakdown}\figpanel{a} and \figpanel{b}, respectively.
The simulations reveal dephasing,
in particular due to sensitivity to flux noise during the two-qubit gates,
as the dominant error mechanism.
In the noiseless case, shown as black markers,
the algorithmic errors due to the limited expressivity of the preparation circuit vanish for $h=0$ and for large $h$.
The largest algorithmic errors, which we observe at $h \approx 0.3$,
are bounded by $\Delta E<0.1$ and $1-F<\SI{3}{\%}$.
These values are substantially smaller than
the average energy difference of $\Delta E_\mathrm{av} = 1.38$
and average infidelity of $1-F_\mathrm{av}= \SI{30.1}{\%}$
in the presence of gate and idling errors.

\section{Joint optimization of the quantum circuit and the classical neural network}\label{app:training}

We train the hybrid neural network on the topological ground state for $h=0$
and a featureless ensemble of product states $\mathcal{P}^{\otimes N}$,
where $\mathcal{P}=\{|0\rangle,|1\rangle,|+\rangle,|-\rangle,|+i\rangle,|-i\rangle \}$
consists of the eigenstates of the Pauli operators.
Measured bit strings $\vec{x}$ are uniformly distributed
when averaged over $\mathcal{P}^{\otimes N}$,
that is, $\mathcal{P}^{\otimes N}$ forms a 1-design equivalent to the $N$-qubit maximally mixed state:
\begin{align}
	\frac{1}{6^N}\sum_{\ket{\psi}\in\mathcal{P}^{\otimes N}} \ket{\psi}\!\bra{\psi} = \bigotimes_{j=1}^N \left(\frac{1}{6}\sum_{\ket{\psi_j}\in\mathcal{P}} \ket{\psi_j}\!\bra{\psi_j} \right) = \frac{I}{2^N},
\end{align}
where $I$ is the identity operator.
Given that the mixed state is invariant under any quantum circuit,
we can emulate it using classically generated $Z$-basis measurement outcomes,
without requiring measurements on the quantum hardware.
In the small-scale implementation for $N=9$ and $N=16$,
we provide all possible bit strings to the classical neural network,
giving them equal weight to emulate the mixed state.
For larger systems, this can be replaced by sampling from
a uniform distribution using a pseudorandom number generator.

We use the binary cross-entropy cost function defined in \cref{eq:bxe}.
For the $i^{\mathrm{th}}$ measured bit string,
the cross-entropy term can be interpreted as a measure of statistical distance between
a Bernoulli distribution with probability $\ptopo_i$ and a one-hot distribution given by the binary label $L$.
If the $i^{\mathrm{th}}$ bit string occurs only for one of the two classes of states,
the optimal prediction~\cite{Arnold2022} of the neural network is $\ptopo_i=L$, yielding zero cost.
If both classes of states can produce the $i^{\mathrm{th}}$ bit string,
the optimal value of $\ptopo_i$ lies between $0$ and $1$,
reflecting the ambiguity and resulting in a nonzero cost.
For the first iteration in \cref{fig:results}\figpanel{a},
the cost is close to $\ln(2) \approx 0.69$.
This value corresponds to
the maximum entropy of a binary random variable
and indicates measurement outcomes that do not carry any information about the class of a given state.

We jointly train the parameterized quantum circuit
and the classical neural network in two nested optimization loops.
As the outer loop, we start from a random initialization of the parameterized quantum circuit
and perform a gradient descent using stochastic gradient estimation as introduced in~\cite{Salimans2017}.
In each iteration, we explore the cost function landscape
around the current parameters of the quantum gates $\pmb{\theta}$
by eight stochastic perturbations $\delta\pmb{\theta}_i$
drawn from a Gaussian distribution with standard deviation $s=0.03\pi$.
We prepare the waveforms corresponding
to these eight perturbations $\delta\pmb{\theta}_i$ and upload them to the
control instruments at once.
The instruments then autonomously perform $8192$ single-shot measurements for each $i$
in an interleaved real-time loop.
On the resulting dataset, we perform leakage rejection as described in \cref{app:device}.

As input data for optimizing the classical neural network in the inner optimization loop,
we use the set of all unique observed bit strings,
with weighting factors given by how often each bit string has occurred.
We train the classical neural network using the Keras interface integrated in TensorFlow.
In the first iteration of the outer loop,
we perform the inner training iteration
for $1000$ epochs starting from randomly initialized parameters.
During all subsequent iterations of the outer loop,
we train the classical neural network
for $10$ epochs starting from the parameters optimized in the previous iteration.

\begin{figure}
	\centering
	\includegraphics[width=\linewidth]{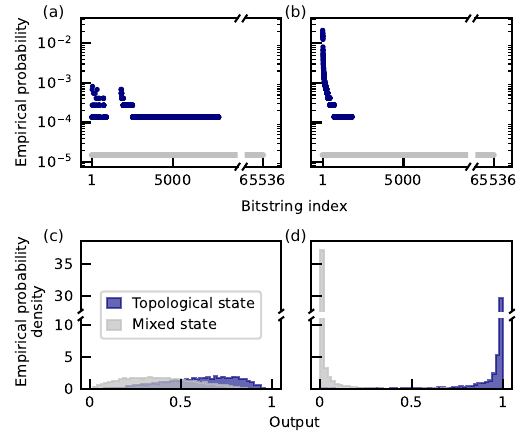}
	\caption{
		Distributions of experimental results during the joint training,
		for the first iteration (left),
		and for iteration 231 (right), which corresponds to the black arrow in \cref{fig:results}\figpanel{a}.
		The training set consists of the topological ground state
		for $h=0$ (blue) and the mixed state (gray).
		\figpanel{a}-\figpanel{b}~Bit string probabilities after the parameterized quantum circuit.
		The bit strings are sorted by the same integer index in both panels.
		\figpanel{c}-\figpanel{d}~Histogrammed output $\ptopo$
		of the trained classical neural network,
		when processing the bit strings in \figpanel{a}-\figpanel{b}.
	}
	\label{fig:app_bitstring_distr}
\end{figure}

From the trained classical neural network,
we obtain a value of the cost function $c_i$
for each set $i$ of parameters of the quantum gates.
After standardizing these values as
$\widetilde{c_i}=(c_i-\mu)/\sigma$ using the ensemble mean $\mu$ and standard deviation $\sigma$,
we estimate the gradient of $\widetilde{c}$ as~\cite{Salimans2017}
\begin{align}
	\vec{\nabla} \widetilde{c}=\frac{1}{ns}\sum_{i=1}^{n} \widetilde{c_i} \frac{\delta\pmb{\theta}_{i}}{s}
\end{align}
where $s=0.03\pi$ is the standard deviation of the stochastic perturbations $\delta\pmb{\theta}_{i}$ used in the measurements.
We then update the parameters of the quantum gates by $-\alpha\vec{\nabla} \widetilde{c}$, where we choose the learning rate $\alpha=2s^2$ to obtain a step size on the order of $2s$.

After $400$ iterations of the outer optimization loop,
we select the gate parameters corresponding to the lowest achieved cost,
as indicated by the black arrow for the best of the $10$ initializations in \cref{fig:results}\figpanel{a}.
The overlap of the empirical bit string distributions
at the input of the classical neural network when using the randomly initialized quantum circuit,
see \cref{fig:app_bitstring_distr}\figpanel{a},
reduces significantly with the optimized circuit,
see \cref{fig:app_bitstring_distr}\figpanel{b}.
This improved distinguishability results in
an increased performance of the classical neural network,
see the reduction of overlap of the output distributions
from \cref{fig:app_bitstring_distr}\figpanel{c} to \cref{fig:app_bitstring_distr}\figpanel{d}.
For the chosen training set, the cost function in \cref{eq:bxe}
is minimized by a quantum circuit that implements
the inverse of the preparation of the topological state
because such a circuit maps the topological state to the $|0 \dots 0\rangle$ product state,
which has support on a single bit string and thus minimal overlap with the maximally mixed state.

\section{Variations of the final training step}\label{app:variants}

During the joint training, the hybrid neural network is trained
on a generic training set consisting of
a topological ground state and a maximally mixed state,
with equal weights in the cost function
as summarized in the first column of \cref{tab:datasets}.
The classical neural network can then be fine-tuned for a concrete use case
by performing a final training step with a modified training set
while keeping the quantum circuit unchanged.
This final training step can be interpreted as transfer learning~\cite{Mari2020}.

\begin{table}
	\begin{tabular}{|l|l|l|l||}
		\cline{2-4}
            \multicolumn{1}{l|}{} & \multicolumn{1}{l|}{\cref{fig:app_variants_datasets}\figpanel{a}-\figpanel{b}} & \multicolumn{1}{l|}{\cref{fig:results}\figpanel{b}-\figpanel{c}} & \multicolumn{1}{l|}{\cref{fig:app_variants_datasets}\figpanel{c}-\figpanel{d}} \\
		\cline{1-4}
		\multicolumn{1}{|l|}{\begin{tabular}{@{}c@{}}Topological\\ state ($h=0$)\end{tabular}} 			& \multicolumn{1}{c|}{\checkmark} 	& \multicolumn{1}{c|}{\checkmark} 	& \multicolumn{1}{c|}{\checkmark}  \\ \cline{1-4}
		\multicolumn{1}{|l|}{Mixed state} 																& \multicolumn{1}{c|}{\checkmark} 	& \multicolumn{1}{c|}{\checkmark} 	& \multicolumn{1}{c|}{}  \\ \cline{1-4}
		\multicolumn{1}{|l|}{\begin{tabular}{@{}c@{}}Paramagnetic\\ state ($h\to\infty$)\end{tabular}} 	& \multicolumn{1}{c|}{} 	    & \multicolumn{1}{c|}{\checkmark}     & \multicolumn{1}{c|}{\checkmark}  \\ \cline{1-4}
	\end{tabular}
	\caption{
		Variations of the training sets (columns)
		used for the final training of the classical neural network.
		The topological ground state belongs to
		the part of the training set labeled as $\mathcal{T}_1$ (topological) in \cref{eq:bxe},
		the other two states to $\mathcal{T}_0$ (trivial).
	}
	\label{tab:datasets}
\end{table}

\begin{figure*}
	\centering
	\includegraphics[trim={0 0.1cm 0 0.cm},clip]{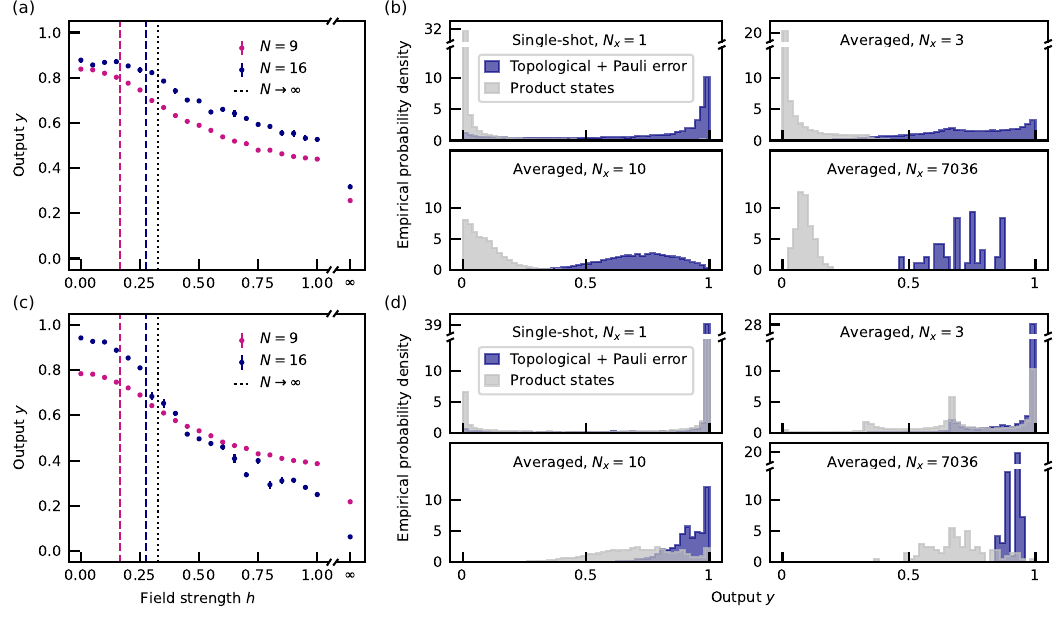}
	\caption{
		Benchmarking results after using a training set
		that contains a single representative of the trivial states
		in the final training step of the classical neural network:
		the maximally mixed state (top) or the paramagnetic ground state (bottom).
		\figpanel{a},~\figpanel{c}~Averaged output
		of the trained hybrid neural networks
		when sweeping the field strength $h$
		that parameterizes the ground states provided at the input,
		plotted for $N=16$ (blue) and $N=9$ (pink).
		The vertical lines mark the expected phase crossover,
		see details in the main text.
		\figpanel{b},~\figpanel{d}~State classification results for $N=16$ qubits.
		Input states are the topological ground state for $h=0$
		with a single-qubit Pauli error (blue) and random product states (gray);
		the output~$\ptopo$ is histogrammed
		after averaging by groups of $\nb\in\{1, 3, 10, 7036\}$ experimental runs.
	}
	\label{fig:app_variants_datasets}
\end{figure*}

Without a transfer learning step, i.e.,
when keeping the training set fixed to the one
shown in the first column of \cref{tab:datasets},
the output of the hybrid neural network smoothly transitions
from $\ptopo\approx 0.9$ for the topological ground state
to $\ptopo\approx 0.3$ for the paramagnetic ground state,
see the sweep of the field strength $h$
in \cref{fig:app_variants_datasets}\figpanel{a}.
The hybrid neural network generalizes well to random product states and topological states subject to local errors,
see \cref{fig:app_variants_datasets}\figpanel{b},
in which case $\nb=8$ measurements are necessary to reach a classification accuracy of $\SI{99}{\%}$.
This performance is achieved without access to any measurements of
the locally-easy phase during the entire training.

To obtain the results presented in the main text,
we perform the transfer learning step by including
measured data for the paramagnetic state in the training set,
as summarized in the second column of \cref{tab:datasets}.
Compared to training without transfer learning,
the contrast of the output as a function of field strength increases significantly,
see \cref{fig:results}\figpanel{b}.
This improvement is achieved while retaining the performance
in distinguishing random product states from topological states subject to local errors,
see \cref{fig:results}\figpanel{c},
with only a marginal increase to $\nb=10$ measurements
necessary to reach a classification accuracy of $\SI{99}{\%}$.

When instead using a training set that includes the paramagnetic state as the only trivial state,
see the last column of \cref{tab:datasets},
the transfer learning step yields a further increased contrast in a sweep of the field strength $h$,
as shown in \cref{fig:app_variants_datasets}\figpanel{c}.
However, we observe a large overlap in
the output distributions when preparing random product states and topological states subject to local errors,
see \cref{fig:app_variants_datasets}\figpanel{d}.
This indicates that the hybrid neural network
has specifically learned the features of the paramagnetic state,
which is beneficial for the task of recognizing quantum phases across the ground-state phase diagram,
but does not generalize well to recognizing other trivial states.

The training set used in the main text
prevents the training procedure from focusing on properties of a particular trivial state,
resulting in a versatile hybrid neural network,
which is capable of both recognizing states in a magnetic field and states subject to local errors.

\section{Hybrid neural network design}\label{app:circuits}

\begin{figure*}
	\centering
	\includegraphics{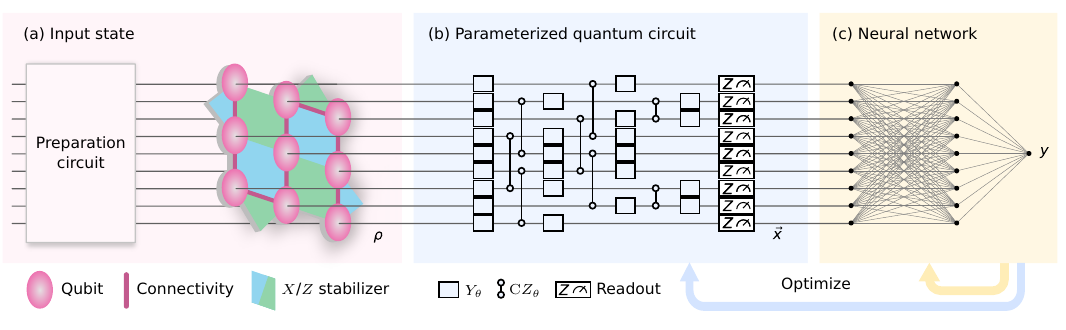}
	\caption{
		Architecture of the hybrid neural network for the $3 \times 3$ surface code.
		\figpanel{a}~The input state $\rho=\ket{\psi}\!\bra{\psi}$ to a hybrid neural network is
		prepared on a lattice of nine qubits (pink circles).
		The connectivity used in the experiment is indicated as pink lines.
		The blue and green polygons indicate the $X$ and $Z$ stabilizers
		of the $3 \times 3$ surface code, whose ground state is
		used as an example of a topological input state.
		\figpanel{b}~Parameterized quantum circuit used in the experiment,
		consisting of arbitrary-angle $Y_\theta$ gates and controlled arbitrary-phase $\cz_\theta$ gates,
		followed by measurements in the computational basis.
		\figpanel{c}~Classical neural network
		mapping the measured bit string $\vec{x}=x_1x_2...x_N$ to the output~$\ptopo$.
		The nested loops of the joint optimization of the quantum circuit and the classical neural network
		are indicated by the two arrows below the blue and yellow boxes.
	}
	\label{fig:concept_d3}
\end{figure*}

In analogy to the hybrid neural network for the $4 \times 4$ surface code described in the main text,
the architecture for the $3 \times 3$ surface code consists of
the inverse of the preparation circuit in \cref{fig:app_prep_circuit_9_cz}
followed by a feedforward neural network.
The latter is constructed with
nine neurons in both the input and hidden layers and a single neuron in the output layer,
as shown in \cref{fig:concept_d3}.

To obtain the extrapolation of the gate sequence duration to a lattice with $15 \times 15$ sites
given in \cref{sec:outlook},
we consider a surface code ground state with an arbitrary odd number of qubits $N$
prepared by the circuit described in the supplementary material of Ref.~\cite{Satzinger2021}.
This circuit can be decomposed into three controlled-NOT gates and one Hadamard gate
for each of the $(\sqrt{N}-1)^2/2$ \mbox{weight-4} $X$ stabilizers,
and one controlled-NOT gate and one Hadamard gate
for each of the $\sqrt{N}-1$ \mbox{weight-2} $X$ stabilizers.
Mapping these operations to
the native gate set used in this paper
yields $(3N -4\sqrt{N} + 1)/2$ two-qubit $\cz_{\pi}$ gates arranged in $(\sqrt{N}+3)/2$ layers.
Including additional $Y_{\theta}$ gates
as in \cref{app:preparation_circuits}
results in $Y_{\theta}$ gates on both involved qubits before each $\cz_{\pi}$ gate
and on each of the $N$ qubits before the readout.
This yields a total of $4N -4\sqrt{N} + 1$ single-qubit $Y_{\theta}$ gates arranged in $(\sqrt{N}+5)/2$ layers.
Assuming typical gate durations of $\SI{40}{\nano\second}$ for single-qubit gates
and $\SI{80}{\nano\second}$ for two-qubit gates,
see \cref{app:device},
the total duration of the gate sequence is $(60\sqrt{N}+220)$\,$\SI{}{\nano\second}$, evaluating to $\SI{1.1}{\micro\second}$ for the example of a $15 \times 15$ surface code.
In practice, this number might increase if there are restrictions on the parallel execution of two-qubit gates.


\begin{thebibliography}{85}%
\makeatletter
\providecommand \@ifxundefined [1]{%
 \@ifx{#1\undefined}
}%
\providecommand \@ifnum [1]{%
 \ifnum #1\expandafter \@firstoftwo
 \else \expandafter \@secondoftwo
 \fi
}%
\providecommand \@ifx [1]{%
 \ifx #1\expandafter \@firstoftwo
 \else \expandafter \@secondoftwo
 \fi
}%
\providecommand \natexlab [1]{#1}%
\providecommand \enquote  [1]{``#1''}%
\providecommand \bibnamefont  [1]{#1}%
\providecommand \bibfnamefont [1]{#1}%
\providecommand \citenamefont [1]{#1}%
\providecommand \href@noop [0]{\@secondoftwo}%
\providecommand \href [0]{\begingroup \@sanitize@url \@href}%
\providecommand \@href[1]{\@@startlink{#1}\@@href}%
\providecommand \@@href[1]{\endgroup#1\@@endlink}%
\providecommand \@sanitize@url [0]{\catcode `\\12\catcode `\$12\catcode
  `\&12\catcode `\#12\catcode `\^12\catcode `\_12\catcode `\%12\relax}%
\providecommand \@@startlink[1]{}%
\providecommand \@@endlink[0]{}%
\providecommand \url  [0]{\begingroup\@sanitize@url \@url }%
\providecommand \@url [1]{\endgroup\@href {#1}{\urlprefix }}%
\providecommand \urlprefix  [0]{URL }%
\providecommand \Eprint [0]{\href }%
\providecommand \doibase [0]{https://doi.org/}%
\providecommand \selectlanguage [0]{\@gobble}%
\providecommand \bibinfo  [0]{\@secondoftwo}%
\providecommand \bibfield  [0]{\@secondoftwo}%
\providecommand \translation [1]{[#1]}%
\providecommand \BibitemOpen [0]{}%
\providecommand \bibitemStop [0]{}%
\providecommand \bibitemNoStop [0]{.\EOS\space}%
\providecommand \EOS [0]{\spacefactor3000\relax}%
\providecommand \BibitemShut  [1]{\csname bibitem#1\endcsname}%
\let\auto@bib@innerbib\@empty
\bibitem [{\citenamefont {Carleo}\ \emph {et~al.}(2019)\citenamefont {Carleo},
  \citenamefont {Cirac}, \citenamefont {Cranmer}, \citenamefont {Daudet},
  \citenamefont {Schuld}, \citenamefont {Tishby}, \citenamefont
  {Vogt-Maranto},\ and\ \citenamefont {Zdeborov\'{a}}}]{Carleo2019}%
  \BibitemOpen
  \bibfield  {author} {\bibinfo {author} {\bibfnamefont {G.}~\bibnamefont
  {Carleo}}, \bibinfo {author} {\bibfnamefont {I.}~\bibnamefont {Cirac}},
  \bibinfo {author} {\bibfnamefont {K.}~\bibnamefont {Cranmer}}, \bibinfo
  {author} {\bibfnamefont {L.}~\bibnamefont {Daudet}}, \bibinfo {author}
  {\bibfnamefont {M.}~\bibnamefont {Schuld}}, \bibinfo {author} {\bibfnamefont
  {N.}~\bibnamefont {Tishby}}, \bibinfo {author} {\bibfnamefont
  {L.}~\bibnamefont {Vogt-Maranto}},\ and\ \bibinfo {author} {\bibfnamefont
  {L.}~\bibnamefont {Zdeborov\'{a}}},\ }\bibfield  {title} {\bibinfo {title}
  {Machine learning and the physical sciences},\ }\href
  {https://doi.org/10.1103/RevModPhys.91.045002} {\bibfield  {journal}
  {\bibinfo  {journal} {Review of Modern Physics}\ }\textbf {\bibinfo {volume}
  {91}},\ \bibinfo {pages} {045002} (\bibinfo {year} {2019})}\BibitemShut
  {NoStop}%
\bibitem [{\citenamefont {Wittek}(2014)}]{Wittek2014}%
  \BibitemOpen
  \bibfield  {author} {\bibinfo {author} {\bibfnamefont {P.}~\bibnamefont
  {Wittek}},\ }\href {https://doi.org/10.1016/C2013-0-19170-2} {\emph {\bibinfo
  {title} {Quantum machine learning: what quantum computing means to data
  mining}}}\ (\bibinfo  {publisher} {Elsevier},\ \bibinfo {year}
  {2014})\BibitemShut {NoStop}%
\bibitem [{\citenamefont {Biamonte}\ \emph {et~al.}(2017)\citenamefont
  {Biamonte}, \citenamefont {Wittek}, \citenamefont {Pancotti}, \citenamefont
  {Rebentrost}, \citenamefont {Wiebe},\ and\ \citenamefont
  {Lloyd}}]{Biamonte2017}%
  \BibitemOpen
  \bibfield  {author} {\bibinfo {author} {\bibfnamefont {J.}~\bibnamefont
  {Biamonte}}, \bibinfo {author} {\bibfnamefont {P.}~\bibnamefont {Wittek}},
  \bibinfo {author} {\bibfnamefont {N.}~\bibnamefont {Pancotti}}, \bibinfo
  {author} {\bibfnamefont {P.}~\bibnamefont {Rebentrost}}, \bibinfo {author}
  {\bibfnamefont {N.}~\bibnamefont {Wiebe}},\ and\ \bibinfo {author}
  {\bibfnamefont {S.}~\bibnamefont {Lloyd}},\ }\bibfield  {title} {\bibinfo
  {title} {Quantum machine learning},\ }\href
  {http://dx.doi.org/10.1038/nature23474} {\bibfield  {journal} {\bibinfo
  {journal} {Nature}\ }\textbf {\bibinfo {volume} {549}},\ \bibinfo {pages}
  {195} (\bibinfo {year} {2017})}\BibitemShut {NoStop}%
\bibitem [{\citenamefont {Cerezo}\ \emph {et~al.}(2022)\citenamefont {Cerezo},
  \citenamefont {Verdon}, \citenamefont {Huang}, \citenamefont {Cincio},\ and\
  \citenamefont {Coles}}]{Cerezo2022a}%
  \BibitemOpen
  \bibfield  {author} {\bibinfo {author} {\bibfnamefont {M.}~\bibnamefont
  {Cerezo}}, \bibinfo {author} {\bibfnamefont {G.}~\bibnamefont {Verdon}},
  \bibinfo {author} {\bibfnamefont {H.-Y.}\ \bibnamefont {Huang}}, \bibinfo
  {author} {\bibfnamefont {L.}~\bibnamefont {Cincio}},\ and\ \bibinfo {author}
  {\bibfnamefont {P.~J.}\ \bibnamefont {Coles}},\ }\bibfield  {title} {\bibinfo
  {title} {Challenges and opportunities in quantum machine learning},\ }\href
  {https://doi.org/10.1038/s43588-022-00311-3} {\bibfield  {journal} {\bibinfo
  {journal} {Nature Computational Science}\ }\textbf {\bibinfo {volume} {2}},\
  \bibinfo {pages} {567} (\bibinfo {year} {2022})}\BibitemShut {NoStop}%
\bibitem [{\citenamefont {Xia}\ and\ \citenamefont {Kais}(2018)}]{Xia2018a}%
  \BibitemOpen
  \bibfield  {author} {\bibinfo {author} {\bibfnamefont {R.}~\bibnamefont
  {Xia}}\ and\ \bibinfo {author} {\bibfnamefont {S.}~\bibnamefont {Kais}},\
  }\bibfield  {title} {\bibinfo {title} {Quantum machine learning for
  electronic structure calculations},\ }\href
  {https://doi.org/10.1038/s41467-018-06598-z} {\bibfield  {journal} {\bibinfo
  {journal} {Nat. Comm.}\ }\textbf {\bibinfo {volume} {9}},\ \bibinfo {pages}
  {4195} (\bibinfo {year} {2018})}\BibitemShut {NoStop}%
\bibitem [{\citenamefont {Guan}\ \emph {et~al.}(2021)\citenamefont {Guan},
  \citenamefont {Perdue}, \citenamefont {Pesah}, \citenamefont {Schuld},
  \citenamefont {Terashi}, \citenamefont {Vallecorsa}, \citenamefont {Vlimant},
  \citenamefont {Guan}, \citenamefont {Perdue}, \citenamefont {Pesah},
  \citenamefont {Schuld}, \citenamefont {Terashi}, \citenamefont {Vallecorsa},\
  and\ \citenamefont {Vlimant}}]{Guan2020}%
  \BibitemOpen
  \bibfield  {author} {\bibinfo {author} {\bibfnamefont {W.}~\bibnamefont
  {Guan}}, \bibinfo {author} {\bibfnamefont {G.}~\bibnamefont {Perdue}},
  \bibinfo {author} {\bibfnamefont {A.}~\bibnamefont {Pesah}}, \bibinfo
  {author} {\bibfnamefont {M.}~\bibnamefont {Schuld}}, \bibinfo {author}
  {\bibfnamefont {K.}~\bibnamefont {Terashi}}, \bibinfo {author} {\bibfnamefont
  {S.}~\bibnamefont {Vallecorsa}}, \bibinfo {author} {\bibfnamefont {J.-R.}\
  \bibnamefont {Vlimant}}, \bibinfo {author} {\bibfnamefont {W.}~\bibnamefont
  {Guan}}, \bibinfo {author} {\bibfnamefont {G.}~\bibnamefont {Perdue}},
  \bibinfo {author} {\bibfnamefont {A.}~\bibnamefont {Pesah}}, \bibinfo
  {author} {\bibfnamefont {M.}~\bibnamefont {Schuld}}, \bibinfo {author}
  {\bibfnamefont {K.}~\bibnamefont {Terashi}}, \bibinfo {author} {\bibfnamefont
  {S.}~\bibnamefont {Vallecorsa}},\ and\ \bibinfo {author} {\bibfnamefont
  {J.-R.}\ \bibnamefont {Vlimant}},\ }\bibfield  {title} {\bibinfo {title}
  {Quantum machine learning in high energy physics},\ }\href
  {https://doi.org/10.1088/2632-2153/abc17d} {\bibfield  {journal} {\bibinfo
  {journal} {Machine Learning: Science and Technology}\ }\textbf {\bibinfo
  {volume} {2}},\ \bibinfo {pages} {011003} (\bibinfo {year}
  {2021})}\BibitemShut {NoStop}%
\bibitem [{\citenamefont {Cong}\ \emph {et~al.}(2019)\citenamefont {Cong},
  \citenamefont {Choi},\ and\ \citenamefont {Lukin}}]{Cong2019}%
  \BibitemOpen
  \bibfield  {author} {\bibinfo {author} {\bibfnamefont {I.}~\bibnamefont
  {Cong}}, \bibinfo {author} {\bibfnamefont {S.}~\bibnamefont {Choi}},\ and\
  \bibinfo {author} {\bibfnamefont {M.~D.}\ \bibnamefont {Lukin}},\ }\bibfield
  {title} {\bibinfo {title} {Quantum convolutional neural networks},\ }\href
  {https://doi.org/https://doi.org/10.1038/s41567-019-0648-8} {\bibfield
  {journal} {\bibinfo  {journal} {Nature Physics}\ }\textbf {\bibinfo {volume}
  {15}},\ \bibinfo {pages} {1273} (\bibinfo {year} {2019})}\BibitemShut
  {NoStop}%
\bibitem [{\citenamefont {Bravo-Prieto}\ \emph {et~al.}(2020)\citenamefont
  {Bravo-Prieto}, \citenamefont {Lumbreras-Zarapico}, \citenamefont
  {Tagliacozzo},\ and\ \citenamefont {Latorre}}]{Bravo-Prieto2020}%
  \BibitemOpen
  \bibfield  {author} {\bibinfo {author} {\bibfnamefont {C.}~\bibnamefont
  {Bravo-Prieto}}, \bibinfo {author} {\bibfnamefont {J.}~\bibnamefont
  {Lumbreras-Zarapico}}, \bibinfo {author} {\bibfnamefont {L.}~\bibnamefont
  {Tagliacozzo}},\ and\ \bibinfo {author} {\bibfnamefont {J.~I.}\ \bibnamefont
  {Latorre}},\ }\bibfield  {title} {\bibinfo {title} {Scaling of variational
  quantum circuit depth for condensed matter systems},\ }\href
  {https://doi.org/10.22331/q-2020-05-28-272} {\bibfield  {journal} {\bibinfo
  {journal} {Quantum}\ }\textbf {\bibinfo {volume} {4}},\ \bibinfo {pages}
  {272} (\bibinfo {year} {2020})}\BibitemShut {NoStop}%
\bibitem [{\citenamefont {Wu}\ \emph {et~al.}(2024)\citenamefont {Wu},
  \citenamefont {Rossi}, \citenamefont {Vicentini}, \citenamefont
  {Astrakhantsev}, \citenamefont {Becca}, \citenamefont {Cao}, \citenamefont
  {Carrasquilla}, \citenamefont {Ferrari}, \citenamefont {Georges},
  \citenamefont {Hibat-Allah}, \citenamefont {Imada}, \citenamefont
  {L{\"a}uchli}, \citenamefont {Mazzola}, \citenamefont {Mezzacapo},
  \citenamefont {Millis}, \citenamefont {Moreno}, \citenamefont {Neupert},
  \citenamefont {Nomura}, \citenamefont {Nys}, \citenamefont {Parcollet},
  \citenamefont {Pohle}, \citenamefont {Romero}, \citenamefont {Schmid},
  \citenamefont {Silvester}, \citenamefont {Sorella}, \citenamefont {Tocchio},
  \citenamefont {Wang}, \citenamefont {White}, \citenamefont {Wietek},
  \citenamefont {Yang}, \citenamefont {Yang}, \citenamefont {Zhang},\ and\
  \citenamefont {Carleo}}]{Wu2023}%
  \BibitemOpen
  \bibfield  {author} {\bibinfo {author} {\bibfnamefont {D.}~\bibnamefont
  {Wu}}, \bibinfo {author} {\bibfnamefont {R.}~\bibnamefont {Rossi}}, \bibinfo
  {author} {\bibfnamefont {F.}~\bibnamefont {Vicentini}}, \bibinfo {author}
  {\bibfnamefont {N.}~\bibnamefont {Astrakhantsev}}, \bibinfo {author}
  {\bibfnamefont {F.}~\bibnamefont {Becca}}, \bibinfo {author} {\bibfnamefont
  {X.}~\bibnamefont {Cao}}, \bibinfo {author} {\bibfnamefont {J.}~\bibnamefont
  {Carrasquilla}}, \bibinfo {author} {\bibfnamefont {F.}~\bibnamefont
  {Ferrari}}, \bibinfo {author} {\bibfnamefont {A.}~\bibnamefont {Georges}},
  \bibinfo {author} {\bibfnamefont {M.}~\bibnamefont {Hibat-Allah}}, \bibinfo
  {author} {\bibfnamefont {M.}~\bibnamefont {Imada}}, \bibinfo {author}
  {\bibfnamefont {A.~M.}\ \bibnamefont {L{\"a}uchli}}, \bibinfo {author}
  {\bibfnamefont {G.}~\bibnamefont {Mazzola}}, \bibinfo {author} {\bibfnamefont
  {A.}~\bibnamefont {Mezzacapo}}, \bibinfo {author} {\bibfnamefont
  {A.}~\bibnamefont {Millis}}, \bibinfo {author} {\bibfnamefont {J.~R.}\
  \bibnamefont {Moreno}}, \bibinfo {author} {\bibfnamefont {T.}~\bibnamefont
  {Neupert}}, \bibinfo {author} {\bibfnamefont {Y.}~\bibnamefont {Nomura}},
  \bibinfo {author} {\bibfnamefont {J.}~\bibnamefont {Nys}}, \bibinfo {author}
  {\bibfnamefont {O.}~\bibnamefont {Parcollet}}, \bibinfo {author}
  {\bibfnamefont {R.}~\bibnamefont {Pohle}}, \bibinfo {author} {\bibfnamefont
  {I.}~\bibnamefont {Romero}}, \bibinfo {author} {\bibfnamefont
  {M.}~\bibnamefont {Schmid}}, \bibinfo {author} {\bibfnamefont {J.~M.}\
  \bibnamefont {Silvester}}, \bibinfo {author} {\bibfnamefont {S.}~\bibnamefont
  {Sorella}}, \bibinfo {author} {\bibfnamefont {L.~F.}\ \bibnamefont
  {Tocchio}}, \bibinfo {author} {\bibfnamefont {L.}~\bibnamefont {Wang}},
  \bibinfo {author} {\bibfnamefont {S.~R.}\ \bibnamefont {White}}, \bibinfo
  {author} {\bibfnamefont {A.}~\bibnamefont {Wietek}}, \bibinfo {author}
  {\bibfnamefont {Q.}~\bibnamefont {Yang}}, \bibinfo {author} {\bibfnamefont
  {Y.}~\bibnamefont {Yang}}, \bibinfo {author} {\bibfnamefont {S.}~\bibnamefont
  {Zhang}},\ and\ \bibinfo {author} {\bibfnamefont {G.}~\bibnamefont
  {Carleo}},\ }\bibfield  {title} {\bibinfo {title} {Variational benchmarks for
  quantum many-body problems},\ }\href
  {https://doi.org/10.1126/science.adg9774} {\bibfield  {journal} {\bibinfo
  {journal} {Science}\ }\textbf {\bibinfo {volume} {386}},\ \bibinfo {pages}
  {296} (\bibinfo {year} {2024})}\BibitemShut {NoStop}%
\bibitem [{\citenamefont {Acampora}\ \emph {et~al.}(2025)\citenamefont
  {Acampora}, \citenamefont {Ambainis}, \citenamefont {Ares}, \citenamefont
  {Banchi}, \citenamefont {Bhardwaj}, \citenamefont {Binosi}, \citenamefont
  {Briggs}, \citenamefont {Calarco}, \citenamefont {Dunjko}, \citenamefont
  {Eisert}, \citenamefont {Ezratty}, \citenamefont {Erker}, \citenamefont
  {Fedele}, \citenamefont {Gil-Fuster}, \citenamefont {G\"arttner},
  \citenamefont {Granath}, \citenamefont {Heyl}, \citenamefont {Kerenidis},
  \citenamefont {Klusch}, \citenamefont {Kockum}, \citenamefont {Kueng},
  \citenamefont {Krenn}, \citenamefont {L\"assig}, \citenamefont {Macaluso},
  \citenamefont {Maniscalco}, \citenamefont {Marquardt}, \citenamefont
  {Michielsen}, \citenamefont {Mu\~noz Gil}, \citenamefont {M\"ussig},
  \citenamefont {Nautrup}, \citenamefont {van Nieuwenburg}, \citenamefont
  {Orus}, \citenamefont {Schmiedmayer}, \citenamefont {Schmitt}, \citenamefont
  {Slusallek}, \citenamefont {Vicentini}, \citenamefont {Weitenberg},\ and\
  \citenamefont {Wilhelm}}]{Acampora2025}%
  \BibitemOpen
  \bibfield  {author} {\bibinfo {author} {\bibfnamefont {G.}~\bibnamefont
  {Acampora}}, \bibinfo {author} {\bibfnamefont {A.}~\bibnamefont {Ambainis}},
  \bibinfo {author} {\bibfnamefont {N.}~\bibnamefont {Ares}}, \bibinfo {author}
  {\bibfnamefont {L.}~\bibnamefont {Banchi}}, \bibinfo {author} {\bibfnamefont
  {P.}~\bibnamefont {Bhardwaj}}, \bibinfo {author} {\bibfnamefont
  {D.}~\bibnamefont {Binosi}}, \bibinfo {author} {\bibfnamefont {G.~A.~D.}\
  \bibnamefont {Briggs}}, \bibinfo {author} {\bibfnamefont {T.}~\bibnamefont
  {Calarco}}, \bibinfo {author} {\bibfnamefont {V.}~\bibnamefont {Dunjko}},
  \bibinfo {author} {\bibfnamefont {J.}~\bibnamefont {Eisert}}, \bibinfo
  {author} {\bibfnamefont {O.}~\bibnamefont {Ezratty}}, \bibinfo {author}
  {\bibfnamefont {P.}~\bibnamefont {Erker}}, \bibinfo {author} {\bibfnamefont
  {F.}~\bibnamefont {Fedele}}, \bibinfo {author} {\bibfnamefont
  {E.}~\bibnamefont {Gil-Fuster}}, \bibinfo {author} {\bibfnamefont
  {M.}~\bibnamefont {G\"arttner}}, \bibinfo {author} {\bibfnamefont
  {M.}~\bibnamefont {Granath}}, \bibinfo {author} {\bibfnamefont
  {M.}~\bibnamefont {Heyl}}, \bibinfo {author} {\bibfnamefont {I.}~\bibnamefont
  {Kerenidis}}, \bibinfo {author} {\bibfnamefont {M.}~\bibnamefont {Klusch}},
  \bibinfo {author} {\bibfnamefont {A.~F.}\ \bibnamefont {Kockum}}, \bibinfo
  {author} {\bibfnamefont {R.}~\bibnamefont {Kueng}}, \bibinfo {author}
  {\bibfnamefont {M.}~\bibnamefont {Krenn}}, \bibinfo {author} {\bibfnamefont
  {J.}~\bibnamefont {L\"assig}}, \bibinfo {author} {\bibfnamefont
  {A.}~\bibnamefont {Macaluso}}, \bibinfo {author} {\bibfnamefont
  {S.}~\bibnamefont {Maniscalco}}, \bibinfo {author} {\bibfnamefont
  {F.}~\bibnamefont {Marquardt}}, \bibinfo {author} {\bibfnamefont
  {K.}~\bibnamefont {Michielsen}}, \bibinfo {author} {\bibfnamefont
  {G.}~\bibnamefont {Mu\~noz Gil}}, \bibinfo {author} {\bibfnamefont
  {D.}~\bibnamefont {M\"ussig}}, \bibinfo {author} {\bibfnamefont {H.~P.}\
  \bibnamefont {Nautrup}}, \bibinfo {author} {\bibfnamefont {E.}~\bibnamefont
  {van Nieuwenburg}}, \bibinfo {author} {\bibfnamefont {R.}~\bibnamefont
  {Orus}}, \bibinfo {author} {\bibfnamefont {J.}~\bibnamefont {Schmiedmayer}},
  \bibinfo {author} {\bibfnamefont {M.}~\bibnamefont {Schmitt}}, \bibinfo
  {author} {\bibfnamefont {P.}~\bibnamefont {Slusallek}}, \bibinfo {author}
  {\bibfnamefont {F.}~\bibnamefont {Vicentini}}, \bibinfo {author}
  {\bibfnamefont {C.}~\bibnamefont {Weitenberg}},\ and\ \bibinfo {author}
  {\bibfnamefont {F.~K.}\ \bibnamefont {Wilhelm}},\ }\bibfield  {title}
  {\bibinfo {title} {Quantum computing and artificial intelligence: status and
  perspectives},\ }\href {https://arxiv.org/abs/2505.23860} {\bibfield
  {journal} {\bibinfo  {journal} {arXiv:2505.23860}\ } (\bibinfo {year}
  {2025})}\BibitemShut {NoStop}%
\bibitem [{\citenamefont {Abbas}\ \emph {et~al.}(2021)\citenamefont {Abbas},
  \citenamefont {Sutter}, \citenamefont {Zoufal}, \citenamefont {Lucchi},
  \citenamefont {Figalli},\ and\ \citenamefont {Woerner}}]{Abbas2021}%
  \BibitemOpen
  \bibfield  {author} {\bibinfo {author} {\bibfnamefont {A.}~\bibnamefont
  {Abbas}}, \bibinfo {author} {\bibfnamefont {D.}~\bibnamefont {Sutter}},
  \bibinfo {author} {\bibfnamefont {C.}~\bibnamefont {Zoufal}}, \bibinfo
  {author} {\bibfnamefont {A.}~\bibnamefont {Lucchi}}, \bibinfo {author}
  {\bibfnamefont {A.}~\bibnamefont {Figalli}},\ and\ \bibinfo {author}
  {\bibfnamefont {S.}~\bibnamefont {Woerner}},\ }\bibfield  {title} {\bibinfo
  {title} {The power of quantum neural networks},\ }\href
  {https://doi.org/10.1038/s43588-021-00084-1} {\bibfield  {journal} {\bibinfo
  {journal} {Nature Computational Science}\ }\textbf {\bibinfo {volume} {1}},\
  \bibinfo {pages} {403} (\bibinfo {year} {2021})}\BibitemShut {NoStop}%
\bibitem [{\citenamefont {Wiersema}\ \emph {et~al.}(2020)\citenamefont
  {Wiersema}, \citenamefont {Zhou}, \citenamefont {de~Sereville}, \citenamefont
  {Carrasquilla}, \citenamefont {Kim},\ and\ \citenamefont
  {Yuen}}]{Wiersema2021}%
  \BibitemOpen
  \bibfield  {author} {\bibinfo {author} {\bibfnamefont {R.}~\bibnamefont
  {Wiersema}}, \bibinfo {author} {\bibfnamefont {C.}~\bibnamefont {Zhou}},
  \bibinfo {author} {\bibfnamefont {Y.}~\bibnamefont {de~Sereville}}, \bibinfo
  {author} {\bibfnamefont {J.~F.}\ \bibnamefont {Carrasquilla}}, \bibinfo
  {author} {\bibfnamefont {Y.~B.}\ \bibnamefont {Kim}},\ and\ \bibinfo {author}
  {\bibfnamefont {H.}~\bibnamefont {Yuen}},\ }\bibfield  {title} {\bibinfo
  {title} {Exploring entanglement and optimization within the {Hamiltonian}
  variational ansatz},\ }\href {https://doi.org/10.1103/PRXQuantum.1.020319}
  {\bibfield  {journal} {\bibinfo  {journal} {PRX Quantum}\ }\textbf {\bibinfo
  {volume} {1}},\ \bibinfo {pages} {020319} (\bibinfo {year}
  {2020})}\BibitemShut {NoStop}%
\bibitem [{\citenamefont {Rudolph}\ \emph {et~al.}(2024)\citenamefont
  {Rudolph}, \citenamefont {Lerch}, \citenamefont {Thanasilp}, \citenamefont
  {Kiss}, \citenamefont {Shaya}, \citenamefont {Vallecorsa}, \citenamefont
  {Grossi},\ and\ \citenamefont {Holmes}}]{Rudolph2024}%
  \BibitemOpen
  \bibfield  {author} {\bibinfo {author} {\bibfnamefont {M.~S.}\ \bibnamefont
  {Rudolph}}, \bibinfo {author} {\bibfnamefont {S.}~\bibnamefont {Lerch}},
  \bibinfo {author} {\bibfnamefont {S.}~\bibnamefont {Thanasilp}}, \bibinfo
  {author} {\bibfnamefont {O.}~\bibnamefont {Kiss}}, \bibinfo {author}
  {\bibfnamefont {O.}~\bibnamefont {Shaya}}, \bibinfo {author} {\bibfnamefont
  {S.}~\bibnamefont {Vallecorsa}}, \bibinfo {author} {\bibfnamefont
  {M.}~\bibnamefont {Grossi}},\ and\ \bibinfo {author} {\bibfnamefont
  {Z.}~\bibnamefont {Holmes}},\ }\bibfield  {title} {\bibinfo {title}
  {Trainability barriers and opportunities in quantum generative modeling},\
  }\href {https://doi.org/10.1038/s41534-024-00902-0} {\bibfield  {journal}
  {\bibinfo  {journal} {npj Quantum Information}\ }\textbf {\bibinfo {volume}
  {10}},\ \bibinfo {pages} {116} (\bibinfo {year} {2024})}\BibitemShut
  {NoStop}%
\bibitem [{\citenamefont {Huang}\ \emph
  {et~al.}(2022{\natexlab{a}})\citenamefont {Huang}, \citenamefont {Broughton},
  \citenamefont {Cotler}, \citenamefont {Chen}, \citenamefont {Li},
  \citenamefont {Mohseni}, \citenamefont {Neven}, \citenamefont {Babbush},
  \citenamefont {Kueng}, \citenamefont {Preskill},\ and\ \citenamefont
  {McClean}}]{Huang2021j}%
  \BibitemOpen
  \bibfield  {author} {\bibinfo {author} {\bibfnamefont {H.-Y.}\ \bibnamefont
  {Huang}}, \bibinfo {author} {\bibfnamefont {M.}~\bibnamefont {Broughton}},
  \bibinfo {author} {\bibfnamefont {J.}~\bibnamefont {Cotler}}, \bibinfo
  {author} {\bibfnamefont {S.}~\bibnamefont {Chen}}, \bibinfo {author}
  {\bibfnamefont {J.}~\bibnamefont {Li}}, \bibinfo {author} {\bibfnamefont
  {M.}~\bibnamefont {Mohseni}}, \bibinfo {author} {\bibfnamefont
  {H.}~\bibnamefont {Neven}}, \bibinfo {author} {\bibfnamefont
  {R.}~\bibnamefont {Babbush}}, \bibinfo {author} {\bibfnamefont
  {R.}~\bibnamefont {Kueng}}, \bibinfo {author} {\bibfnamefont
  {J.}~\bibnamefont {Preskill}},\ and\ \bibinfo {author} {\bibfnamefont
  {J.~R.}\ \bibnamefont {McClean}},\ }\bibfield  {title} {\bibinfo {title}
  {Quantum advantage in learning from experiments},\ }\href
  {https://doi.org/10.1126/science.abn7293} {\bibfield  {journal} {\bibinfo
  {journal} {Science}\ }\textbf {\bibinfo {volume} {376}},\ \bibinfo {pages}
  {1182} (\bibinfo {year} {2022}{\natexlab{a}})}\BibitemShut {NoStop}%
\bibitem [{\citenamefont {Huang}\ \emph {et~al.}(2021)\citenamefont {Huang},
  \citenamefont {Kueng},\ and\ \citenamefont {Preskill}}]{Huang2021f}%
  \BibitemOpen
  \bibfield  {author} {\bibinfo {author} {\bibfnamefont {H.-Y.}\ \bibnamefont
  {Huang}}, \bibinfo {author} {\bibfnamefont {R.}~\bibnamefont {Kueng}},\ and\
  \bibinfo {author} {\bibfnamefont {J.}~\bibnamefont {Preskill}},\ }\bibfield
  {title} {\bibinfo {title} {Information-theoretic bounds on quantum advantage
  in machine learning},\ }\href
  {https://doi.org/10.1103/PhysRevLett.126.190505} {\bibfield  {journal}
  {\bibinfo  {journal} {Phys. Rev. Lett.}\ }\textbf {\bibinfo {volume} {126}},\
  \bibinfo {pages} {190505} (\bibinfo {year} {2021})}\BibitemShut {NoStop}%
\bibitem [{\citenamefont {Bravyi}\ and\ \citenamefont
  {Kitaev}(1998)}]{Bravyi1998}%
  \BibitemOpen
  \bibfield  {author} {\bibinfo {author} {\bibfnamefont {S.~B.}\ \bibnamefont
  {Bravyi}}\ and\ \bibinfo {author} {\bibfnamefont {A.~Y.}\ \bibnamefont
  {Kitaev}},\ }\bibfield  {title} {\bibinfo {title} {Quantum codes on a lattice
  with boundary},\ }\href {https://arxiv.org/abs/quant-ph/9811052} {\bibfield
  {journal} {\bibinfo  {journal} {arXiv:quant-ph/9811052}\ } (\bibinfo {year}
  {1998})}\BibitemShut {NoStop}%
\bibitem [{\citenamefont {Kitaev}(2003)}]{Kitaev2003}%
  \BibitemOpen
  \bibfield  {author} {\bibinfo {author} {\bibfnamefont {A.~Y.}\ \bibnamefont
  {Kitaev}},\ }\bibfield  {title} {\bibinfo {title} {Fault-tolerant quantum
  computation by anyons},\ }\href
  {https://doi.org/10.1016/S0003-4916(02)00018-0} {\bibfield  {journal}
  {\bibinfo  {journal} {Annals of Physics}\ }\textbf {\bibinfo {volume}
  {303}},\ \bibinfo {pages} {2} (\bibinfo {year} {2003})}\BibitemShut {NoStop}%
\bibitem [{\citenamefont {Carrasquilla}\ and\ \citenamefont
  {Melko}(2017)}]{Carrasquilla2017}%
  \BibitemOpen
  \bibfield  {author} {\bibinfo {author} {\bibfnamefont {J.}~\bibnamefont
  {Carrasquilla}}\ and\ \bibinfo {author} {\bibfnamefont {R.~G.}\ \bibnamefont
  {Melko}},\ }\bibfield  {title} {\bibinfo {title} {Machine learning phases of
  matter},\ }\href {https://doi.org/10.1038/nphys4035} {\bibfield  {journal}
  {\bibinfo  {journal} {Nature Physics}\ }\textbf {\bibinfo {volume} {13}},\
  \bibinfo {pages} {431} (\bibinfo {year} {2017})}\BibitemShut {NoStop}%
\bibitem [{\citenamefont {van Nieuwenburg}\ \emph {et~al.}(2017)\citenamefont
  {van Nieuwenburg}, \citenamefont {Liu},\ and\ \citenamefont
  {Huber}}]{Nieuwenburg2017}%
  \BibitemOpen
  \bibfield  {author} {\bibinfo {author} {\bibfnamefont {E.}~\bibnamefont {van
  Nieuwenburg}}, \bibinfo {author} {\bibfnamefont {Y.-H.}\ \bibnamefont
  {Liu}},\ and\ \bibinfo {author} {\bibfnamefont {S.}~\bibnamefont {Huber}},\
  }\bibfield  {title} {\bibinfo {title} {Learning phase transitions by
  confusion},\ }\href {https://doi.org/10.1038/nphys4037} {\bibfield  {journal}
  {\bibinfo  {journal} {Nature Physics}\ }\textbf {\bibinfo {volume} {13}},\
  \bibinfo {pages} {435} (\bibinfo {year} {2017})}\BibitemShut {NoStop}%
\bibitem [{\citenamefont {Huang}\ \emph
  {et~al.}(2022{\natexlab{b}})\citenamefont {Huang}, \citenamefont {Kueng},
  \citenamefont {Torlai}, \citenamefont {Albert},\ and\ \citenamefont
  {Preskill}}]{Huang2021g}%
  \BibitemOpen
  \bibfield  {author} {\bibinfo {author} {\bibfnamefont {H.-Y.}\ \bibnamefont
  {Huang}}, \bibinfo {author} {\bibfnamefont {R.}~\bibnamefont {Kueng}},
  \bibinfo {author} {\bibfnamefont {G.}~\bibnamefont {Torlai}}, \bibinfo
  {author} {\bibfnamefont {V.~V.}\ \bibnamefont {Albert}},\ and\ \bibinfo
  {author} {\bibfnamefont {J.}~\bibnamefont {Preskill}},\ }\bibfield  {title}
  {\bibinfo {title} {Provably efficient machine learning for quantum many-body
  problems},\ }\href {https://doi.org/10.1126/science.abk3333} {\bibfield
  {journal} {\bibinfo  {journal} {Science}\ }\textbf {\bibinfo {volume}
  {377}},\ \bibinfo {pages} {eabk3333} (\bibinfo {year}
  {2022}{\natexlab{b}})}\BibitemShut {NoStop}%
\bibitem [{\citenamefont {Cong}\ \emph {et~al.}(2024)\citenamefont {Cong},
  \citenamefont {Maskara}, \citenamefont {Tran}, \citenamefont {Pichler},
  \citenamefont {Semeghini}, \citenamefont {Yelin}, \citenamefont {Choi},\ and\
  \citenamefont {Lukin}}]{Cong2024}%
  \BibitemOpen
  \bibfield  {author} {\bibinfo {author} {\bibfnamefont {I.}~\bibnamefont
  {Cong}}, \bibinfo {author} {\bibfnamefont {N.}~\bibnamefont {Maskara}},
  \bibinfo {author} {\bibfnamefont {M.~C.}\ \bibnamefont {Tran}}, \bibinfo
  {author} {\bibfnamefont {H.}~\bibnamefont {Pichler}}, \bibinfo {author}
  {\bibfnamefont {G.}~\bibnamefont {Semeghini}}, \bibinfo {author}
  {\bibfnamefont {S.~F.}\ \bibnamefont {Yelin}}, \bibinfo {author}
  {\bibfnamefont {S.}~\bibnamefont {Choi}},\ and\ \bibinfo {author}
  {\bibfnamefont {M.~D.}\ \bibnamefont {Lukin}},\ }\bibfield  {title} {\bibinfo
  {title} {Enhancing detection of topological order by local error
  correction},\ }\href {https://doi.org/10.1038/s41467-024-45584-6} {\bibfield
  {journal} {\bibinfo  {journal} {Nat. Commun.}\ }\textbf {\bibinfo {volume}
  {15}} (\bibinfo {year} {2024})}\BibitemShut {NoStop}%
\bibitem [{\citenamefont {Liu}\ \emph {et~al.}(2024)\citenamefont {Liu},
  \citenamefont {Shtengel},\ and\ \citenamefont {Pollmann}}]{Liu2024s}%
  \BibitemOpen
  \bibfield  {author} {\bibinfo {author} {\bibfnamefont {Y.-J.}\ \bibnamefont
  {Liu}}, \bibinfo {author} {\bibfnamefont {K.}~\bibnamefont {Shtengel}},\ and\
  \bibinfo {author} {\bibfnamefont {F.}~\bibnamefont {Pollmann}},\ }\bibfield
  {title} {\bibinfo {title} {Simulating two-dimensional topological quantum
  phase transitions on a digital quantum computer},\ }\href
  {https://doi.org/10.1103/PhysRevResearch.6.043256} {\bibfield  {journal}
  {\bibinfo  {journal} {Phys. Rev. Research}\ }\textbf {\bibinfo {volume}
  {6}},\ \bibinfo {pages} {043256} (\bibinfo {year} {2024})}\BibitemShut
  {NoStop}%
\bibitem [{\citenamefont {Wahl}\ \emph {et~al.}(2025)\citenamefont {Wahl},
  \citenamefont {Jankowski}, \citenamefont {Bouhon}, \citenamefont
  {Chaudhary},\ and\ \citenamefont {Slager}}]{Wahl2025}%
  \BibitemOpen
  \bibfield  {author} {\bibinfo {author} {\bibfnamefont {T.~B.}\ \bibnamefont
  {Wahl}}, \bibinfo {author} {\bibfnamefont {W.~J.}\ \bibnamefont {Jankowski}},
  \bibinfo {author} {\bibfnamefont {A.}~\bibnamefont {Bouhon}}, \bibinfo
  {author} {\bibfnamefont {G.}~\bibnamefont {Chaudhary}},\ and\ \bibinfo
  {author} {\bibfnamefont {R.-J.}\ \bibnamefont {Slager}},\ }\bibfield  {title}
  {\bibinfo {title} {Exact projected entangled pair ground states with
  topological {Euler} invariant},\ }\href
  {https://doi.org/10.1038/s41467-024-55484-4} {\bibfield  {journal} {\bibinfo
  {journal} {Nat. Commun.}\ }\textbf {\bibinfo {volume} {16}},\ \bibinfo
  {pages} {284} (\bibinfo {year} {2025})}\BibitemShut {NoStop}%
\bibitem [{\citenamefont {Teng}\ \emph {et~al.}(2025)\citenamefont {Teng},
  \citenamefont {Samajdar}, \citenamefont {Van~Kirk}, \citenamefont {Wilde},
  \citenamefont {Sachdev}, \citenamefont {Eisert}, \citenamefont {Sweke},\ and\
  \citenamefont {Najafi}}]{Teng2025}%
  \BibitemOpen
  \bibfield  {author} {\bibinfo {author} {\bibfnamefont {Y.}~\bibnamefont
  {Teng}}, \bibinfo {author} {\bibfnamefont {R.}~\bibnamefont {Samajdar}},
  \bibinfo {author} {\bibfnamefont {K.}~\bibnamefont {Van~Kirk}}, \bibinfo
  {author} {\bibfnamefont {F.}~\bibnamefont {Wilde}}, \bibinfo {author}
  {\bibfnamefont {S.}~\bibnamefont {Sachdev}}, \bibinfo {author} {\bibfnamefont
  {J.}~\bibnamefont {Eisert}}, \bibinfo {author} {\bibfnamefont
  {R.}~\bibnamefont {Sweke}},\ and\ \bibinfo {author} {\bibfnamefont
  {K.}~\bibnamefont {Najafi}},\ }\bibfield  {title} {\bibinfo {title} {Learning
  topological states from randomized measurements using variational
  tensor-network tomography},\ }\href {https://doi.org/10.1103/qm7q-w9qj}
  {\bibfield  {journal} {\bibinfo  {journal} {PRX Quantum}\ }\textbf {\bibinfo
  {volume} {6}},\ \bibinfo {pages} {040303} (\bibinfo {year}
  {2025})}\BibitemShut {NoStop}%
\bibitem [{\citenamefont {Sancho-Lorente}\ \emph {et~al.}(2022)\citenamefont
  {Sancho-Lorente}, \citenamefont {Rom\'an-Roche},\ and\ \citenamefont
  {Zueco}}]{SanchoLorente2022}%
  \BibitemOpen
  \bibfield  {author} {\bibinfo {author} {\bibfnamefont {T.}~\bibnamefont
  {Sancho-Lorente}}, \bibinfo {author} {\bibfnamefont {J.}~\bibnamefont
  {Rom\'an-Roche}},\ and\ \bibinfo {author} {\bibfnamefont {D.}~\bibnamefont
  {Zueco}},\ }\bibfield  {title} {\bibinfo {title} {Quantum kernels to learn
  the phases of quantum matter},\ }\href
  {https://doi.org/10.1103/PhysRevA.105.042432} {\bibfield  {journal} {\bibinfo
   {journal} {Phys. Rev. A}\ }\textbf {\bibinfo {volume} {105}},\ \bibinfo
  {pages} {042432} (\bibinfo {year} {2022})}\BibitemShut {NoStop}%
\bibitem [{\citenamefont {Tanji}\ \emph {et~al.}(2026)\citenamefont {Tanji},
  \citenamefont {Yano},\ and\ \citenamefont {Yamamoto}}]{Tanji2026}%
  \BibitemOpen
  \bibfield  {author} {\bibinfo {author} {\bibfnamefont {A.}~\bibnamefont
  {Tanji}}, \bibinfo {author} {\bibfnamefont {H.}~\bibnamefont {Yano}},\ and\
  \bibinfo {author} {\bibfnamefont {N.}~\bibnamefont {Yamamoto}},\ }\bibfield
  {title} {\bibinfo {title} {Quantum phase classification via partial
  tomography-based quantum hypothesis testing},\ }\href
  {https://doi.org/10.1038/s41598-025-34610-2} {\bibfield  {journal} {\bibinfo
  {journal} {Scientific Reports}\ }\textbf {\bibinfo {volume} {16}},\ \bibinfo
  {pages} {4555} (\bibinfo {year} {2026})}\BibitemShut {NoStop}%
\bibitem [{\citenamefont {Chen}\ \emph {et~al.}(2025)\citenamefont {Chen},
  \citenamefont {Wu}, \citenamefont {Yang}, \citenamefont {Xu}, \citenamefont
  {Ye}, \citenamefont {Li}, \citenamefont {Wang}, \citenamefont {Zhang},
  \citenamefont {Jin}, \citenamefont {Zhu}, \citenamefont {Gao}, \citenamefont
  {Tan}, \citenamefont {Cui}, \citenamefont {Zhang}, \citenamefont {Wang},
  \citenamefont {Zou}, \citenamefont {Li}, \citenamefont {Shen}, \citenamefont
  {Zhong}, \citenamefont {Bao}, \citenamefont {Zhu}, \citenamefont {Song},
  \citenamefont {Deng}, \citenamefont {Dong}, \citenamefont {Zhang},
  \citenamefont {Zhang}, \citenamefont {Li}, \citenamefont {Guo}, \citenamefont
  {Wang}, \citenamefont {Li}, \citenamefont {Wang}, \citenamefont {Song},\ and\
  \citenamefont {Wang}}]{Chen2025e}%
  \BibitemOpen
  \bibfield  {author} {\bibinfo {author} {\bibfnamefont {J.}~\bibnamefont
  {Chen}}, \bibinfo {author} {\bibfnamefont {Y.}~\bibnamefont {Wu}}, \bibinfo
  {author} {\bibfnamefont {Z.}~\bibnamefont {Yang}}, \bibinfo {author}
  {\bibfnamefont {S.}~\bibnamefont {Xu}}, \bibinfo {author} {\bibfnamefont
  {X.}~\bibnamefont {Ye}}, \bibinfo {author} {\bibfnamefont {D.}~\bibnamefont
  {Li}}, \bibinfo {author} {\bibfnamefont {K.}~\bibnamefont {Wang}}, \bibinfo
  {author} {\bibfnamefont {C.}~\bibnamefont {Zhang}}, \bibinfo {author}
  {\bibfnamefont {F.}~\bibnamefont {Jin}}, \bibinfo {author} {\bibfnamefont
  {X.}~\bibnamefont {Zhu}}, \bibinfo {author} {\bibfnamefont {Y.}~\bibnamefont
  {Gao}}, \bibinfo {author} {\bibfnamefont {Z.}~\bibnamefont {Tan}}, \bibinfo
  {author} {\bibfnamefont {Z.}~\bibnamefont {Cui}}, \bibinfo {author}
  {\bibfnamefont {A.}~\bibnamefont {Zhang}}, \bibinfo {author} {\bibfnamefont
  {N.}~\bibnamefont {Wang}}, \bibinfo {author} {\bibfnamefont {Y.}~\bibnamefont
  {Zou}}, \bibinfo {author} {\bibfnamefont {T.}~\bibnamefont {Li}}, \bibinfo
  {author} {\bibfnamefont {F.}~\bibnamefont {Shen}}, \bibinfo {author}
  {\bibfnamefont {J.}~\bibnamefont {Zhong}}, \bibinfo {author} {\bibfnamefont
  {Z.}~\bibnamefont {Bao}}, \bibinfo {author} {\bibfnamefont {Z.}~\bibnamefont
  {Zhu}}, \bibinfo {author} {\bibfnamefont {Z.}~\bibnamefont {Song}}, \bibinfo
  {author} {\bibfnamefont {J.}~\bibnamefont {Deng}}, \bibinfo {author}
  {\bibfnamefont {H.}~\bibnamefont {Dong}}, \bibinfo {author} {\bibfnamefont
  {P.}~\bibnamefont {Zhang}}, \bibinfo {author} {\bibfnamefont
  {W.}~\bibnamefont {Zhang}}, \bibinfo {author} {\bibfnamefont
  {H.}~\bibnamefont {Li}}, \bibinfo {author} {\bibfnamefont {Q.}~\bibnamefont
  {Guo}}, \bibinfo {author} {\bibfnamefont {Z.}~\bibnamefont {Wang}}, \bibinfo
  {author} {\bibfnamefont {Y.}~\bibnamefont {Li}}, \bibinfo {author}
  {\bibfnamefont {X.}~\bibnamefont {Wang}}, \bibinfo {author} {\bibfnamefont
  {C.}~\bibnamefont {Song}},\ and\ \bibinfo {author} {\bibfnamefont
  {H.}~\bibnamefont {Wang}},\ }\bibfield  {title} {\bibinfo {title} {Quantum
  ensemble learning with a programmable superconducting processor},\ }\href
  {https://doi.org/10.1038/s41534-025-01037-6} {\bibfield  {journal} {\bibinfo
  {journal} {npj Quantum Information}\ }\textbf {\bibinfo {volume} {11}}
  (\bibinfo {year} {2025})}\BibitemShut {NoStop}%
\bibitem [{\citenamefont {Havlicek}\ \emph {et~al.}(2019)\citenamefont
  {Havlicek}, \citenamefont {Corcoles}, \citenamefont {Temme}, \citenamefont
  {Harrow}, \citenamefont {Kandala}, \citenamefont {Chow},\ and\ \citenamefont
  {Gambetta}}]{Havlicek2019}%
  \BibitemOpen
  \bibfield  {author} {\bibinfo {author} {\bibfnamefont {V.}~\bibnamefont
  {Havlicek}}, \bibinfo {author} {\bibfnamefont {A.~D.}\ \bibnamefont
  {Corcoles}}, \bibinfo {author} {\bibfnamefont {K.}~\bibnamefont {Temme}},
  \bibinfo {author} {\bibfnamefont {A.~W.}\ \bibnamefont {Harrow}}, \bibinfo
  {author} {\bibfnamefont {A.}~\bibnamefont {Kandala}}, \bibinfo {author}
  {\bibfnamefont {J.~M.}\ \bibnamefont {Chow}},\ and\ \bibinfo {author}
  {\bibfnamefont {J.~M.}\ \bibnamefont {Gambetta}},\ }\bibfield  {title}
  {\bibinfo {title} {Supervised learning with quantum-enhanced feature
  spaces},\ }\href {https://doi.org/10.1038/s41586-019-0980-2} {\bibfield
  {journal} {\bibinfo  {journal} {Nature}\ }\textbf {\bibinfo {volume} {567}},\
  \bibinfo {pages} {209} (\bibinfo {year} {2019})}\BibitemShut {NoStop}%
\bibitem [{\citenamefont {Henderson}\ \emph {et~al.}(2020)\citenamefont
  {Henderson}, \citenamefont {Shakya}, \citenamefont {Pradhan},\ and\
  \citenamefont {Cook}}]{Henderson2020}%
  \BibitemOpen
  \bibfield  {author} {\bibinfo {author} {\bibfnamefont {M.}~\bibnamefont
  {Henderson}}, \bibinfo {author} {\bibfnamefont {S.}~\bibnamefont {Shakya}},
  \bibinfo {author} {\bibfnamefont {S.}~\bibnamefont {Pradhan}},\ and\ \bibinfo
  {author} {\bibfnamefont {T.}~\bibnamefont {Cook}},\ }\bibfield  {title}
  {\bibinfo {title} {Quanvolutional neural networks: powering image recognition
  with quantum circuits},\ }\href {https://doi.org/10.1007/s42484-020-00012-y}
  {\bibfield  {journal} {\bibinfo  {journal} {Quantum Machine Intelligence}\
  }\textbf {\bibinfo {volume} {2}},\ \bibinfo {pages} {2} (\bibinfo {year}
  {2020})}\BibitemShut {NoStop}%
\bibitem [{\citenamefont {Liu}\ \emph {et~al.}(2021)\citenamefont {Liu},
  \citenamefont {Lim}, \citenamefont {Wood}, \citenamefont {Huang},
  \citenamefont {Guo},\ and\ \citenamefont {Huang}}]{Liu2021s}%
  \BibitemOpen
  \bibfield  {author} {\bibinfo {author} {\bibfnamefont {J.}~\bibnamefont
  {Liu}}, \bibinfo {author} {\bibfnamefont {K.~H.}\ \bibnamefont {Lim}},
  \bibinfo {author} {\bibfnamefont {K.~L.}\ \bibnamefont {Wood}}, \bibinfo
  {author} {\bibfnamefont {W.}~\bibnamefont {Huang}}, \bibinfo {author}
  {\bibfnamefont {C.}~\bibnamefont {Guo}},\ and\ \bibinfo {author}
  {\bibfnamefont {H.-L.}\ \bibnamefont {Huang}},\ }\bibfield  {title} {\bibinfo
  {title} {Hybrid quantum-classical convolutional neural networks},\ }\href
  {https://doi.org/10.1007/s11433-021-1734-3} {\bibfield  {journal} {\bibinfo
  {journal} {Science China Physics, Mechanics \& Astronomy}\ }\textbf {\bibinfo
  {volume} {64}},\ \bibinfo {pages} {290311} (\bibinfo {year}
  {2021})}\BibitemShut {NoStop}%
\bibitem [{\citenamefont {Peters}\ \emph {et~al.}(2021)\citenamefont {Peters},
  \citenamefont {Caldeira}, \citenamefont {Ho}, \citenamefont {Leichenauer},
  \citenamefont {Mohseni}, \citenamefont {Neven}, \citenamefont {Spentzouris},
  \citenamefont {Strain},\ and\ \citenamefont {Perdue}}]{Peters2021}%
  \BibitemOpen
  \bibfield  {author} {\bibinfo {author} {\bibfnamefont {E.}~\bibnamefont
  {Peters}}, \bibinfo {author} {\bibfnamefont {J.}~\bibnamefont {Caldeira}},
  \bibinfo {author} {\bibfnamefont {A.}~\bibnamefont {Ho}}, \bibinfo {author}
  {\bibfnamefont {S.}~\bibnamefont {Leichenauer}}, \bibinfo {author}
  {\bibfnamefont {M.}~\bibnamefont {Mohseni}}, \bibinfo {author} {\bibfnamefont
  {H.}~\bibnamefont {Neven}}, \bibinfo {author} {\bibfnamefont
  {P.}~\bibnamefont {Spentzouris}}, \bibinfo {author} {\bibfnamefont
  {D.}~\bibnamefont {Strain}},\ and\ \bibinfo {author} {\bibfnamefont {G.~N.}\
  \bibnamefont {Perdue}},\ }\bibfield  {title} {\bibinfo {title} {Machine
  learning of high dimensional data on a noisy quantum processor},\ }\href
  {https://doi.org/10.1038/s41534-021-00498-9} {\bibfield  {journal} {\bibinfo
  {journal} {npj Quantum Information}\ }\textbf {\bibinfo {volume} {7}},\
  \bibinfo {pages} {161} (\bibinfo {year} {2021})}\BibitemShut {NoStop}%
\bibitem [{\citenamefont {Senokosov}\ \emph {et~al.}(2024)\citenamefont
  {Senokosov}, \citenamefont {Sedykh}, \citenamefont {Sagingalieva},
  \citenamefont {Kyriacou},\ and\ \citenamefont {Melnikov}}]{Senokosov2024}%
  \BibitemOpen
  \bibfield  {author} {\bibinfo {author} {\bibfnamefont {A.}~\bibnamefont
  {Senokosov}}, \bibinfo {author} {\bibfnamefont {A.}~\bibnamefont {Sedykh}},
  \bibinfo {author} {\bibfnamefont {A.}~\bibnamefont {Sagingalieva}}, \bibinfo
  {author} {\bibfnamefont {B.}~\bibnamefont {Kyriacou}},\ and\ \bibinfo
  {author} {\bibfnamefont {A.}~\bibnamefont {Melnikov}},\ }\bibfield  {title}
  {\bibinfo {title} {Quantum machine learning for image classification},\
  }\href {https://doi.org/10.1088/2632-2153/ad2aef} {\bibfield  {journal}
  {\bibinfo  {journal} {Machine Learning: Science and Technology}\ }\textbf
  {\bibinfo {volume} {5}},\ \bibinfo {pages} {015040} (\bibinfo {year}
  {2024})}\BibitemShut {NoStop}%
\bibitem [{\citenamefont {Romero}\ \emph {et~al.}(2017)\citenamefont {Romero},
  \citenamefont {Olson},\ and\ \citenamefont {Aspuru-Guzik}}]{Romero2017b}%
  \BibitemOpen
  \bibfield  {author} {\bibinfo {author} {\bibfnamefont {J.}~\bibnamefont
  {Romero}}, \bibinfo {author} {\bibfnamefont {J.~P.}\ \bibnamefont {Olson}},\
  and\ \bibinfo {author} {\bibfnamefont {A.}~\bibnamefont {Aspuru-Guzik}},\
  }\bibfield  {title} {\bibinfo {title} {Quantum autoencoders for efficient
  compression of quantum data},\ }\href
  {https://doi.org/10.1088/2058-9565/aa8072} {\bibfield  {journal} {\bibinfo
  {journal} {Quantum Science and Technology}\ }\textbf {\bibinfo {volume}
  {2}},\ \bibinfo {pages} {045001} (\bibinfo {year} {2017})}\BibitemShut
  {NoStop}%
\bibitem [{\citenamefont {Beckey}\ \emph {et~al.}(2022)\citenamefont {Beckey},
  \citenamefont {Cerezo}, \citenamefont {Sone},\ and\ \citenamefont
  {Coles}}]{Beckey2022}%
  \BibitemOpen
  \bibfield  {author} {\bibinfo {author} {\bibfnamefont {J.~L.}\ \bibnamefont
  {Beckey}}, \bibinfo {author} {\bibfnamefont {M.}~\bibnamefont {Cerezo}},
  \bibinfo {author} {\bibfnamefont {A.}~\bibnamefont {Sone}},\ and\ \bibinfo
  {author} {\bibfnamefont {P.~J.}\ \bibnamefont {Coles}},\ }\bibfield  {title}
  {\bibinfo {title} {Variational quantum algorithm for estimating the quantum
  {Fisher} information},\ }\href
  {https://doi.org/10.1103/PhysRevResearch.4.013083} {\bibfield  {journal}
  {\bibinfo  {journal} {Phys. Rev. Research}\ }\textbf {\bibinfo {volume}
  {4}},\ \bibinfo {pages} {013083} (\bibinfo {year} {2022})}\BibitemShut
  {NoStop}%
\bibitem [{\citenamefont {Geller}\ \emph {et~al.}(2021)\citenamefont {Geller},
  \citenamefont {Holmes}, \citenamefont {Coles},\ and\ \citenamefont
  {Sornborger}}]{Gellerc2021}%
  \BibitemOpen
  \bibfield  {author} {\bibinfo {author} {\bibfnamefont {M.~R.}\ \bibnamefont
  {Geller}}, \bibinfo {author} {\bibfnamefont {Z.}~\bibnamefont {Holmes}},
  \bibinfo {author} {\bibfnamefont {P.~J.}\ \bibnamefont {Coles}},\ and\
  \bibinfo {author} {\bibfnamefont {A.}~\bibnamefont {Sornborger}},\ }\bibfield
   {title} {\bibinfo {title} {Experimental quantum learning of a spectral
  decomposition},\ }\href {https://doi.org/10.1103/PhysRevResearch.3.033200}
  {\bibfield  {journal} {\bibinfo  {journal} {Phys. Rev. Research}\ }\textbf
  {\bibinfo {volume} {3}},\ \bibinfo {pages} {033200} (\bibinfo {year}
  {2021})}\BibitemShut {NoStop}%
\bibitem [{\citenamefont {Gong}\ \emph {et~al.}(2023)\citenamefont {Gong},
  \citenamefont {Huang}, \citenamefont {Wang}, \citenamefont {Guo},
  \citenamefont {Li}, \citenamefont {Wu}, \citenamefont {Zhu}, \citenamefont
  {Zhao}, \citenamefont {Guo}, \citenamefont {Qian}, \citenamefont {Ye},
  \citenamefont {Zha}, \citenamefont {Chen}, \citenamefont {Ying},
  \citenamefont {Yu}, \citenamefont {Fan}, \citenamefont {Wu}, \citenamefont
  {Su}, \citenamefont {Deng}, \citenamefont {Rong}, \citenamefont {Zhang},
  \citenamefont {Cao}, \citenamefont {Lin}, \citenamefont {Xu}, \citenamefont
  {Sun}, \citenamefont {Guo}, \citenamefont {Li}, \citenamefont {Liang},
  \citenamefont {Sakurai}, \citenamefont {Nemoto}, \citenamefont {Munro},
  \citenamefont {Huo}, \citenamefont {Lu}, \citenamefont {Peng}, \citenamefont
  {Zhu},\ and\ \citenamefont {Pan}}]{Gong2023}%
  \BibitemOpen
  \bibfield  {author} {\bibinfo {author} {\bibfnamefont {M.}~\bibnamefont
  {Gong}}, \bibinfo {author} {\bibfnamefont {H.-L.}\ \bibnamefont {Huang}},
  \bibinfo {author} {\bibfnamefont {S.}~\bibnamefont {Wang}}, \bibinfo {author}
  {\bibfnamefont {C.}~\bibnamefont {Guo}}, \bibinfo {author} {\bibfnamefont
  {S.}~\bibnamefont {Li}}, \bibinfo {author} {\bibfnamefont {Y.}~\bibnamefont
  {Wu}}, \bibinfo {author} {\bibfnamefont {Q.}~\bibnamefont {Zhu}}, \bibinfo
  {author} {\bibfnamefont {Y.}~\bibnamefont {Zhao}}, \bibinfo {author}
  {\bibfnamefont {S.}~\bibnamefont {Guo}}, \bibinfo {author} {\bibfnamefont
  {H.}~\bibnamefont {Qian}}, \bibinfo {author} {\bibfnamefont {Y.}~\bibnamefont
  {Ye}}, \bibinfo {author} {\bibfnamefont {C.}~\bibnamefont {Zha}}, \bibinfo
  {author} {\bibfnamefont {F.}~\bibnamefont {Chen}}, \bibinfo {author}
  {\bibfnamefont {C.}~\bibnamefont {Ying}}, \bibinfo {author} {\bibfnamefont
  {J.}~\bibnamefont {Yu}}, \bibinfo {author} {\bibfnamefont {D.}~\bibnamefont
  {Fan}}, \bibinfo {author} {\bibfnamefont {D.}~\bibnamefont {Wu}}, \bibinfo
  {author} {\bibfnamefont {H.}~\bibnamefont {Su}}, \bibinfo {author}
  {\bibfnamefont {H.}~\bibnamefont {Deng}}, \bibinfo {author} {\bibfnamefont
  {H.}~\bibnamefont {Rong}}, \bibinfo {author} {\bibfnamefont {K.}~\bibnamefont
  {Zhang}}, \bibinfo {author} {\bibfnamefont {S.}~\bibnamefont {Cao}}, \bibinfo
  {author} {\bibfnamefont {J.}~\bibnamefont {Lin}}, \bibinfo {author}
  {\bibfnamefont {Y.}~\bibnamefont {Xu}}, \bibinfo {author} {\bibfnamefont
  {L.}~\bibnamefont {Sun}}, \bibinfo {author} {\bibfnamefont {C.}~\bibnamefont
  {Guo}}, \bibinfo {author} {\bibfnamefont {N.}~\bibnamefont {Li}}, \bibinfo
  {author} {\bibfnamefont {F.}~\bibnamefont {Liang}}, \bibinfo {author}
  {\bibfnamefont {A.}~\bibnamefont {Sakurai}}, \bibinfo {author} {\bibfnamefont
  {K.}~\bibnamefont {Nemoto}}, \bibinfo {author} {\bibfnamefont {W.~J.}\
  \bibnamefont {Munro}}, \bibinfo {author} {\bibfnamefont {Y.-H.}\ \bibnamefont
  {Huo}}, \bibinfo {author} {\bibfnamefont {C.-Y.}\ \bibnamefont {Lu}},
  \bibinfo {author} {\bibfnamefont {C.-Z.}\ \bibnamefont {Peng}}, \bibinfo
  {author} {\bibfnamefont {X.}~\bibnamefont {Zhu}},\ and\ \bibinfo {author}
  {\bibfnamefont {J.-W.}\ \bibnamefont {Pan}},\ }\bibfield  {title} {\bibinfo
  {title} {Quantum neuronal sensing of quantum many-body states on a 61-qubit
  programmable superconducting processor},\ }\href
  {https://doi.org/https://doi.org/10.1016/j.scib.2023.04.003} {\bibfield
  {journal} {\bibinfo  {journal} {Science Bulletin}\ }\textbf {\bibinfo
  {volume} {68}},\ \bibinfo {pages} {906} (\bibinfo {year} {2023})}\BibitemShut
  {NoStop}%
\bibitem [{\citenamefont {Caro}\ \emph {et~al.}(2022)\citenamefont {Caro},
  \citenamefont {Huang}, \citenamefont {Cerezo}, \citenamefont {Sharma},
  \citenamefont {Sornborger}, \citenamefont {Cincio},\ and\ \citenamefont
  {Coles}}]{Caro2022}%
  \BibitemOpen
  \bibfield  {author} {\bibinfo {author} {\bibfnamefont {M.~C.}\ \bibnamefont
  {Caro}}, \bibinfo {author} {\bibfnamefont {H.-Y.}\ \bibnamefont {Huang}},
  \bibinfo {author} {\bibfnamefont {M.}~\bibnamefont {Cerezo}}, \bibinfo
  {author} {\bibfnamefont {K.}~\bibnamefont {Sharma}}, \bibinfo {author}
  {\bibfnamefont {A.}~\bibnamefont {Sornborger}}, \bibinfo {author}
  {\bibfnamefont {L.}~\bibnamefont {Cincio}},\ and\ \bibinfo {author}
  {\bibfnamefont {P.~J.}\ \bibnamefont {Coles}},\ }\bibfield  {title} {\bibinfo
  {title} {Generalization in quantum machine learning from few training data},\
  }\href {https://doi.org/10.1038/s41467-022-32550-3} {\bibfield  {journal}
  {\bibinfo  {journal} {Nat. Commun.}\ }\textbf {\bibinfo {volume} {13}},\
  \bibinfo {pages} {4919} (\bibinfo {year} {2022})}\BibitemShut {NoStop}%
\bibitem [{\citenamefont {Caro}\ \emph {et~al.}(2023)\citenamefont {Caro},
  \citenamefont {Huang}, \citenamefont {Ezzell}, \citenamefont {Gibbs},
  \citenamefont {Sornborger}, \citenamefont {Cincio}, \citenamefont {Coles},\
  and\ \citenamefont {Holmes}}]{Caro2023}%
  \BibitemOpen
  \bibfield  {author} {\bibinfo {author} {\bibfnamefont {M.~C.}\ \bibnamefont
  {Caro}}, \bibinfo {author} {\bibfnamefont {H.-Y.}\ \bibnamefont {Huang}},
  \bibinfo {author} {\bibfnamefont {N.}~\bibnamefont {Ezzell}}, \bibinfo
  {author} {\bibfnamefont {J.}~\bibnamefont {Gibbs}}, \bibinfo {author}
  {\bibfnamefont {A.~T.}\ \bibnamefont {Sornborger}}, \bibinfo {author}
  {\bibfnamefont {L.}~\bibnamefont {Cincio}}, \bibinfo {author} {\bibfnamefont
  {P.~J.}\ \bibnamefont {Coles}},\ and\ \bibinfo {author} {\bibfnamefont
  {Z.}~\bibnamefont {Holmes}},\ }\bibfield  {title} {\bibinfo {title}
  {Out-of-distribution generalization for learning quantum dynamics},\ }\href
  {https://doi.org/10.1038/s41467-023-39381-w} {\bibfield  {journal} {\bibinfo
  {journal} {Nat. Commun.}\ }\textbf {\bibinfo {volume} {14}},\ \bibinfo
  {pages} {3751} (\bibinfo {year} {2023})}\BibitemShut {NoStop}%
\bibitem [{\citenamefont {Schuld}\ and\ \citenamefont
  {Petruccione}(2018)}]{Schuld2018c}%
  \BibitemOpen
  \bibfield  {author} {\bibinfo {author} {\bibfnamefont {M.}~\bibnamefont
  {Schuld}}\ and\ \bibinfo {author} {\bibfnamefont {F.}~\bibnamefont
  {Petruccione}},\ }\href
  {https://link.springer.com/book/10.1007/978-3-319-96424-9} {\emph {\bibinfo
  {title} {Supervised Learning with Quantum Computers}}}\ (\bibinfo
  {publisher} {Springer-Verlag GmbH},\ \bibinfo {year} {2018})\BibitemShut
  {NoStop}%
\bibitem [{\citenamefont {Farhi}\ and\ \citenamefont
  {Neven}(2018)}]{Farhi2018}%
  \BibitemOpen
  \bibfield  {author} {\bibinfo {author} {\bibfnamefont {E.}~\bibnamefont
  {Farhi}}\ and\ \bibinfo {author} {\bibfnamefont {H.}~\bibnamefont {Neven}},\
  }\bibfield  {title} {\bibinfo {title} {Classification with quantum neural
  networks on near term processors},\ }\href {https://arxiv.org/abs/1802.06002}
  {\bibfield  {journal} {\bibinfo  {journal} {arXiv:1802.06002}\ } (\bibinfo
  {year} {2018})}\BibitemShut {NoStop}%
\bibitem [{\citenamefont {Beer}\ \emph {et~al.}(2020)\citenamefont {Beer},
  \citenamefont {Bondarenko}, \citenamefont {Farrelly}, \citenamefont
  {Osborne}, \citenamefont {Salzmann}, \citenamefont {Scheiermann},\ and\
  \citenamefont {Wolf}}]{Beer2020}%
  \BibitemOpen
  \bibfield  {author} {\bibinfo {author} {\bibfnamefont {K.}~\bibnamefont
  {Beer}}, \bibinfo {author} {\bibfnamefont {D.}~\bibnamefont {Bondarenko}},
  \bibinfo {author} {\bibfnamefont {T.}~\bibnamefont {Farrelly}}, \bibinfo
  {author} {\bibfnamefont {T.~J.}\ \bibnamefont {Osborne}}, \bibinfo {author}
  {\bibfnamefont {R.}~\bibnamefont {Salzmann}}, \bibinfo {author}
  {\bibfnamefont {D.}~\bibnamefont {Scheiermann}},\ and\ \bibinfo {author}
  {\bibfnamefont {R.}~\bibnamefont {Wolf}},\ }\bibfield  {title} {\bibinfo
  {title} {Training deep quantum neural networks},\ }\href
  {https://doi.org/10.1038/s41467-020-14454-2} {\bibfield  {journal} {\bibinfo
  {journal} {Nat. Commun.}\ }\textbf {\bibinfo {volume} {11}},\ \bibinfo
  {pages} {808} (\bibinfo {year} {2020})}\BibitemShut {NoStop}%
\bibitem [{\citenamefont {Broughton}\ \emph {et~al.}(2020)\citenamefont
  {Broughton}, \citenamefont {Verdon}, \citenamefont {McCourt}, \citenamefont
  {Martinez}, \citenamefont {Yoo}, \citenamefont {Isakov}, \citenamefont
  {Massey}, \citenamefont {Niu}, \citenamefont {Halavati}, \citenamefont
  {Peters}, \citenamefont {Leib}, \citenamefont {Skolik}, \citenamefont
  {Streif}, \citenamefont {Von~Dollen}, \citenamefont {McClean}, \citenamefont
  {Boixo}, \citenamefont {Bacon}, \citenamefont {Ho}, \citenamefont {Neven},
  \citenamefont {Mohseni}, \citenamefont {Kim}, \citenamefont {Yoder},
  \citenamefont {Orlando}, \citenamefont {Gustavsson},\ and\ \citenamefont
  {Oliver}}]{Broughton2020}%
  \BibitemOpen
  \bibfield  {author} {\bibinfo {author} {\bibfnamefont {M.}~\bibnamefont
  {Broughton}}, \bibinfo {author} {\bibfnamefont {G.}~\bibnamefont {Verdon}},
  \bibinfo {author} {\bibfnamefont {T.}~\bibnamefont {McCourt}}, \bibinfo
  {author} {\bibfnamefont {A.~J.}\ \bibnamefont {Martinez}}, \bibinfo {author}
  {\bibfnamefont {J.~H.}\ \bibnamefont {Yoo}}, \bibinfo {author} {\bibfnamefont
  {S.~V.}\ \bibnamefont {Isakov}}, \bibinfo {author} {\bibfnamefont
  {P.}~\bibnamefont {Massey}}, \bibinfo {author} {\bibfnamefont {M.~Y.}\
  \bibnamefont {Niu}}, \bibinfo {author} {\bibfnamefont {R.}~\bibnamefont
  {Halavati}}, \bibinfo {author} {\bibfnamefont {E.}~\bibnamefont {Peters}},
  \bibinfo {author} {\bibfnamefont {M.}~\bibnamefont {Leib}}, \bibinfo {author}
  {\bibfnamefont {A.}~\bibnamefont {Skolik}}, \bibinfo {author} {\bibfnamefont
  {M.}~\bibnamefont {Streif}}, \bibinfo {author} {\bibfnamefont
  {D.}~\bibnamefont {Von~Dollen}}, \bibinfo {author} {\bibfnamefont {J.~R.}\
  \bibnamefont {McClean}}, \bibinfo {author} {\bibfnamefont {S.}~\bibnamefont
  {Boixo}}, \bibinfo {author} {\bibfnamefont {D.}~\bibnamefont {Bacon}},
  \bibinfo {author} {\bibfnamefont {A.~K.}\ \bibnamefont {Ho}}, \bibinfo
  {author} {\bibfnamefont {H.}~\bibnamefont {Neven}}, \bibinfo {author}
  {\bibfnamefont {M.~E.}\ \bibnamefont {Mohseni}}, \bibinfo {author}
  {\bibfnamefont {D.~K.}\ \bibnamefont {Kim}}, \bibinfo {author} {\bibfnamefont
  {J.~L.}\ \bibnamefont {Yoder}}, \bibinfo {author} {\bibfnamefont {T.~P.}\
  \bibnamefont {Orlando}}, \bibinfo {author} {\bibfnamefont {S.}~\bibnamefont
  {Gustavsson}},\ and\ \bibinfo {author} {\bibfnamefont {W.~D.}\ \bibnamefont
  {Oliver}},\ }\bibfield  {title} {\bibinfo {title} {Tensorflow quantum: A
  software framework for quantum machine learning},\ }\href
  {https://arxiv.org/abs/2003.02989} {\bibfield  {journal} {\bibinfo  {journal}
  {arXiv:2003.02989}\ } (\bibinfo {year} {2020})}\BibitemShut {NoStop}%
\bibitem [{\citenamefont {Uvarov}\ \emph {et~al.}(2020)\citenamefont {Uvarov},
  \citenamefont {Kardashin},\ and\ \citenamefont {Biamonte}}]{Uvarov2020}%
  \BibitemOpen
  \bibfield  {author} {\bibinfo {author} {\bibfnamefont {A.~V.}\ \bibnamefont
  {Uvarov}}, \bibinfo {author} {\bibfnamefont {A.~S.}\ \bibnamefont
  {Kardashin}},\ and\ \bibinfo {author} {\bibfnamefont {J.~D.}\ \bibnamefont
  {Biamonte}},\ }\bibfield  {title} {\bibinfo {title} {Machine learning phase
  transitions with a quantum processor},\ }\href
  {https://doi.org/10.1103/PhysRevA.102.012415} {\bibfield  {journal} {\bibinfo
   {journal} {Phys. Rev. A}\ }\textbf {\bibinfo {volume} {102}},\ \bibinfo
  {pages} {012415} (\bibinfo {year} {2020})}\BibitemShut {NoStop}%
\bibitem [{\citenamefont {Herrmann}\ \emph {et~al.}(2022)\citenamefont
  {Herrmann}, \citenamefont {Llima}, \citenamefont {Remm}, \citenamefont
  {Zapletal}, \citenamefont {McMahon}, \citenamefont {Scarato}, \citenamefont
  {Swiadek}, \citenamefont {Andersen}, \citenamefont {Hellings}, \citenamefont
  {Krinner}, \citenamefont {Lacroix}, \citenamefont {Laz\u{a}r}, \citenamefont
  {Kerschbaum}, \citenamefont {Zanuz}, \citenamefont {Norris}, \citenamefont
  {Hartmann}, \citenamefont {Wallraff},\ and\ \citenamefont
  {Eichler}}]{Herrmann2022}%
  \BibitemOpen
  \bibfield  {author} {\bibinfo {author} {\bibfnamefont {J.}~\bibnamefont
  {Herrmann}}, \bibinfo {author} {\bibfnamefont {S.~M.}\ \bibnamefont {Llima}},
  \bibinfo {author} {\bibfnamefont {A.}~\bibnamefont {Remm}}, \bibinfo {author}
  {\bibfnamefont {P.}~\bibnamefont {Zapletal}}, \bibinfo {author}
  {\bibfnamefont {N.~A.}\ \bibnamefont {McMahon}}, \bibinfo {author}
  {\bibfnamefont {C.}~\bibnamefont {Scarato}}, \bibinfo {author} {\bibfnamefont
  {F.}~\bibnamefont {Swiadek}}, \bibinfo {author} {\bibfnamefont {C.~K.}\
  \bibnamefont {Andersen}}, \bibinfo {author} {\bibfnamefont {C.}~\bibnamefont
  {Hellings}}, \bibinfo {author} {\bibfnamefont {S.}~\bibnamefont {Krinner}},
  \bibinfo {author} {\bibfnamefont {N.}~\bibnamefont {Lacroix}}, \bibinfo
  {author} {\bibfnamefont {S.}~\bibnamefont {Laz\u{a}r}}, \bibinfo {author}
  {\bibfnamefont {M.}~\bibnamefont {Kerschbaum}}, \bibinfo {author}
  {\bibfnamefont {D.~C.}\ \bibnamefont {Zanuz}}, \bibinfo {author}
  {\bibfnamefont {G.~J.}\ \bibnamefont {Norris}}, \bibinfo {author}
  {\bibfnamefont {M.~J.}\ \bibnamefont {Hartmann}}, \bibinfo {author}
  {\bibfnamefont {A.}~\bibnamefont {Wallraff}},\ and\ \bibinfo {author}
  {\bibfnamefont {C.}~\bibnamefont {Eichler}},\ }\bibfield  {title} {\bibinfo
  {title} {Realizing quantum convolutional neural networks on a superconducting
  quantum processor to recognize quantum phases},\ }\href
  {https://doi.org/10.1038/s41467-022-31679-5} {\bibfield  {journal} {\bibinfo
  {journal} {Nat. Commun.}\ }\textbf {\bibinfo {volume} {13}},\ \bibinfo
  {pages} {4144} (\bibinfo {year} {2022})}\BibitemShut {NoStop}%
\bibitem [{\citenamefont {Liu}\ \emph {et~al.}(2023)\citenamefont {Liu},
  \citenamefont {Smith}, \citenamefont {Knap},\ and\ \citenamefont
  {Pollmann}}]{Liu2022h}%
  \BibitemOpen
  \bibfield  {author} {\bibinfo {author} {\bibfnamefont {Y.-J.}\ \bibnamefont
  {Liu}}, \bibinfo {author} {\bibfnamefont {A.}~\bibnamefont {Smith}}, \bibinfo
  {author} {\bibfnamefont {M.}~\bibnamefont {Knap}},\ and\ \bibinfo {author}
  {\bibfnamefont {F.}~\bibnamefont {Pollmann}},\ }\bibfield  {title} {\bibinfo
  {title} {Model-independent learning of quantum phases of matter with quantum
  convolutional neural networks},\ }\href
  {https://doi.org/10.1103/PhysRevLett.130.220603} {\bibfield  {journal}
  {\bibinfo  {journal} {Phys. Rev. Lett.}\ }\textbf {\bibinfo {volume} {130}},\
  \bibinfo {pages} {220603} (\bibinfo {year} {2023})}\BibitemShut {NoStop}%
\bibitem [{\citenamefont {Sander}\ \emph {et~al.}(2025)\citenamefont {Sander},
  \citenamefont {McMahon}, \citenamefont {Zapletal},\ and\ \citenamefont
  {Hartmann}}]{Sander2025}%
  \BibitemOpen
  \bibfield  {author} {\bibinfo {author} {\bibfnamefont {L.~C.}\ \bibnamefont
  {Sander}}, \bibinfo {author} {\bibfnamefont {N.~A.}\ \bibnamefont {McMahon}},
  \bibinfo {author} {\bibfnamefont {P.}~\bibnamefont {Zapletal}},\ and\
  \bibinfo {author} {\bibfnamefont {M.~J.}\ \bibnamefont {Hartmann}},\
  }\bibfield  {title} {\bibinfo {title} {Quantum convolutional neural network
  for phase recognition in two dimensions},\ }\href
  {https://doi.org/10.1103/vpyw-nszk} {\bibfield  {journal} {\bibinfo
  {journal} {Phys. Rev. Research}\ }\textbf {\bibinfo {volume} {7}},\ \bibinfo
  {pages} {L042032} (\bibinfo {year} {2025})}\BibitemShut {NoStop}%
\bibitem [{\citenamefont {Aktar}\ \emph {et~al.}(2025)\citenamefont {Aktar},
  \citenamefont {Bhardwaj}, \citenamefont {B\"artschi}, \citenamefont
  {Bhattacharya},\ and\ \citenamefont {Eidenbenz}}]{Aktar2025}%
  \BibitemOpen
  \bibfield  {author} {\bibinfo {author} {\bibfnamefont {S.}~\bibnamefont
  {Aktar}}, \bibinfo {author} {\bibfnamefont {R.}~\bibnamefont {Bhardwaj}},
  \bibinfo {author} {\bibfnamefont {A.}~\bibnamefont {B\"artschi}}, \bibinfo
  {author} {\bibfnamefont {T.}~\bibnamefont {Bhattacharya}},\ and\ \bibinfo
  {author} {\bibfnamefont {S.}~\bibnamefont {Eidenbenz}},\ }\bibfield  {title}
  {\bibinfo {title} {Quantum data learning of topological-to-ferromagnetic
  phase transitions in the 2+1{D} toric code loop gas model},\ }\href
  {https://arxiv.org/abs/2511.16851} {\bibfield  {journal} {\bibinfo  {journal}
  {arXiv:2511.16851}\ } (\bibinfo {year} {2025})}\BibitemShut {NoStop}%
\bibitem [{\citenamefont {Ren}\ \emph {et~al.}(2022)\citenamefont {Ren},
  \citenamefont {Li}, \citenamefont {Xu}, \citenamefont {Wang}, \citenamefont
  {Jiang}, \citenamefont {Jin}, \citenamefont {Zhu}, \citenamefont {Chen},
  \citenamefont {Song}, \citenamefont {Zhang}, \citenamefont {Dong},
  \citenamefont {Zhang}, \citenamefont {Deng}, \citenamefont {Gao},
  \citenamefont {Zhang}, \citenamefont {Wu}, \citenamefont {Zhang},
  \citenamefont {Guo}, \citenamefont {Li}, \citenamefont {Wang}, \citenamefont
  {Biamonte}, \citenamefont {Song}, \citenamefont {Deng},\ and\ \citenamefont
  {Wang}}]{Ren2022a}%
  \BibitemOpen
  \bibfield  {author} {\bibinfo {author} {\bibfnamefont {W.}~\bibnamefont
  {Ren}}, \bibinfo {author} {\bibfnamefont {W.}~\bibnamefont {Li}}, \bibinfo
  {author} {\bibfnamefont {S.}~\bibnamefont {Xu}}, \bibinfo {author}
  {\bibfnamefont {K.}~\bibnamefont {Wang}}, \bibinfo {author} {\bibfnamefont
  {W.}~\bibnamefont {Jiang}}, \bibinfo {author} {\bibfnamefont
  {F.}~\bibnamefont {Jin}}, \bibinfo {author} {\bibfnamefont {X.}~\bibnamefont
  {Zhu}}, \bibinfo {author} {\bibfnamefont {J.}~\bibnamefont {Chen}}, \bibinfo
  {author} {\bibfnamefont {Z.}~\bibnamefont {Song}}, \bibinfo {author}
  {\bibfnamefont {P.}~\bibnamefont {Zhang}}, \bibinfo {author} {\bibfnamefont
  {H.}~\bibnamefont {Dong}}, \bibinfo {author} {\bibfnamefont {X.}~\bibnamefont
  {Zhang}}, \bibinfo {author} {\bibfnamefont {J.}~\bibnamefont {Deng}},
  \bibinfo {author} {\bibfnamefont {Y.}~\bibnamefont {Gao}}, \bibinfo {author}
  {\bibfnamefont {C.}~\bibnamefont {Zhang}}, \bibinfo {author} {\bibfnamefont
  {Y.}~\bibnamefont {Wu}}, \bibinfo {author} {\bibfnamefont {B.}~\bibnamefont
  {Zhang}}, \bibinfo {author} {\bibfnamefont {Q.}~\bibnamefont {Guo}}, \bibinfo
  {author} {\bibfnamefont {H.}~\bibnamefont {Li}}, \bibinfo {author}
  {\bibfnamefont {Z.}~\bibnamefont {Wang}}, \bibinfo {author} {\bibfnamefont
  {J.}~\bibnamefont {Biamonte}}, \bibinfo {author} {\bibfnamefont
  {C.}~\bibnamefont {Song}}, \bibinfo {author} {\bibfnamefont {D.-L.}\
  \bibnamefont {Deng}},\ and\ \bibinfo {author} {\bibfnamefont
  {H.}~\bibnamefont {Wang}},\ }\bibfield  {title} {\bibinfo {title}
  {Experimental quantum adversarial learning with programmable superconducting
  qubits},\ }\href {https://doi.org/10.1038/s43588-022-00351-9} {\bibfield
  {journal} {\bibinfo  {journal} {Nature Computational Science}\ }\textbf
  {\bibinfo {volume} {2}},\ \bibinfo {pages} {711} (\bibinfo {year}
  {2022})}\BibitemShut {NoStop}%
\bibitem [{\citenamefont {Trebst}\ \emph {et~al.}(2007)\citenamefont {Trebst},
  \citenamefont {Werner}, \citenamefont {Troyer}, \citenamefont {Shtengel},\
  and\ \citenamefont {Nayak}}]{Trebst2007}%
  \BibitemOpen
  \bibfield  {author} {\bibinfo {author} {\bibfnamefont {S.}~\bibnamefont
  {Trebst}}, \bibinfo {author} {\bibfnamefont {P.}~\bibnamefont {Werner}},
  \bibinfo {author} {\bibfnamefont {M.}~\bibnamefont {Troyer}}, \bibinfo
  {author} {\bibfnamefont {K.}~\bibnamefont {Shtengel}},\ and\ \bibinfo
  {author} {\bibfnamefont {C.}~\bibnamefont {Nayak}},\ }\bibfield  {title}
  {\bibinfo {title} {Breakdown of a topological phase: Quantum phase transition
  in a loop gas model with tension},\ }\href
  {https://doi.org/10.1103/PhysRevLett.98.070602} {\bibfield  {journal}
  {\bibinfo  {journal} {Phys. Rev. Lett.}\ }\textbf {\bibinfo {volume} {98}},\
  \bibinfo {pages} {070602} (\bibinfo {year} {2007})}\BibitemShut {NoStop}%
\bibitem [{\citenamefont {Lacroix}\ \emph {et~al.}(2020)\citenamefont
  {Lacroix}, \citenamefont {Hellings}, \citenamefont {Andersen}, \citenamefont
  {Di~Paolo}, \citenamefont {Remm}, \citenamefont {Laz\u{a}r}, \citenamefont
  {Krinner}, \citenamefont {Norris}, \citenamefont {Gabureac}, \citenamefont
  {Heinsoo}, \citenamefont {Blais}, \citenamefont {Eichler},\ and\
  \citenamefont {Wallraff}}]{Lacroix2020}%
  \BibitemOpen
  \bibfield  {author} {\bibinfo {author} {\bibfnamefont {N.}~\bibnamefont
  {Lacroix}}, \bibinfo {author} {\bibfnamefont {C.}~\bibnamefont {Hellings}},
  \bibinfo {author} {\bibfnamefont {C.~K.}\ \bibnamefont {Andersen}}, \bibinfo
  {author} {\bibfnamefont {A.}~\bibnamefont {Di~Paolo}}, \bibinfo {author}
  {\bibfnamefont {A.}~\bibnamefont {Remm}}, \bibinfo {author} {\bibfnamefont
  {S.}~\bibnamefont {Laz\u{a}r}}, \bibinfo {author} {\bibfnamefont
  {S.}~\bibnamefont {Krinner}}, \bibinfo {author} {\bibfnamefont {G.~J.}\
  \bibnamefont {Norris}}, \bibinfo {author} {\bibfnamefont {M.}~\bibnamefont
  {Gabureac}}, \bibinfo {author} {\bibfnamefont {J.}~\bibnamefont {Heinsoo}},
  \bibinfo {author} {\bibfnamefont {A.}~\bibnamefont {Blais}}, \bibinfo
  {author} {\bibfnamefont {C.}~\bibnamefont {Eichler}},\ and\ \bibinfo {author}
  {\bibfnamefont {A.}~\bibnamefont {Wallraff}},\ }\bibfield  {title} {\bibinfo
  {title} {Improving the performance of deep quantum optimization algorithms
  with continuous gate sets},\ }\href
  {https://doi.org/10.1103/PRXQuantum.1.020304} {\bibfield  {journal} {\bibinfo
   {journal} {PRX Quantum}\ }\textbf {\bibinfo {volume} {1}},\ \bibinfo {pages}
  {020304} (\bibinfo {year} {2020})}\BibitemShut {NoStop}%
\bibitem [{\citenamefont {Cirstoiu}\ \emph {et~al.}(2020)\citenamefont
  {Cirstoiu}, \citenamefont {Holmes}, \citenamefont {Iosue}, \citenamefont
  {Cincio}, \citenamefont {Coles},\ and\ \citenamefont
  {Sornborger}}]{Cirstoiu2020}%
  \BibitemOpen
  \bibfield  {author} {\bibinfo {author} {\bibfnamefont {C.}~\bibnamefont
  {Cirstoiu}}, \bibinfo {author} {\bibfnamefont {Z.}~\bibnamefont {Holmes}},
  \bibinfo {author} {\bibfnamefont {J.}~\bibnamefont {Iosue}}, \bibinfo
  {author} {\bibfnamefont {L.}~\bibnamefont {Cincio}}, \bibinfo {author}
  {\bibfnamefont {P.~J.}\ \bibnamefont {Coles}},\ and\ \bibinfo {author}
  {\bibfnamefont {A.}~\bibnamefont {Sornborger}},\ }\bibfield  {title}
  {\bibinfo {title} {Variational fast forwarding for quantum simulation beyond
  the coherence time},\ }\href {https://doi.org/10.1038/s41534-020-00302-0}
  {\bibfield  {journal} {\bibinfo  {journal} {npj Quantum Information}\
  }\textbf {\bibinfo {volume} {6}},\ \bibinfo {pages} {82} (\bibinfo {year}
  {2020})}\BibitemShut {NoStop}%
\bibitem [{\citenamefont {Scarato}\ \emph {et~al.}(2025)\citenamefont
  {Scarato}, \citenamefont {Hanke}, \citenamefont {Remm}, \citenamefont
  {Lazăr}, \citenamefont {Lacroix}, \citenamefont {Colao~Zanuz}, \citenamefont
  {Flasby}, \citenamefont {Wallraff},\ and\ \citenamefont
  {Hellings}}]{Scarato2025}%
  \BibitemOpen
  \bibfield  {author} {\bibinfo {author} {\bibfnamefont {C.}~\bibnamefont
  {Scarato}}, \bibinfo {author} {\bibfnamefont {K.}~\bibnamefont {Hanke}},
  \bibinfo {author} {\bibfnamefont {A.}~\bibnamefont {Remm}}, \bibinfo {author}
  {\bibfnamefont {S.}~\bibnamefont {Lazăr}}, \bibinfo {author} {\bibfnamefont
  {N.}~\bibnamefont {Lacroix}}, \bibinfo {author} {\bibfnamefont
  {D.}~\bibnamefont {Colao~Zanuz}}, \bibinfo {author} {\bibfnamefont
  {A.}~\bibnamefont {Flasby}}, \bibinfo {author} {\bibfnamefont
  {A.}~\bibnamefont {Wallraff}},\ and\ \bibinfo {author} {\bibfnamefont
  {C.}~\bibnamefont {Hellings}},\ }\bibfield  {title} {\bibinfo {title}
  {Realizing a continuous set of two-qubit gates parameterized by an idle
  time},\ }\href {https://doi.org/10.1103/h7cv-xgw2} {\bibfield  {journal}
  {\bibinfo  {journal} {PRX Quantum}\ }\textbf {\bibinfo {volume} {6}},\
  \bibinfo {pages} {040317} (\bibinfo {year} {2025})}\BibitemShut {NoStop}%
\bibitem [{\citenamefont {Bermejo}\ \emph {et~al.}(2026)\citenamefont
  {Bermejo}, \citenamefont {Braccia}, \citenamefont {Rudolph}, \citenamefont
  {Holmes}, \citenamefont {Cincio},\ and\ \citenamefont
  {Cerezo}}]{Bermejo2026}%
  \BibitemOpen
  \bibfield  {author} {\bibinfo {author} {\bibfnamefont {P.}~\bibnamefont
  {Bermejo}}, \bibinfo {author} {\bibfnamefont {P.}~\bibnamefont {Braccia}},
  \bibinfo {author} {\bibfnamefont {M.~S.}\ \bibnamefont {Rudolph}}, \bibinfo
  {author} {\bibfnamefont {Z.}~\bibnamefont {Holmes}}, \bibinfo {author}
  {\bibfnamefont {L.}~\bibnamefont {Cincio}},\ and\ \bibinfo {author}
  {\bibfnamefont {M.}~\bibnamefont {Cerezo}},\ }\bibfield  {title} {\bibinfo
  {title} {Quantum convolutional neural networks are effectively classically
  simulable},\ }\href {https://doi.org/10.1103/8qt9-72ts} {\bibfield  {journal}
  {\bibinfo  {journal} {PRX Quantum}\ }\textbf {\bibinfo {volume} {7}},\
  \bibinfo {pages} {020304} (\bibinfo {year} {2026})}\BibitemShut {NoStop}%
\bibitem [{\citenamefont {Hoffmann}\ \emph {et~al.}(2026)\citenamefont
  {Hoffmann}, \citenamefont {Sander}, \citenamefont {Scarato}, \citenamefont
  {Hellings}, \citenamefont {Kn{\"o}rzer}, \citenamefont {Hartmann},\ and\
  \citenamefont {Zapletal}}]{Hoffmann2026}%
  \BibitemOpen
  \bibfield  {author} {\bibinfo {author} {\bibfnamefont {M.~K.}\ \bibnamefont
  {Hoffmann}}, \bibinfo {author} {\bibfnamefont {L.~C.}\ \bibnamefont
  {Sander}}, \bibinfo {author} {\bibfnamefont {C.}~\bibnamefont {Scarato}},
  \bibinfo {author} {\bibfnamefont {C.}~\bibnamefont {Hellings}}, \bibinfo
  {author} {\bibfnamefont {J.}~\bibnamefont {Kn{\"o}rzer}}, \bibinfo {author}
  {\bibfnamefont {M.~J.}\ \bibnamefont {Hartmann}},\ and\ \bibinfo {author}
  {\bibfnamefont {P.}~\bibnamefont {Zapletal}},\ }\bibfield  {title} {\bibinfo
  {title} {Hybrid quantum-classical neural network for sample-efficient
  recognition of topological phases},\ }\href@noop {} {\bibfield  {journal}
  {\bibinfo  {journal} {companion paper}\ } (\bibinfo {year}
  {2026})}\BibitemShut {NoStop}%
\bibitem [{\citenamefont {Hamma}\ \emph {et~al.}(2005)\citenamefont {Hamma},
  \citenamefont {Ionicioiu},\ and\ \citenamefont {Zanardi}}]{Hamma2005}%
  \BibitemOpen
  \bibfield  {author} {\bibinfo {author} {\bibfnamefont {A.}~\bibnamefont
  {Hamma}}, \bibinfo {author} {\bibfnamefont {R.}~\bibnamefont {Ionicioiu}},\
  and\ \bibinfo {author} {\bibfnamefont {P.}~\bibnamefont {Zanardi}},\
  }\bibfield  {title} {\bibinfo {title} {Ground state entanglement and
  geometric entropy in the {Kitaev} model},\ }\href
  {https://doi.org/https://doi.org/10.1016/j.physleta.2005.01.060} {\bibfield
  {journal} {\bibinfo  {journal} {Physics Letters A}\ }\textbf {\bibinfo
  {volume} {337}},\ \bibinfo {pages} {22} (\bibinfo {year} {2005})}\BibitemShut
  {NoStop}%
\bibitem [{\citenamefont {Bluvstein}\ \emph {et~al.}(2022)\citenamefont
  {Bluvstein}, \citenamefont {Levine}, \citenamefont {Semeghini}, \citenamefont
  {Wang}, \citenamefont {Ebadi}, \citenamefont {Kalinowski}, \citenamefont
  {Keesling}, \citenamefont {Maskara}, \citenamefont {Pichler}, \citenamefont
  {Greiner}, \citenamefont {Vuleti{\'{c}}},\ and\ \citenamefont
  {Lukin}}]{Bluvstein2022}%
  \BibitemOpen
  \bibfield  {author} {\bibinfo {author} {\bibfnamefont {D.}~\bibnamefont
  {Bluvstein}}, \bibinfo {author} {\bibfnamefont {H.}~\bibnamefont {Levine}},
  \bibinfo {author} {\bibfnamefont {G.}~\bibnamefont {Semeghini}}, \bibinfo
  {author} {\bibfnamefont {T.~T.}\ \bibnamefont {Wang}}, \bibinfo {author}
  {\bibfnamefont {S.}~\bibnamefont {Ebadi}}, \bibinfo {author} {\bibfnamefont
  {M.}~\bibnamefont {Kalinowski}}, \bibinfo {author} {\bibfnamefont
  {A.}~\bibnamefont {Keesling}}, \bibinfo {author} {\bibfnamefont
  {N.}~\bibnamefont {Maskara}}, \bibinfo {author} {\bibfnamefont
  {H.}~\bibnamefont {Pichler}}, \bibinfo {author} {\bibfnamefont
  {M.}~\bibnamefont {Greiner}}, \bibinfo {author} {\bibfnamefont
  {V.}~\bibnamefont {Vuleti{\'{c}}}},\ and\ \bibinfo {author} {\bibfnamefont
  {M.~D.}\ \bibnamefont {Lukin}},\ }\bibfield  {title} {\bibinfo {title} {A
  quantum processor based on coherent transport of entangled atom arrays},\
  }\href {https://doi.org/10.1038/s41586-022-04592-6} {\bibfield  {journal}
  {\bibinfo  {journal} {Nature}\ }\textbf {\bibinfo {volume} {604}},\ \bibinfo
  {pages} {451} (\bibinfo {year} {2022})}\BibitemShut {NoStop}%
\bibitem [{\citenamefont {Iqbal}\ \emph {et~al.}(2024)\citenamefont {Iqbal},
  \citenamefont {Tantivasadakarn}, \citenamefont {Gatterman}, \citenamefont
  {Gerber}, \citenamefont {Gilmore}, \citenamefont {Gresh}, \citenamefont
  {Hankin}, \citenamefont {Hewitt}, \citenamefont {Horst}, \citenamefont
  {Matheny}, \citenamefont {Mengle}, \citenamefont {Neyenhuis}, \citenamefont
  {Vishwanath}, \citenamefont {Foss-Feig}, \citenamefont {Verresen},\ and\
  \citenamefont {Dreyer}}]{Iqbal2024a}%
  \BibitemOpen
  \bibfield  {author} {\bibinfo {author} {\bibfnamefont {M.}~\bibnamefont
  {Iqbal}}, \bibinfo {author} {\bibfnamefont {N.}~\bibnamefont
  {Tantivasadakarn}}, \bibinfo {author} {\bibfnamefont {T.~M.}\ \bibnamefont
  {Gatterman}}, \bibinfo {author} {\bibfnamefont {J.~A.}\ \bibnamefont
  {Gerber}}, \bibinfo {author} {\bibfnamefont {K.}~\bibnamefont {Gilmore}},
  \bibinfo {author} {\bibfnamefont {D.}~\bibnamefont {Gresh}}, \bibinfo
  {author} {\bibfnamefont {A.}~\bibnamefont {Hankin}}, \bibinfo {author}
  {\bibfnamefont {N.}~\bibnamefont {Hewitt}}, \bibinfo {author} {\bibfnamefont
  {C.~V.}\ \bibnamefont {Horst}}, \bibinfo {author} {\bibfnamefont
  {M.}~\bibnamefont {Matheny}}, \bibinfo {author} {\bibfnamefont
  {T.}~\bibnamefont {Mengle}}, \bibinfo {author} {\bibfnamefont
  {B.}~\bibnamefont {Neyenhuis}}, \bibinfo {author} {\bibfnamefont
  {A.}~\bibnamefont {Vishwanath}}, \bibinfo {author} {\bibfnamefont
  {M.}~\bibnamefont {Foss-Feig}}, \bibinfo {author} {\bibfnamefont
  {R.}~\bibnamefont {Verresen}},\ and\ \bibinfo {author} {\bibfnamefont
  {H.}~\bibnamefont {Dreyer}},\ }\bibfield  {title} {\bibinfo {title}
  {Topological order from measurements and feed-forward on a trapped ion
  quantum computer},\ }\href {https://doi.org/10.1038/s42005-024-01698-3}
  {\bibfield  {journal} {\bibinfo  {journal} {Communications Physics}\ }\textbf
  {\bibinfo {volume} {7}},\ \bibinfo {pages} {205} (\bibinfo {year}
  {2024})}\BibitemShut {NoStop}%
\bibitem [{\citenamefont {Cochran}\ \emph {et~al.}(2025)\citenamefont
  {Cochran}, \citenamefont {Jobst}, \citenamefont {Rosenberg}, \citenamefont
  {Lensky}, \citenamefont {Gyawali}, \citenamefont {Eassa}, \citenamefont
  {Will}, \citenamefont {Szasz}, \citenamefont {Abanin}, \citenamefont
  {Acharya}, \citenamefont {Aghababaie~Beni}, \citenamefont {Andersen},
  \citenamefont {Ansmann}, \citenamefont {Arute}, \citenamefont {Arya},
  \citenamefont {Asfaw}, \citenamefont {Atalaya}, \citenamefont {Babbush},
  \citenamefont {Ballard}, \citenamefont {Bardin}, \citenamefont {Bengtsson},
  \citenamefont {Bilmes}, \citenamefont {Bourassa}, \citenamefont {Bovaird},
  \citenamefont {Broughton}, \citenamefont {Browne}, \citenamefont {Buchea},
  \citenamefont {Buckley}, \citenamefont {Burger}, \citenamefont {Burkett},
  \citenamefont {Bushnell}, \citenamefont {Cabrera}, \citenamefont {Campero},
  \citenamefont {Chang}, \citenamefont {Chen}, \citenamefont {Chiaro},
  \citenamefont {Claes}, \citenamefont {Cleland}, \citenamefont {Cogan},
  \citenamefont {Collins}, \citenamefont {Conner}, \citenamefont {Courtney},
  \citenamefont {Crook}, \citenamefont {Curtin}, \citenamefont {Das},
  \citenamefont {Demura}, \citenamefont {De~Lorenzo}, \citenamefont {Di~Paolo},
  \citenamefont {Donohoe}, \citenamefont {Drozdov}, \citenamefont {Dunsworth},
  \citenamefont {Eickbusch}, \citenamefont {Elbag}, \citenamefont {Elzouka},
  \citenamefont {Erickson}, \citenamefont {Ferreira}, \citenamefont {Burgos},
  \citenamefont {Forati}, \citenamefont {Fowler}, \citenamefont {Foxen},
  \citenamefont {Ganjam}, \citenamefont {Gasca}, \citenamefont {Genois},
  \citenamefont {Giang}, \citenamefont {Gilboa}, \citenamefont {Gosula},
  \citenamefont {Grajales~Dau}, \citenamefont {Graumann}, \citenamefont
  {Greene}, \citenamefont {Gross}, \citenamefont {Habegger}, \citenamefont
  {Hansen}, \citenamefont {Harrigan}, \citenamefont {Harrington}, \citenamefont
  {Heu}, \citenamefont {Higgott}, \citenamefont {Hilton}, \citenamefont
  {Huang}, \citenamefont {Huff}, \citenamefont {Huggins}, \citenamefont
  {Jeffrey}, \citenamefont {Jiang}, \citenamefont {Jones}, \citenamefont
  {Joshi}, \citenamefont {Juhas}, \citenamefont {Kafri}, \citenamefont {Kang},
  \citenamefont {Karamlou}, \citenamefont {Kechedzhi}, \citenamefont {Khaire},
  \citenamefont {Khattar}, \citenamefont {Khezri}, \citenamefont {Kim},
  \citenamefont {Klimov}, \citenamefont {Kobrin}, \citenamefont {Korotkov},
  \citenamefont {Kostritsa}, \citenamefont {Kreikebaum}, \citenamefont
  {Kurilovich}, \citenamefont {Landhuis}, \citenamefont {Lange-Dei},
  \citenamefont {Langley}, \citenamefont {Lau}, \citenamefont {Ledford},
  \citenamefont {Lee}, \citenamefont {Lester}, \citenamefont {Le~Guevel},
  \citenamefont {Li}, \citenamefont {Lill}, \citenamefont {Livingston},
  \citenamefont {Locharla}, \citenamefont {Lundahl}, \citenamefont {Lunt},
  \citenamefont {Madhuk}, \citenamefont {Maloney}, \citenamefont {Mandrà},
  \citenamefont {Martin}, \citenamefont {Martin}, \citenamefont {Maxfield},
  \citenamefont {McClean}, \citenamefont {McEwen}, \citenamefont {Meeks},
  \citenamefont {Megrant}, \citenamefont {Miao}, \citenamefont {Molavi},
  \citenamefont {Molina}, \citenamefont {Montazeri}, \citenamefont {Movassagh},
  \citenamefont {Neill}, \citenamefont {Newman}, \citenamefont {Nguyen},
  \citenamefont {Nguyen}, \citenamefont {Ni}, \citenamefont {Ottosson},
  \citenamefont {Pizzuto}, \citenamefont {Potter}, \citenamefont {Pritchard},
  \citenamefont {Quintana}, \citenamefont {Ramachandran}, \citenamefont
  {Reagor}, \citenamefont {Rhodes}, \citenamefont {Roberts}, \citenamefont
  {Sankaragomathi}, \citenamefont {Satzinger}, \citenamefont {Schurkus},
  \citenamefont {Shearn}, \citenamefont {Shorter}, \citenamefont {Shutty},
  \citenamefont {Shvarts}, \citenamefont {Sivak}, \citenamefont {Small},
  \citenamefont {Smith}, \citenamefont {Springer}, \citenamefont {Sterling},
  \citenamefont {Suchard}, \citenamefont {Sztein}, \citenamefont {Thor},
  \citenamefont {Torunbalci}, \citenamefont {Vaishnav}, \citenamefont {Vargas},
  \citenamefont {Vdovichev}, \citenamefont {Vidal}, \citenamefont
  {Vollgraff~Heidweiller}, \citenamefont {Waltman}, \citenamefont {Wang},
  \citenamefont {Ware}, \citenamefont {White}, \citenamefont {Wong},
  \citenamefont {Woo}, \citenamefont {Xing}, \citenamefont {Yao}, \citenamefont
  {Yeh}, \citenamefont {Ying}, \citenamefont {Yoo}, \citenamefont {Yosri},
  \citenamefont {Young}, \citenamefont {Zalcman}, \citenamefont {Zhang},
  \citenamefont {Zhu}, \citenamefont {Zobrist}, \citenamefont {Boixo},
  \citenamefont {Kelly}, \citenamefont {Lucero}, \citenamefont {Chen},
  \citenamefont {Smelyanskiy}, \citenamefont {Neven}, \citenamefont
  {Gammon-Smith}, \citenamefont {Pollmann}, \citenamefont {Knap},\ and\
  \citenamefont {Roushan}}]{Cochran2024}%
  \BibitemOpen
  \bibfield  {author} {\bibinfo {author} {\bibfnamefont {T.~A.}\ \bibnamefont
  {Cochran}}, \bibinfo {author} {\bibfnamefont {B.}~\bibnamefont {Jobst}},
  \bibinfo {author} {\bibfnamefont {E.}~\bibnamefont {Rosenberg}}, \bibinfo
  {author} {\bibfnamefont {Y.~D.}\ \bibnamefont {Lensky}}, \bibinfo {author}
  {\bibfnamefont {G.}~\bibnamefont {Gyawali}}, \bibinfo {author} {\bibfnamefont
  {N.}~\bibnamefont {Eassa}}, \bibinfo {author} {\bibfnamefont
  {M.}~\bibnamefont {Will}}, \bibinfo {author} {\bibfnamefont {A.}~\bibnamefont
  {Szasz}}, \bibinfo {author} {\bibfnamefont {D.}~\bibnamefont {Abanin}},
  \bibinfo {author} {\bibfnamefont {R.}~\bibnamefont {Acharya}}, \bibinfo
  {author} {\bibfnamefont {L.}~\bibnamefont {Aghababaie~Beni}}, \bibinfo
  {author} {\bibfnamefont {T.~I.}\ \bibnamefont {Andersen}}, \bibinfo {author}
  {\bibfnamefont {M.}~\bibnamefont {Ansmann}}, \bibinfo {author} {\bibfnamefont
  {F.}~\bibnamefont {Arute}}, \bibinfo {author} {\bibfnamefont
  {K.}~\bibnamefont {Arya}}, \bibinfo {author} {\bibfnamefont {A.}~\bibnamefont
  {Asfaw}}, \bibinfo {author} {\bibfnamefont {J.}~\bibnamefont {Atalaya}},
  \bibinfo {author} {\bibfnamefont {R.}~\bibnamefont {Babbush}}, \bibinfo
  {author} {\bibfnamefont {B.}~\bibnamefont {Ballard}}, \bibinfo {author}
  {\bibfnamefont {J.~C.}\ \bibnamefont {Bardin}}, \bibinfo {author}
  {\bibfnamefont {A.}~\bibnamefont {Bengtsson}}, \bibinfo {author}
  {\bibfnamefont {A.}~\bibnamefont {Bilmes}}, \bibinfo {author} {\bibfnamefont
  {A.}~\bibnamefont {Bourassa}}, \bibinfo {author} {\bibfnamefont
  {J.}~\bibnamefont {Bovaird}}, \bibinfo {author} {\bibfnamefont
  {M.}~\bibnamefont {Broughton}}, \bibinfo {author} {\bibfnamefont {D.~A.}\
  \bibnamefont {Browne}}, \bibinfo {author} {\bibfnamefont {B.}~\bibnamefont
  {Buchea}}, \bibinfo {author} {\bibfnamefont {B.~B.}\ \bibnamefont {Buckley}},
  \bibinfo {author} {\bibfnamefont {T.}~\bibnamefont {Burger}}, \bibinfo
  {author} {\bibfnamefont {B.}~\bibnamefont {Burkett}}, \bibinfo {author}
  {\bibfnamefont {N.}~\bibnamefont {Bushnell}}, \bibinfo {author}
  {\bibfnamefont {A.}~\bibnamefont {Cabrera}}, \bibinfo {author} {\bibfnamefont
  {J.}~\bibnamefont {Campero}}, \bibinfo {author} {\bibfnamefont {H.-S.}\
  \bibnamefont {Chang}}, \bibinfo {author} {\bibfnamefont {Z.}~\bibnamefont
  {Chen}}, \bibinfo {author} {\bibfnamefont {B.}~\bibnamefont {Chiaro}},
  \bibinfo {author} {\bibfnamefont {J.}~\bibnamefont {Claes}}, \bibinfo
  {author} {\bibfnamefont {A.~Y.}\ \bibnamefont {Cleland}}, \bibinfo {author}
  {\bibfnamefont {J.}~\bibnamefont {Cogan}}, \bibinfo {author} {\bibfnamefont
  {R.}~\bibnamefont {Collins}}, \bibinfo {author} {\bibfnamefont
  {P.}~\bibnamefont {Conner}}, \bibinfo {author} {\bibfnamefont
  {W.}~\bibnamefont {Courtney}}, \bibinfo {author} {\bibfnamefont {A.~L.}\
  \bibnamefont {Crook}}, \bibinfo {author} {\bibfnamefont {B.}~\bibnamefont
  {Curtin}}, \bibinfo {author} {\bibfnamefont {S.}~\bibnamefont {Das}},
  \bibinfo {author} {\bibfnamefont {S.}~\bibnamefont {Demura}}, \bibinfo
  {author} {\bibfnamefont {L.}~\bibnamefont {De~Lorenzo}}, \bibinfo {author}
  {\bibfnamefont {A.}~\bibnamefont {Di~Paolo}}, \bibinfo {author}
  {\bibfnamefont {P.}~\bibnamefont {Donohoe}}, \bibinfo {author} {\bibfnamefont
  {I.}~\bibnamefont {Drozdov}}, \bibinfo {author} {\bibfnamefont
  {A.}~\bibnamefont {Dunsworth}}, \bibinfo {author} {\bibfnamefont
  {A.}~\bibnamefont {Eickbusch}}, \bibinfo {author} {\bibfnamefont {A.~M.}\
  \bibnamefont {Elbag}}, \bibinfo {author} {\bibfnamefont {M.}~\bibnamefont
  {Elzouka}}, \bibinfo {author} {\bibfnamefont {C.}~\bibnamefont {Erickson}},
  \bibinfo {author} {\bibfnamefont {V.~S.}\ \bibnamefont {Ferreira}}, \bibinfo
  {author} {\bibfnamefont {L.~F.}\ \bibnamefont {Burgos}}, \bibinfo {author}
  {\bibfnamefont {E.}~\bibnamefont {Forati}}, \bibinfo {author} {\bibfnamefont
  {A.~G.}\ \bibnamefont {Fowler}}, \bibinfo {author} {\bibfnamefont
  {B.}~\bibnamefont {Foxen}}, \bibinfo {author} {\bibfnamefont
  {S.}~\bibnamefont {Ganjam}}, \bibinfo {author} {\bibfnamefont
  {R.}~\bibnamefont {Gasca}}, \bibinfo {author} {\bibfnamefont
  {E.}~\bibnamefont {Genois}}, \bibinfo {author} {\bibfnamefont
  {W.}~\bibnamefont {Giang}}, \bibinfo {author} {\bibfnamefont
  {D.}~\bibnamefont {Gilboa}}, \bibinfo {author} {\bibfnamefont
  {R.}~\bibnamefont {Gosula}}, \bibinfo {author} {\bibfnamefont
  {A.}~\bibnamefont {Grajales~Dau}}, \bibinfo {author} {\bibfnamefont
  {D.}~\bibnamefont {Graumann}}, \bibinfo {author} {\bibfnamefont
  {A.}~\bibnamefont {Greene}}, \bibinfo {author} {\bibfnamefont {J.~A.}\
  \bibnamefont {Gross}}, \bibinfo {author} {\bibfnamefont {S.}~\bibnamefont
  {Habegger}}, \bibinfo {author} {\bibfnamefont {M.}~\bibnamefont {Hansen}},
  \bibinfo {author} {\bibfnamefont {M.~P.}\ \bibnamefont {Harrigan}}, \bibinfo
  {author} {\bibfnamefont {S.~D.}\ \bibnamefont {Harrington}}, \bibinfo
  {author} {\bibfnamefont {P.}~\bibnamefont {Heu}}, \bibinfo {author}
  {\bibfnamefont {O.}~\bibnamefont {Higgott}}, \bibinfo {author} {\bibfnamefont
  {J.}~\bibnamefont {Hilton}}, \bibinfo {author} {\bibfnamefont {H.-Y.}\
  \bibnamefont {Huang}}, \bibinfo {author} {\bibfnamefont {A.}~\bibnamefont
  {Huff}}, \bibinfo {author} {\bibfnamefont {W.}~\bibnamefont {Huggins}},
  \bibinfo {author} {\bibfnamefont {E.}~\bibnamefont {Jeffrey}}, \bibinfo
  {author} {\bibfnamefont {Z.}~\bibnamefont {Jiang}}, \bibinfo {author}
  {\bibfnamefont {C.}~\bibnamefont {Jones}}, \bibinfo {author} {\bibfnamefont
  {C.}~\bibnamefont {Joshi}}, \bibinfo {author} {\bibfnamefont
  {P.}~\bibnamefont {Juhas}}, \bibinfo {author} {\bibfnamefont
  {D.}~\bibnamefont {Kafri}}, \bibinfo {author} {\bibfnamefont
  {H.}~\bibnamefont {Kang}}, \bibinfo {author} {\bibfnamefont {A.~H.}\
  \bibnamefont {Karamlou}}, \bibinfo {author} {\bibfnamefont {K.}~\bibnamefont
  {Kechedzhi}}, \bibinfo {author} {\bibfnamefont {T.}~\bibnamefont {Khaire}},
  \bibinfo {author} {\bibfnamefont {T.}~\bibnamefont {Khattar}}, \bibinfo
  {author} {\bibfnamefont {M.}~\bibnamefont {Khezri}}, \bibinfo {author}
  {\bibfnamefont {S.}~\bibnamefont {Kim}}, \bibinfo {author} {\bibfnamefont
  {P.}~\bibnamefont {Klimov}}, \bibinfo {author} {\bibfnamefont
  {B.}~\bibnamefont {Kobrin}}, \bibinfo {author} {\bibfnamefont
  {A.}~\bibnamefont {Korotkov}}, \bibinfo {author} {\bibfnamefont
  {F.}~\bibnamefont {Kostritsa}}, \bibinfo {author} {\bibfnamefont
  {J.}~\bibnamefont {Kreikebaum}}, \bibinfo {author} {\bibfnamefont
  {V.}~\bibnamefont {Kurilovich}}, \bibinfo {author} {\bibfnamefont
  {D.}~\bibnamefont {Landhuis}}, \bibinfo {author} {\bibfnamefont
  {T.}~\bibnamefont {Lange-Dei}}, \bibinfo {author} {\bibfnamefont
  {B.}~\bibnamefont {Langley}}, \bibinfo {author} {\bibfnamefont {K.-M.}\
  \bibnamefont {Lau}}, \bibinfo {author} {\bibfnamefont {J.}~\bibnamefont
  {Ledford}}, \bibinfo {author} {\bibfnamefont {K.}~\bibnamefont {Lee}},
  \bibinfo {author} {\bibfnamefont {B.}~\bibnamefont {Lester}}, \bibinfo
  {author} {\bibfnamefont {L.}~\bibnamefont {Le~Guevel}}, \bibinfo {author}
  {\bibfnamefont {W.}~\bibnamefont {Li}}, \bibinfo {author} {\bibfnamefont
  {A.~T.}\ \bibnamefont {Lill}}, \bibinfo {author} {\bibfnamefont
  {W.}~\bibnamefont {Livingston}}, \bibinfo {author} {\bibfnamefont
  {A.}~\bibnamefont {Locharla}}, \bibinfo {author} {\bibfnamefont
  {D.}~\bibnamefont {Lundahl}}, \bibinfo {author} {\bibfnamefont
  {A.}~\bibnamefont {Lunt}}, \bibinfo {author} {\bibfnamefont {S.}~\bibnamefont
  {Madhuk}}, \bibinfo {author} {\bibfnamefont {A.}~\bibnamefont {Maloney}},
  \bibinfo {author} {\bibfnamefont {S.}~\bibnamefont {Mandrà}}, \bibinfo
  {author} {\bibfnamefont {L.}~\bibnamefont {Martin}}, \bibinfo {author}
  {\bibfnamefont {O.}~\bibnamefont {Martin}}, \bibinfo {author} {\bibfnamefont
  {C.}~\bibnamefont {Maxfield}}, \bibinfo {author} {\bibfnamefont
  {J.}~\bibnamefont {McClean}}, \bibinfo {author} {\bibfnamefont
  {M.}~\bibnamefont {McEwen}}, \bibinfo {author} {\bibfnamefont
  {S.}~\bibnamefont {Meeks}}, \bibinfo {author} {\bibfnamefont
  {A.}~\bibnamefont {Megrant}}, \bibinfo {author} {\bibfnamefont
  {K.}~\bibnamefont {Miao}}, \bibinfo {author} {\bibfnamefont {R.}~\bibnamefont
  {Molavi}}, \bibinfo {author} {\bibfnamefont {S.}~\bibnamefont {Molina}},
  \bibinfo {author} {\bibfnamefont {S.}~\bibnamefont {Montazeri}}, \bibinfo
  {author} {\bibfnamefont {R.}~\bibnamefont {Movassagh}}, \bibinfo {author}
  {\bibfnamefont {C.}~\bibnamefont {Neill}}, \bibinfo {author} {\bibfnamefont
  {M.}~\bibnamefont {Newman}}, \bibinfo {author} {\bibfnamefont
  {A.}~\bibnamefont {Nguyen}}, \bibinfo {author} {\bibfnamefont
  {M.}~\bibnamefont {Nguyen}}, \bibinfo {author} {\bibfnamefont {C.-H.}\
  \bibnamefont {Ni}}, \bibinfo {author} {\bibfnamefont {K.}~\bibnamefont
  {Ottosson}}, \bibinfo {author} {\bibfnamefont {A.}~\bibnamefont {Pizzuto}},
  \bibinfo {author} {\bibfnamefont {R.}~\bibnamefont {Potter}}, \bibinfo
  {author} {\bibfnamefont {O.}~\bibnamefont {Pritchard}}, \bibinfo {author}
  {\bibfnamefont {C.}~\bibnamefont {Quintana}}, \bibinfo {author}
  {\bibfnamefont {G.}~\bibnamefont {Ramachandran}}, \bibinfo {author}
  {\bibfnamefont {M.}~\bibnamefont {Reagor}}, \bibinfo {author} {\bibfnamefont
  {D.}~\bibnamefont {Rhodes}}, \bibinfo {author} {\bibfnamefont
  {G.}~\bibnamefont {Roberts}}, \bibinfo {author} {\bibfnamefont
  {K.}~\bibnamefont {Sankaragomathi}}, \bibinfo {author} {\bibfnamefont
  {K.}~\bibnamefont {Satzinger}}, \bibinfo {author} {\bibfnamefont
  {H.}~\bibnamefont {Schurkus}}, \bibinfo {author} {\bibfnamefont
  {M.}~\bibnamefont {Shearn}}, \bibinfo {author} {\bibfnamefont
  {A.}~\bibnamefont {Shorter}}, \bibinfo {author} {\bibfnamefont
  {N.}~\bibnamefont {Shutty}}, \bibinfo {author} {\bibfnamefont
  {V.}~\bibnamefont {Shvarts}}, \bibinfo {author} {\bibfnamefont
  {V.}~\bibnamefont {Sivak}}, \bibinfo {author} {\bibfnamefont
  {S.}~\bibnamefont {Small}}, \bibinfo {author} {\bibfnamefont {W.~C.}\
  \bibnamefont {Smith}}, \bibinfo {author} {\bibfnamefont {S.}~\bibnamefont
  {Springer}}, \bibinfo {author} {\bibfnamefont {G.}~\bibnamefont {Sterling}},
  \bibinfo {author} {\bibfnamefont {J.}~\bibnamefont {Suchard}}, \bibinfo
  {author} {\bibfnamefont {A.}~\bibnamefont {Sztein}}, \bibinfo {author}
  {\bibfnamefont {D.}~\bibnamefont {Thor}}, \bibinfo {author} {\bibfnamefont
  {M.}~\bibnamefont {Torunbalci}}, \bibinfo {author} {\bibfnamefont
  {A.}~\bibnamefont {Vaishnav}}, \bibinfo {author} {\bibfnamefont
  {J.}~\bibnamefont {Vargas}}, \bibinfo {author} {\bibfnamefont
  {S.}~\bibnamefont {Vdovichev}}, \bibinfo {author} {\bibfnamefont
  {G.}~\bibnamefont {Vidal}}, \bibinfo {author} {\bibfnamefont
  {C.}~\bibnamefont {Vollgraff~Heidweiller}}, \bibinfo {author} {\bibfnamefont
  {S.}~\bibnamefont {Waltman}}, \bibinfo {author} {\bibfnamefont {S.~X.}\
  \bibnamefont {Wang}}, \bibinfo {author} {\bibfnamefont {B.}~\bibnamefont
  {Ware}}, \bibinfo {author} {\bibfnamefont {T.}~\bibnamefont {White}},
  \bibinfo {author} {\bibfnamefont {K.}~\bibnamefont {Wong}}, \bibinfo {author}
  {\bibfnamefont {B.~W.~K.}\ \bibnamefont {Woo}}, \bibinfo {author}
  {\bibfnamefont {C.}~\bibnamefont {Xing}}, \bibinfo {author} {\bibfnamefont
  {Z.~J.}\ \bibnamefont {Yao}}, \bibinfo {author} {\bibfnamefont
  {P.}~\bibnamefont {Yeh}}, \bibinfo {author} {\bibfnamefont {B.}~\bibnamefont
  {Ying}}, \bibinfo {author} {\bibfnamefont {J.}~\bibnamefont {Yoo}}, \bibinfo
  {author} {\bibfnamefont {N.}~\bibnamefont {Yosri}}, \bibinfo {author}
  {\bibfnamefont {G.}~\bibnamefont {Young}}, \bibinfo {author} {\bibfnamefont
  {A.}~\bibnamefont {Zalcman}}, \bibinfo {author} {\bibfnamefont
  {Y.}~\bibnamefont {Zhang}}, \bibinfo {author} {\bibfnamefont
  {N.}~\bibnamefont {Zhu}}, \bibinfo {author} {\bibfnamefont {N.}~\bibnamefont
  {Zobrist}}, \bibinfo {author} {\bibfnamefont {S.}~\bibnamefont {Boixo}},
  \bibinfo {author} {\bibfnamefont {J.}~\bibnamefont {Kelly}}, \bibinfo
  {author} {\bibfnamefont {E.}~\bibnamefont {Lucero}}, \bibinfo {author}
  {\bibfnamefont {Y.}~\bibnamefont {Chen}}, \bibinfo {author} {\bibfnamefont
  {V.}~\bibnamefont {Smelyanskiy}}, \bibinfo {author} {\bibfnamefont
  {H.}~\bibnamefont {Neven}}, \bibinfo {author} {\bibfnamefont
  {A.}~\bibnamefont {Gammon-Smith}}, \bibinfo {author} {\bibfnamefont
  {F.}~\bibnamefont {Pollmann}}, \bibinfo {author} {\bibfnamefont
  {M.}~\bibnamefont {Knap}},\ and\ \bibinfo {author} {\bibfnamefont
  {P.}~\bibnamefont {Roushan}},\ }\bibfield  {title} {\bibinfo {title}
  {Visualizing dynamics of charges and strings in (2 + 1){D} lattice gauge
  theories},\ }\href {https://doi.org/10.1038/s41586-025-08999-9} {\bibfield
  {journal} {\bibinfo  {journal} {Nature}\ }\textbf {\bibinfo {volume} {642}},\
  \bibinfo {pages} {315} (\bibinfo {year} {2025})}\BibitemShut {NoStop}%
\bibitem [{\citenamefont {Satzinger}\ \emph {et~al.}(2021)\citenamefont
  {Satzinger}, \citenamefont {Liu}, \citenamefont {Smith}, \citenamefont
  {Knapp}, \citenamefont {Newman}, \citenamefont {Jones}, \citenamefont {Chen},
  \citenamefont {Quintana}, \citenamefont {Mi}, \citenamefont {Dunsworth},
  \citenamefont {Gidney}, \citenamefont {Aleiner}, \citenamefont {Arute},
  \citenamefont {Arya}, \citenamefont {Atalaya}, \citenamefont {Babbush},
  \citenamefont {Bardin}, \citenamefont {Barends}, \citenamefont {Basso},
  \citenamefont {Bengtsson}, \citenamefont {Bilmes}, \citenamefont {Broughton},
  \citenamefont {Buckley}, \citenamefont {Buell}, \citenamefont {Burkett},
  \citenamefont {Bushnell}, \citenamefont {Chiaro}, \citenamefont {Collins},
  \citenamefont {Courtney}, \citenamefont {Demura}, \citenamefont {Derk},
  \citenamefont {Eppens}, \citenamefont {Erickson}, \citenamefont {Faoro},
  \citenamefont {Farhi}, \citenamefont {Fowler}, \citenamefont {Foxen},
  \citenamefont {Giustina}, \citenamefont {Greene}, \citenamefont {Gross},
  \citenamefont {Harrigan}, \citenamefont {Harrington}, \citenamefont {Hilton},
  \citenamefont {Hong}, \citenamefont {Huang}, \citenamefont {Huggins},
  \citenamefont {Ioffe}, \citenamefont {Isakov}, \citenamefont {Jeffrey},
  \citenamefont {Jiang}, \citenamefont {Kafri}, \citenamefont {Kechedzhi},
  \citenamefont {Khattar}, \citenamefont {Kim}, \citenamefont {Klimov},
  \citenamefont {Korotkov}, \citenamefont {Kostritsa}, \citenamefont
  {Landhuis}, \citenamefont {Laptev}, \citenamefont {Locharla}, \citenamefont
  {Lucero}, \citenamefont {Martin}, \citenamefont {McClean}, \citenamefont
  {McEwen}, \citenamefont {Miao}, \citenamefont {Mohseni}, \citenamefont
  {Montazeri}, \citenamefont {Mruczkiewicz}, \citenamefont {Mutus},
  \citenamefont {Naaman}, \citenamefont {Neeley}, \citenamefont {Neill},
  \citenamefont {Niu}, \citenamefont {O’Brien}, \citenamefont {Opremcak},
  \citenamefont {Pató}, \citenamefont {Petukhov}, \citenamefont {Rubin},
  \citenamefont {Sank}, \citenamefont {Shvarts}, \citenamefont {Strain},
  \citenamefont {Szalay}, \citenamefont {Villalonga}, \citenamefont {White},
  \citenamefont {Yao}, \citenamefont {Yeh}, \citenamefont {Yoo}, \citenamefont
  {Zalcman}, \citenamefont {Neven}, \citenamefont {Boixo}, \citenamefont
  {Megrant}, \citenamefont {Chen}, \citenamefont {Kelly}, \citenamefont
  {Smelyanskiy}, \citenamefont {Kitaev}, \citenamefont {Knap}, \citenamefont
  {Pollmann},\ and\ \citenamefont {Roushan}}]{Satzinger2021}%
  \BibitemOpen
  \bibfield  {author} {\bibinfo {author} {\bibfnamefont {K.~J.}\ \bibnamefont
  {Satzinger}}, \bibinfo {author} {\bibfnamefont {Y.-J.}\ \bibnamefont {Liu}},
  \bibinfo {author} {\bibfnamefont {A.}~\bibnamefont {Smith}}, \bibinfo
  {author} {\bibfnamefont {C.}~\bibnamefont {Knapp}}, \bibinfo {author}
  {\bibfnamefont {M.}~\bibnamefont {Newman}}, \bibinfo {author} {\bibfnamefont
  {C.}~\bibnamefont {Jones}}, \bibinfo {author} {\bibfnamefont
  {Z.}~\bibnamefont {Chen}}, \bibinfo {author} {\bibfnamefont {C.}~\bibnamefont
  {Quintana}}, \bibinfo {author} {\bibfnamefont {X.}~\bibnamefont {Mi}},
  \bibinfo {author} {\bibfnamefont {A.}~\bibnamefont {Dunsworth}}, \bibinfo
  {author} {\bibfnamefont {C.}~\bibnamefont {Gidney}}, \bibinfo {author}
  {\bibfnamefont {I.}~\bibnamefont {Aleiner}}, \bibinfo {author} {\bibfnamefont
  {F.}~\bibnamefont {Arute}}, \bibinfo {author} {\bibfnamefont
  {K.}~\bibnamefont {Arya}}, \bibinfo {author} {\bibfnamefont {J.}~\bibnamefont
  {Atalaya}}, \bibinfo {author} {\bibfnamefont {R.}~\bibnamefont {Babbush}},
  \bibinfo {author} {\bibfnamefont {J.~C.}\ \bibnamefont {Bardin}}, \bibinfo
  {author} {\bibfnamefont {R.}~\bibnamefont {Barends}}, \bibinfo {author}
  {\bibfnamefont {J.}~\bibnamefont {Basso}}, \bibinfo {author} {\bibfnamefont
  {A.}~\bibnamefont {Bengtsson}}, \bibinfo {author} {\bibfnamefont
  {A.}~\bibnamefont {Bilmes}}, \bibinfo {author} {\bibfnamefont
  {M.}~\bibnamefont {Broughton}}, \bibinfo {author} {\bibfnamefont {B.~B.}\
  \bibnamefont {Buckley}}, \bibinfo {author} {\bibfnamefont {D.~A.}\
  \bibnamefont {Buell}}, \bibinfo {author} {\bibfnamefont {B.}~\bibnamefont
  {Burkett}}, \bibinfo {author} {\bibfnamefont {N.}~\bibnamefont {Bushnell}},
  \bibinfo {author} {\bibfnamefont {B.}~\bibnamefont {Chiaro}}, \bibinfo
  {author} {\bibfnamefont {R.}~\bibnamefont {Collins}}, \bibinfo {author}
  {\bibfnamefont {W.}~\bibnamefont {Courtney}}, \bibinfo {author}
  {\bibfnamefont {S.}~\bibnamefont {Demura}}, \bibinfo {author} {\bibfnamefont
  {A.~R.}\ \bibnamefont {Derk}}, \bibinfo {author} {\bibfnamefont
  {D.}~\bibnamefont {Eppens}}, \bibinfo {author} {\bibfnamefont
  {C.}~\bibnamefont {Erickson}}, \bibinfo {author} {\bibfnamefont
  {L.}~\bibnamefont {Faoro}}, \bibinfo {author} {\bibfnamefont
  {E.}~\bibnamefont {Farhi}}, \bibinfo {author} {\bibfnamefont {A.~G.}\
  \bibnamefont {Fowler}}, \bibinfo {author} {\bibfnamefont {B.}~\bibnamefont
  {Foxen}}, \bibinfo {author} {\bibfnamefont {M.}~\bibnamefont {Giustina}},
  \bibinfo {author} {\bibfnamefont {A.}~\bibnamefont {Greene}}, \bibinfo
  {author} {\bibfnamefont {J.~A.}\ \bibnamefont {Gross}}, \bibinfo {author}
  {\bibfnamefont {M.~P.}\ \bibnamefont {Harrigan}}, \bibinfo {author}
  {\bibfnamefont {S.~D.}\ \bibnamefont {Harrington}}, \bibinfo {author}
  {\bibfnamefont {J.}~\bibnamefont {Hilton}}, \bibinfo {author} {\bibfnamefont
  {S.}~\bibnamefont {Hong}}, \bibinfo {author} {\bibfnamefont {T.}~\bibnamefont
  {Huang}}, \bibinfo {author} {\bibfnamefont {W.~J.}\ \bibnamefont {Huggins}},
  \bibinfo {author} {\bibfnamefont {L.~B.}\ \bibnamefont {Ioffe}}, \bibinfo
  {author} {\bibfnamefont {S.~V.}\ \bibnamefont {Isakov}}, \bibinfo {author}
  {\bibfnamefont {E.}~\bibnamefont {Jeffrey}}, \bibinfo {author} {\bibfnamefont
  {Z.}~\bibnamefont {Jiang}}, \bibinfo {author} {\bibfnamefont
  {D.}~\bibnamefont {Kafri}}, \bibinfo {author} {\bibfnamefont
  {K.}~\bibnamefont {Kechedzhi}}, \bibinfo {author} {\bibfnamefont
  {T.}~\bibnamefont {Khattar}}, \bibinfo {author} {\bibfnamefont
  {S.}~\bibnamefont {Kim}}, \bibinfo {author} {\bibfnamefont {P.~V.}\
  \bibnamefont {Klimov}}, \bibinfo {author} {\bibfnamefont {A.~N.}\
  \bibnamefont {Korotkov}}, \bibinfo {author} {\bibfnamefont {F.}~\bibnamefont
  {Kostritsa}}, \bibinfo {author} {\bibfnamefont {D.}~\bibnamefont {Landhuis}},
  \bibinfo {author} {\bibfnamefont {P.}~\bibnamefont {Laptev}}, \bibinfo
  {author} {\bibfnamefont {A.}~\bibnamefont {Locharla}}, \bibinfo {author}
  {\bibfnamefont {E.}~\bibnamefont {Lucero}}, \bibinfo {author} {\bibfnamefont
  {O.}~\bibnamefont {Martin}}, \bibinfo {author} {\bibfnamefont {J.~R.}\
  \bibnamefont {McClean}}, \bibinfo {author} {\bibfnamefont {M.}~\bibnamefont
  {McEwen}}, \bibinfo {author} {\bibfnamefont {K.~C.}\ \bibnamefont {Miao}},
  \bibinfo {author} {\bibfnamefont {M.}~\bibnamefont {Mohseni}}, \bibinfo
  {author} {\bibfnamefont {S.}~\bibnamefont {Montazeri}}, \bibinfo {author}
  {\bibfnamefont {W.}~\bibnamefont {Mruczkiewicz}}, \bibinfo {author}
  {\bibfnamefont {J.}~\bibnamefont {Mutus}}, \bibinfo {author} {\bibfnamefont
  {O.}~\bibnamefont {Naaman}}, \bibinfo {author} {\bibfnamefont
  {M.}~\bibnamefont {Neeley}}, \bibinfo {author} {\bibfnamefont
  {C.}~\bibnamefont {Neill}}, \bibinfo {author} {\bibfnamefont {M.~Y.}\
  \bibnamefont {Niu}}, \bibinfo {author} {\bibfnamefont {T.~E.}\ \bibnamefont
  {O’Brien}}, \bibinfo {author} {\bibfnamefont {A.}~\bibnamefont {Opremcak}},
  \bibinfo {author} {\bibfnamefont {B.}~\bibnamefont {Pató}}, \bibinfo
  {author} {\bibfnamefont {A.}~\bibnamefont {Petukhov}}, \bibinfo {author}
  {\bibfnamefont {N.~C.}\ \bibnamefont {Rubin}}, \bibinfo {author}
  {\bibfnamefont {D.}~\bibnamefont {Sank}}, \bibinfo {author} {\bibfnamefont
  {V.}~\bibnamefont {Shvarts}}, \bibinfo {author} {\bibfnamefont
  {D.}~\bibnamefont {Strain}}, \bibinfo {author} {\bibfnamefont
  {M.}~\bibnamefont {Szalay}}, \bibinfo {author} {\bibfnamefont
  {B.}~\bibnamefont {Villalonga}}, \bibinfo {author} {\bibfnamefont {T.~C.}\
  \bibnamefont {White}}, \bibinfo {author} {\bibfnamefont {Z.}~\bibnamefont
  {Yao}}, \bibinfo {author} {\bibfnamefont {P.}~\bibnamefont {Yeh}}, \bibinfo
  {author} {\bibfnamefont {J.}~\bibnamefont {Yoo}}, \bibinfo {author}
  {\bibfnamefont {A.}~\bibnamefont {Zalcman}}, \bibinfo {author} {\bibfnamefont
  {H.}~\bibnamefont {Neven}}, \bibinfo {author} {\bibfnamefont
  {S.}~\bibnamefont {Boixo}}, \bibinfo {author} {\bibfnamefont
  {A.}~\bibnamefont {Megrant}}, \bibinfo {author} {\bibfnamefont
  {Y.}~\bibnamefont {Chen}}, \bibinfo {author} {\bibfnamefont {J.}~\bibnamefont
  {Kelly}}, \bibinfo {author} {\bibfnamefont {V.}~\bibnamefont {Smelyanskiy}},
  \bibinfo {author} {\bibfnamefont {A.}~\bibnamefont {Kitaev}}, \bibinfo
  {author} {\bibfnamefont {M.}~\bibnamefont {Knap}}, \bibinfo {author}
  {\bibfnamefont {F.}~\bibnamefont {Pollmann}},\ and\ \bibinfo {author}
  {\bibfnamefont {P.}~\bibnamefont {Roushan}},\ }\bibfield  {title} {\bibinfo
  {title} {Realizing topologically ordered states on a quantum processor},\
  }\href {https://doi.org/10.1126/science.abi8378} {\bibfield  {journal}
  {\bibinfo  {journal} {Science}\ }\textbf {\bibinfo {volume} {374}},\ \bibinfo
  {pages} {1237} (\bibinfo {year} {2021})}\BibitemShut {NoStop}%
\bibitem [{\citenamefont {Sun}\ \emph {et~al.}(2023)\citenamefont {Sun},
  \citenamefont {Shirakawa},\ and\ \citenamefont {Yunoki}}]{Sun2023f}%
  \BibitemOpen
  \bibfield  {author} {\bibinfo {author} {\bibfnamefont {R.-Y.}\ \bibnamefont
  {Sun}}, \bibinfo {author} {\bibfnamefont {T.}~\bibnamefont {Shirakawa}},\
  and\ \bibinfo {author} {\bibfnamefont {S.}~\bibnamefont {Yunoki}},\
  }\bibfield  {title} {\bibinfo {title} {Parametrized quantum circuit for
  weight-adjustable quantum loop gas},\ }\href
  {https://doi.org/10.1103/PhysRevB.107.L041109} {\bibfield  {journal}
  {\bibinfo  {journal} {Phys. Rev. B}\ }\textbf {\bibinfo {volume} {107}},\
  \bibinfo {pages} {L041109} (\bibinfo {year} {2023})}\BibitemShut {NoStop}%
\bibitem [{\citenamefont {Dusuel}\ \emph {et~al.}(2011)\citenamefont {Dusuel},
  \citenamefont {Kamfor}, \citenamefont {Or\'us}, \citenamefont {Schmidt},\
  and\ \citenamefont {Vidal}}]{Dusuel2011}%
  \BibitemOpen
  \bibfield  {author} {\bibinfo {author} {\bibfnamefont {S.}~\bibnamefont
  {Dusuel}}, \bibinfo {author} {\bibfnamefont {M.}~\bibnamefont {Kamfor}},
  \bibinfo {author} {\bibfnamefont {R.}~\bibnamefont {Or\'us}}, \bibinfo
  {author} {\bibfnamefont {K.~P.}\ \bibnamefont {Schmidt}},\ and\ \bibinfo
  {author} {\bibfnamefont {J.}~\bibnamefont {Vidal}},\ }\bibfield  {title}
  {\bibinfo {title} {Robustness of a perturbed topological phase},\ }\href
  {https://doi.org/10.1103/PhysRevLett.106.107203} {\bibfield  {journal}
  {\bibinfo  {journal} {Phys. Rev. Lett.}\ }\textbf {\bibinfo {volume} {106}},\
  \bibinfo {pages} {107203} (\bibinfo {year} {2011})}\BibitemShut {NoStop}%
\bibitem [{\citenamefont {Kitaev}\ and\ \citenamefont
  {Preskill}(2006)}]{Kitaev2006}%
  \BibitemOpen
  \bibfield  {author} {\bibinfo {author} {\bibfnamefont {A.}~\bibnamefont
  {Kitaev}}\ and\ \bibinfo {author} {\bibfnamefont {J.}~\bibnamefont
  {Preskill}},\ }\bibfield  {title} {\bibinfo {title} {Topological entanglement
  entropy},\ }\href {https://doi.org/10.1103/PhysRevLett.96.110404} {\bibfield
  {journal} {\bibinfo  {journal} {Phys. Rev. Lett.}\ }\textbf {\bibinfo
  {volume} {96}},\ \bibinfo {pages} {110404} (\bibinfo {year}
  {2006})}\BibitemShut {NoStop}%
\bibitem [{\citenamefont {Brydges}\ \emph {et~al.}(2019)\citenamefont
  {Brydges}, \citenamefont {Elben}, \citenamefont {Jurcevic}, \citenamefont
  {Vermersch}, \citenamefont {Maier}, \citenamefont {Lanyon}, \citenamefont
  {Zoller}, \citenamefont {Blatt},\ and\ \citenamefont {Roos}}]{Brydges2019}%
  \BibitemOpen
  \bibfield  {author} {\bibinfo {author} {\bibfnamefont {T.}~\bibnamefont
  {Brydges}}, \bibinfo {author} {\bibfnamefont {A.}~\bibnamefont {Elben}},
  \bibinfo {author} {\bibfnamefont {P.}~\bibnamefont {Jurcevic}}, \bibinfo
  {author} {\bibfnamefont {B.}~\bibnamefont {Vermersch}}, \bibinfo {author}
  {\bibfnamefont {C.}~\bibnamefont {Maier}}, \bibinfo {author} {\bibfnamefont
  {B.~P.}\ \bibnamefont {Lanyon}}, \bibinfo {author} {\bibfnamefont
  {P.}~\bibnamefont {Zoller}}, \bibinfo {author} {\bibfnamefont
  {R.}~\bibnamefont {Blatt}},\ and\ \bibinfo {author} {\bibfnamefont {C.~F.}\
  \bibnamefont {Roos}},\ }\bibfield  {title} {\bibinfo {title} {Probing
  {R}{\'e}nyi entanglement entropy via randomized measurements},\ }\href
  {https://doi.org/10.1126/science.aau4963} {\bibfield  {journal} {\bibinfo
  {journal} {Science}\ }\textbf {\bibinfo {volume} {364}},\ \bibinfo {pages}
  {260} (\bibinfo {year} {2019})}\BibitemShut {NoStop}%
\bibitem [{\citenamefont {Zhou}\ \emph {et~al.}(2021)\citenamefont {Zhou},
  \citenamefont {Zhang}, \citenamefont {Yin}, \citenamefont {Huai},
  \citenamefont {Gu}, \citenamefont {Xu}, \citenamefont {Allcock},
  \citenamefont {Liu}, \citenamefont {Xi}, \citenamefont {Yu}, \citenamefont
  {Zhang}, \citenamefont {Zhang}, \citenamefont {Li}, \citenamefont {Song},
  \citenamefont {Wang}, \citenamefont {Zheng}, \citenamefont {An},
  \citenamefont {Zheng},\ and\ \citenamefont {Zhang}}]{Zhou2021}%
  \BibitemOpen
  \bibfield  {author} {\bibinfo {author} {\bibfnamefont {Y.}~\bibnamefont
  {Zhou}}, \bibinfo {author} {\bibfnamefont {Z.}~\bibnamefont {Zhang}},
  \bibinfo {author} {\bibfnamefont {Z.}~\bibnamefont {Yin}}, \bibinfo {author}
  {\bibfnamefont {S.}~\bibnamefont {Huai}}, \bibinfo {author} {\bibfnamefont
  {X.}~\bibnamefont {Gu}}, \bibinfo {author} {\bibfnamefont {X.}~\bibnamefont
  {Xu}}, \bibinfo {author} {\bibfnamefont {J.}~\bibnamefont {Allcock}},
  \bibinfo {author} {\bibfnamefont {F.}~\bibnamefont {Liu}}, \bibinfo {author}
  {\bibfnamefont {G.}~\bibnamefont {Xi}}, \bibinfo {author} {\bibfnamefont
  {Q.}~\bibnamefont {Yu}}, \bibinfo {author} {\bibfnamefont {H.}~\bibnamefont
  {Zhang}}, \bibinfo {author} {\bibfnamefont {M.}~\bibnamefont {Zhang}},
  \bibinfo {author} {\bibfnamefont {H.}~\bibnamefont {Li}}, \bibinfo {author}
  {\bibfnamefont {X.}~\bibnamefont {Song}}, \bibinfo {author} {\bibfnamefont
  {Z.}~\bibnamefont {Wang}}, \bibinfo {author} {\bibfnamefont {D.}~\bibnamefont
  {Zheng}}, \bibinfo {author} {\bibfnamefont {S.}~\bibnamefont {An}}, \bibinfo
  {author} {\bibfnamefont {Y.}~\bibnamefont {Zheng}},\ and\ \bibinfo {author}
  {\bibfnamefont {S.}~\bibnamefont {Zhang}},\ }\bibfield  {title} {\bibinfo
  {title} {Rapid and unconditional parametric reset protocol for tunable
  superconducting qubits},\ }\href {https://doi.org/10.1038/s41467-021-26205-y}
  {\bibfield  {journal} {\bibinfo  {journal} {Nat. Commun.}\ }\textbf {\bibinfo
  {volume} {12}},\ \bibinfo {pages} {5924} (\bibinfo {year}
  {2021})}\BibitemShut {NoStop}%
\bibitem [{\citenamefont {Dawid}\ \emph {et~al.}(2025)\citenamefont {Dawid},
  \citenamefont {Arnold}, \citenamefont {Requena}, \citenamefont {Gresch},
  \citenamefont {P\l{}odzie\'{n}}, \citenamefont {Donatella}, \citenamefont
  {Nicoli}, \citenamefont {Stornati}, \citenamefont {Koch}, \citenamefont
  {B\"{u}ttner}, \citenamefont {Oku\l{}a}, \citenamefont {Mu{\~n}oz-Gil},
  \citenamefont {Vargas-Hern\'{a}ndez}, \citenamefont {Cervera-Lierta},
  \citenamefont {Carrasquilla}, \citenamefont {Dunjko}, \citenamefont
  {Gabri\'{e}}, \citenamefont {Huembeli}, \citenamefont {van Nieuwenburg},
  \citenamefont {Vicentini}, \citenamefont {Wang}, \citenamefont {Wetzel},
  \citenamefont {Carleo}, \citenamefont {Greplov\'{a}}, \citenamefont {Krems},
  \citenamefont {Marquardt}, \citenamefont {Tomza}, \citenamefont
  {Lewenstein},\ and\ \citenamefont {Dauphin}}]{Dawid2022}%
  \BibitemOpen
  \bibfield  {author} {\bibinfo {author} {\bibfnamefont {A.}~\bibnamefont
  {Dawid}}, \bibinfo {author} {\bibfnamefont {J.}~\bibnamefont {Arnold}},
  \bibinfo {author} {\bibfnamefont {B.}~\bibnamefont {Requena}}, \bibinfo
  {author} {\bibfnamefont {A.}~\bibnamefont {Gresch}}, \bibinfo {author}
  {\bibfnamefont {M.}~\bibnamefont {P\l{}odzie\'{n}}}, \bibinfo {author}
  {\bibfnamefont {K.}~\bibnamefont {Donatella}}, \bibinfo {author}
  {\bibfnamefont {K.~A.}\ \bibnamefont {Nicoli}}, \bibinfo {author}
  {\bibfnamefont {P.}~\bibnamefont {Stornati}}, \bibinfo {author}
  {\bibfnamefont {R.}~\bibnamefont {Koch}}, \bibinfo {author} {\bibfnamefont
  {M.}~\bibnamefont {B\"{u}ttner}}, \bibinfo {author} {\bibfnamefont
  {R.}~\bibnamefont {Oku\l{}a}}, \bibinfo {author} {\bibfnamefont
  {G.}~\bibnamefont {Mu{\~n}oz-Gil}}, \bibinfo {author} {\bibfnamefont {R.~A.}\
  \bibnamefont {Vargas-Hern\'{a}ndez}}, \bibinfo {author} {\bibfnamefont
  {A.}~\bibnamefont {Cervera-Lierta}}, \bibinfo {author} {\bibfnamefont
  {J.}~\bibnamefont {Carrasquilla}}, \bibinfo {author} {\bibfnamefont
  {V.}~\bibnamefont {Dunjko}}, \bibinfo {author} {\bibfnamefont
  {M.}~\bibnamefont {Gabri\'{e}}}, \bibinfo {author} {\bibfnamefont
  {P.}~\bibnamefont {Huembeli}}, \bibinfo {author} {\bibfnamefont
  {E.}~\bibnamefont {van Nieuwenburg}}, \bibinfo {author} {\bibfnamefont
  {F.}~\bibnamefont {Vicentini}}, \bibinfo {author} {\bibfnamefont
  {L.}~\bibnamefont {Wang}}, \bibinfo {author} {\bibfnamefont {S.~J.}\
  \bibnamefont {Wetzel}}, \bibinfo {author} {\bibfnamefont {G.}~\bibnamefont
  {Carleo}}, \bibinfo {author} {\bibfnamefont {E.}~\bibnamefont
  {Greplov\'{a}}}, \bibinfo {author} {\bibfnamefont {R.}~\bibnamefont {Krems}},
  \bibinfo {author} {\bibfnamefont {F.}~\bibnamefont {Marquardt}}, \bibinfo
  {author} {\bibfnamefont {M.}~\bibnamefont {Tomza}}, \bibinfo {author}
  {\bibfnamefont {M.}~\bibnamefont {Lewenstein}},\ and\ \bibinfo {author}
  {\bibfnamefont {A.}~\bibnamefont {Dauphin}},\ }\href
  {https://arxiv.org/abs/2204.04198} {\emph {\bibinfo {title} {Machine Learning
  in Quantum Sciences}}}\ (\bibinfo  {publisher} {Cambridge University Press},\
  \bibinfo {year} {2025})\BibitemShut {NoStop}%
\bibitem [{\citenamefont {Kingma}\ and\ \citenamefont {Ba}(2014)}]{Kingma2017}%
  \BibitemOpen
  \bibfield  {author} {\bibinfo {author} {\bibfnamefont {D.~P.}\ \bibnamefont
  {Kingma}}\ and\ \bibinfo {author} {\bibfnamefont {J.}~\bibnamefont {Ba}},\
  }\bibfield  {title} {\bibinfo {title} {Adam: A method for stochastic
  optimization},\ }\href {https://arxiv.org/abs/1412.6980} {\bibfield
  {journal} {\bibinfo  {journal} {arXiv:1412.6980}\ } (\bibinfo {year}
  {2014})}\BibitemShut {NoStop}%
\bibitem [{\citenamefont {Salimans}\ \emph {et~al.}(2017)\citenamefont
  {Salimans}, \citenamefont {Ho}, \citenamefont {Chen}, \citenamefont {Sidor},\
  and\ \citenamefont {Sutskever}}]{Salimans2017}%
  \BibitemOpen
  \bibfield  {author} {\bibinfo {author} {\bibfnamefont {T.}~\bibnamefont
  {Salimans}}, \bibinfo {author} {\bibfnamefont {J.}~\bibnamefont {Ho}},
  \bibinfo {author} {\bibfnamefont {X.}~\bibnamefont {Chen}}, \bibinfo {author}
  {\bibfnamefont {S.}~\bibnamefont {Sidor}},\ and\ \bibinfo {author}
  {\bibfnamefont {I.}~\bibnamefont {Sutskever}},\ }\bibfield  {title} {\bibinfo
  {title} {Evolution strategies as a scalable alternative to reinforcement
  learning},\ }\href {https://arxiv.org/abs/1703.03864} {\bibfield  {journal}
  {\bibinfo  {journal} {arXiv:1703.03864}\ } (\bibinfo {year}
  {2017})}\BibitemShut {NoStop}%
\bibitem [{\citenamefont {Mari}\ \emph {et~al.}(2020)\citenamefont {Mari},
  \citenamefont {Bromley}, \citenamefont {Izaac}, \citenamefont {Schuld},\ and\
  \citenamefont {Killoran}}]{Mari2020}%
  \BibitemOpen
  \bibfield  {author} {\bibinfo {author} {\bibfnamefont {A.}~\bibnamefont
  {Mari}}, \bibinfo {author} {\bibfnamefont {T.~R.}\ \bibnamefont {Bromley}},
  \bibinfo {author} {\bibfnamefont {J.}~\bibnamefont {Izaac}}, \bibinfo
  {author} {\bibfnamefont {M.}~\bibnamefont {Schuld}},\ and\ \bibinfo {author}
  {\bibfnamefont {N.}~\bibnamefont {Killoran}},\ }\bibfield  {title} {\bibinfo
  {title} {Transfer learning in hybrid classical-quantum neural networks},\
  }\href {https://doi.org/10.22331/q-2020-10-09-340} {\bibfield  {journal}
  {\bibinfo  {journal} {{Quantum}}\ }\textbf {\bibinfo {volume} {4}},\ \bibinfo
  {pages} {340} (\bibinfo {year} {2020})}\BibitemShut {NoStop}%
\bibitem [{\citenamefont {Hosmer}\ \emph {et~al.}(2000)\citenamefont {Hosmer},
  \citenamefont {Lemeshow},\ and\ \citenamefont {Sturdivant}}]{HosmerJr2000}%
  \BibitemOpen
  \bibfield  {author} {\bibinfo {author} {\bibfnamefont {D.~W.}\ \bibnamefont
  {Hosmer}}, \bibinfo {author} {\bibfnamefont {S.}~\bibnamefont {Lemeshow}},\
  and\ \bibinfo {author} {\bibfnamefont {R.~X.}\ \bibnamefont {Sturdivant}},\
  }\href {https://onlinelibrary.wiley.com/doi/book/10.1002/0471722146} {\emph
  {\bibinfo {title} {Applied logistic regression}}}\ (\bibinfo  {publisher}
  {John Wiley \& Sons},\ \bibinfo {year} {2000})\BibitemShut {NoStop}%
\bibitem [{\citenamefont {Haller}\ \emph {et~al.}(2023)\citenamefont {Haller},
  \citenamefont {Xu}, \citenamefont {Liu},\ and\ \citenamefont
  {Pollmann}}]{Haller2023}%
  \BibitemOpen
  \bibfield  {author} {\bibinfo {author} {\bibfnamefont {L.}~\bibnamefont
  {Haller}}, \bibinfo {author} {\bibfnamefont {W.-T.}\ \bibnamefont {Xu}},
  \bibinfo {author} {\bibfnamefont {Y.-J.}\ \bibnamefont {Liu}},\ and\ \bibinfo
  {author} {\bibfnamefont {F.}~\bibnamefont {Pollmann}},\ }\bibfield  {title}
  {\bibinfo {title} {Quantum phase transition between symmetry enriched
  topological phases in tensor-network states},\ }\href
  {https://doi.org/10.1103/PhysRevResearch.5.043078} {\bibfield  {journal}
  {\bibinfo  {journal} {Phys. Rev. Research}\ }\textbf {\bibinfo {volume}
  {5}},\ \bibinfo {pages} {043078} (\bibinfo {year} {2023})}\BibitemShut
  {NoStop}%
\bibitem [{\citenamefont {Holmes}\ \emph {et~al.}(2022)\citenamefont {Holmes},
  \citenamefont {Sharma}, \citenamefont {Cerezo},\ and\ \citenamefont
  {Coles}}]{Holmes2021}%
  \BibitemOpen
  \bibfield  {author} {\bibinfo {author} {\bibfnamefont {Z.}~\bibnamefont
  {Holmes}}, \bibinfo {author} {\bibfnamefont {K.}~\bibnamefont {Sharma}},
  \bibinfo {author} {\bibfnamefont {M.}~\bibnamefont {Cerezo}},\ and\ \bibinfo
  {author} {\bibfnamefont {P.~J.}\ \bibnamefont {Coles}},\ }\bibfield  {title}
  {\bibinfo {title} {Connecting ansatz expressibility to gradient magnitudes
  and barren plateaus},\ }\href {https://doi.org/10.1103/PRXQuantum.3.010313}
  {\bibfield  {journal} {\bibinfo  {journal} {PRX Quantum}\ }\textbf {\bibinfo
  {volume} {3}},\ \bibinfo {pages} {010313} (\bibinfo {year}
  {2022})}\BibitemShut {NoStop}%
\bibitem [{\citenamefont {Larocca}\ \emph {et~al.}(2022)\citenamefont
  {Larocca}, \citenamefont {Czarnik}, \citenamefont {Sharma}, \citenamefont
  {Muraleedharan}, \citenamefont {Coles},\ and\ \citenamefont
  {Cerezo}}]{Larocca2021}%
  \BibitemOpen
  \bibfield  {author} {\bibinfo {author} {\bibfnamefont {M.}~\bibnamefont
  {Larocca}}, \bibinfo {author} {\bibfnamefont {P.}~\bibnamefont {Czarnik}},
  \bibinfo {author} {\bibfnamefont {K.}~\bibnamefont {Sharma}}, \bibinfo
  {author} {\bibfnamefont {G.}~\bibnamefont {Muraleedharan}}, \bibinfo {author}
  {\bibfnamefont {P.~J.}\ \bibnamefont {Coles}},\ and\ \bibinfo {author}
  {\bibfnamefont {M.}~\bibnamefont {Cerezo}},\ }\bibfield  {title} {\bibinfo
  {title} {Diagnosing barren plateaus with tools from quantum optimal
  control},\ }\href {https://doi.org/10.22331/q-2022-09-29-824} {\bibfield
  {journal} {\bibinfo  {journal} {Quantum}\ }\textbf {\bibinfo {volume} {6}},\
  \bibinfo {pages} {824} (\bibinfo {year} {2022})}\BibitemShut {NoStop}%
\bibitem [{\citenamefont {Ragone}\ \emph {et~al.}(2024)\citenamefont {Ragone},
  \citenamefont {Bakalov}, \citenamefont {Sauvage}, \citenamefont {Kemper},
  \citenamefont {Ortiz~Marrero}, \citenamefont {Larocca},\ and\ \citenamefont
  {Cerezo}}]{Ragone2024}%
  \BibitemOpen
  \bibfield  {author} {\bibinfo {author} {\bibfnamefont {M.}~\bibnamefont
  {Ragone}}, \bibinfo {author} {\bibfnamefont {B.~N.}\ \bibnamefont {Bakalov}},
  \bibinfo {author} {\bibfnamefont {F.}~\bibnamefont {Sauvage}}, \bibinfo
  {author} {\bibfnamefont {A.~F.}\ \bibnamefont {Kemper}}, \bibinfo {author}
  {\bibfnamefont {C.}~\bibnamefont {Ortiz~Marrero}}, \bibinfo {author}
  {\bibfnamefont {M.}~\bibnamefont {Larocca}},\ and\ \bibinfo {author}
  {\bibfnamefont {M.}~\bibnamefont {Cerezo}},\ }\bibfield  {title} {\bibinfo
  {title} {A {Lie} algebraic theory of barren plateaus for deep parameterized
  quantum circuits},\ }\href {https://doi.org/10.1038/s41467-024-49909-3}
  {\bibfield  {journal} {\bibinfo  {journal} {Nat. Commun.}\ }\textbf {\bibinfo
  {volume} {15}},\ \bibinfo {pages} {7172} (\bibinfo {year}
  {2024})}\BibitemShut {NoStop}%
\bibitem [{\citenamefont {Krinner}\ \emph {et~al.}(2022)\citenamefont
  {Krinner}, \citenamefont {Lacroix}, \citenamefont {Remm}, \citenamefont
  {Paolo}, \citenamefont {Genois}, \citenamefont {Leroux}, \citenamefont
  {Hellings}, \citenamefont {Laz\u{a}r}, \citenamefont {Swiadek}, \citenamefont
  {Herrmann}, \citenamefont {Norris}, \citenamefont {Andersen}, \citenamefont
  {M\"{u}ller}, \citenamefont {Blais}, \citenamefont {Eichler},\ and\
  \citenamefont {Wallraff}}]{Krinner2022}%
  \BibitemOpen
  \bibfield  {author} {\bibinfo {author} {\bibfnamefont {S.}~\bibnamefont
  {Krinner}}, \bibinfo {author} {\bibfnamefont {N.}~\bibnamefont {Lacroix}},
  \bibinfo {author} {\bibfnamefont {A.}~\bibnamefont {Remm}}, \bibinfo {author}
  {\bibfnamefont {A.~D.}\ \bibnamefont {Paolo}}, \bibinfo {author}
  {\bibfnamefont {E.}~\bibnamefont {Genois}}, \bibinfo {author} {\bibfnamefont
  {C.}~\bibnamefont {Leroux}}, \bibinfo {author} {\bibfnamefont
  {C.}~\bibnamefont {Hellings}}, \bibinfo {author} {\bibfnamefont
  {S.}~\bibnamefont {Laz\u{a}r}}, \bibinfo {author} {\bibfnamefont
  {F.}~\bibnamefont {Swiadek}}, \bibinfo {author} {\bibfnamefont
  {J.}~\bibnamefont {Herrmann}}, \bibinfo {author} {\bibfnamefont {G.~J.}\
  \bibnamefont {Norris}}, \bibinfo {author} {\bibfnamefont {C.~K.}\
  \bibnamefont {Andersen}}, \bibinfo {author} {\bibfnamefont {M.}~\bibnamefont
  {M\"{u}ller}}, \bibinfo {author} {\bibfnamefont {A.}~\bibnamefont {Blais}},
  \bibinfo {author} {\bibfnamefont {C.}~\bibnamefont {Eichler}},\ and\ \bibinfo
  {author} {\bibfnamefont {A.}~\bibnamefont {Wallraff}},\ }\bibfield  {title}
  {\bibinfo {title} {Realizing repeated quantum error correction in a
  distance-three surface code},\ }\href
  {https://doi.org/10.1038/s41586-022-04566-8} {\bibfield  {journal} {\bibinfo
  {journal} {Nature}\ }\textbf {\bibinfo {volume} {605}},\ \bibinfo {pages}
  {669} (\bibinfo {year} {2022})}\BibitemShut {NoStop}%
\bibitem [{\citenamefont {Lacroix}\ \emph {et~al.}(2025)\citenamefont
  {Lacroix}, \citenamefont {Hofele}, \citenamefont {Remm}, \citenamefont
  {Benhayoune-Khadraoui}, \citenamefont {McDonald}, \citenamefont {Shillito},
  \citenamefont {Laz\u{a}r}, \citenamefont {Hellings}, \citenamefont {Swiadek},
  \citenamefont {Colao-Zanuz}, \citenamefont {Flasby}, \citenamefont {Panah},
  \citenamefont {Kerschbaum}, \citenamefont {Norris}, \citenamefont {Blais},
  \citenamefont {Wallraff},\ and\ \citenamefont {Krinner}}]{Lacroix2025}%
  \BibitemOpen
  \bibfield  {author} {\bibinfo {author} {\bibfnamefont {N.}~\bibnamefont
  {Lacroix}}, \bibinfo {author} {\bibfnamefont {L.}~\bibnamefont {Hofele}},
  \bibinfo {author} {\bibfnamefont {A.}~\bibnamefont {Remm}}, \bibinfo {author}
  {\bibfnamefont {O.}~\bibnamefont {Benhayoune-Khadraoui}}, \bibinfo {author}
  {\bibfnamefont {A.}~\bibnamefont {McDonald}}, \bibinfo {author}
  {\bibfnamefont {R.}~\bibnamefont {Shillito}}, \bibinfo {author}
  {\bibfnamefont {S.}~\bibnamefont {Laz\u{a}r}}, \bibinfo {author}
  {\bibfnamefont {C.}~\bibnamefont {Hellings}}, \bibinfo {author}
  {\bibfnamefont {F.}~\bibnamefont {Swiadek}}, \bibinfo {author} {\bibfnamefont
  {D.}~\bibnamefont {Colao-Zanuz}}, \bibinfo {author} {\bibfnamefont
  {A.}~\bibnamefont {Flasby}}, \bibinfo {author} {\bibfnamefont {M.~B.}\
  \bibnamefont {Panah}}, \bibinfo {author} {\bibfnamefont {M.}~\bibnamefont
  {Kerschbaum}}, \bibinfo {author} {\bibfnamefont {G.~J.}\ \bibnamefont
  {Norris}}, \bibinfo {author} {\bibfnamefont {A.}~\bibnamefont {Blais}},
  \bibinfo {author} {\bibfnamefont {A.}~\bibnamefont {Wallraff}},\ and\
  \bibinfo {author} {\bibfnamefont {S.}~\bibnamefont {Krinner}},\ }\bibfield
  {title} {\bibinfo {title} {Fast flux-activated leakage reduction for
  superconducting quantum circuits},\ }\href
  {https://doi.org/10.1103/PhysRevLett.134.120601} {\bibfield  {journal}
  {\bibinfo  {journal} {Phys. Rev. Lett.}\ }\textbf {\bibinfo {volume} {134}},\
  \bibinfo {pages} {120601} (\bibinfo {year} {2025})}\BibitemShut {NoStop}%
\bibitem [{\citenamefont {Bilmes}\ \emph {et~al.}(2021)\citenamefont {Bilmes},
  \citenamefont {H\"andel}, \citenamefont {Volosheniuk}, \citenamefont
  {Ustinov},\ and\ \citenamefont {Lisenfeld}}]{Bilmes2021}%
  \BibitemOpen
  \bibfield  {author} {\bibinfo {author} {\bibfnamefont {A.}~\bibnamefont
  {Bilmes}}, \bibinfo {author} {\bibfnamefont {A.~K.}\ \bibnamefont
  {H\"andel}}, \bibinfo {author} {\bibfnamefont {S.}~\bibnamefont
  {Volosheniuk}}, \bibinfo {author} {\bibfnamefont {A.~V.}\ \bibnamefont
  {Ustinov}},\ and\ \bibinfo {author} {\bibfnamefont {J.}~\bibnamefont
  {Lisenfeld}},\ }\bibfield  {title} {\bibinfo {title} {In-situ bandaged
  {Josephson} junctions for superconducting quantum processors},\ }\href
  {https://doi.org/10.1088/1361-6668/ac2a6d} {\bibfield  {journal} {\bibinfo
  {journal} {Superconductor Science and Technology}\ }\textbf {\bibinfo
  {volume} {34}},\ \bibinfo {pages} {125011} (\bibinfo {year}
  {2021})}\BibitemShut {NoStop}%
\bibitem [{\citenamefont {Hellings}\ \emph {et~al.}(2025)\citenamefont
  {Hellings}, \citenamefont {Lacroix}, \citenamefont {Remm}, \citenamefont
  {Boell}, \citenamefont {Herrmann}, \citenamefont {Laz\u{a}r}, \citenamefont
  {Krinner}, \citenamefont {Swiadek}, \citenamefont {Andersen}, \citenamefont
  {Eichler},\ and\ \citenamefont {Wallraff}}]{Hellings2025}%
  \BibitemOpen
  \bibfield  {author} {\bibinfo {author} {\bibfnamefont {C.}~\bibnamefont
  {Hellings}}, \bibinfo {author} {\bibfnamefont {N.}~\bibnamefont {Lacroix}},
  \bibinfo {author} {\bibfnamefont {A.}~\bibnamefont {Remm}}, \bibinfo {author}
  {\bibfnamefont {R.}~\bibnamefont {Boell}}, \bibinfo {author} {\bibfnamefont
  {J.}~\bibnamefont {Herrmann}}, \bibinfo {author} {\bibfnamefont
  {S.}~\bibnamefont {Laz\u{a}r}}, \bibinfo {author} {\bibfnamefont
  {S.}~\bibnamefont {Krinner}}, \bibinfo {author} {\bibfnamefont
  {F.}~\bibnamefont {Swiadek}}, \bibinfo {author} {\bibfnamefont {C.~K.}\
  \bibnamefont {Andersen}}, \bibinfo {author} {\bibfnamefont {C.}~\bibnamefont
  {Eichler}},\ and\ \bibinfo {author} {\bibfnamefont {A.}~\bibnamefont
  {Wallraff}},\ }\bibfield  {title} {\bibinfo {title} {Calibrating magnetic
  flux control in superconducting circuits by compensating distortions on
  timescales from nanoseconds up to tens of microseconds},\ }\href
  {https://doi.org/10.1103/1qhb-r4fb} {\bibfield  {journal} {\bibinfo
  {journal} {Phys. Rev. Research}\ }\textbf {\bibinfo {volume} {7}},\ \bibinfo
  {pages} {043142} (\bibinfo {year} {2025})}\BibitemShut {NoStop}%
\bibitem [{\citenamefont {Motzoi}\ \emph {et~al.}(2009)\citenamefont {Motzoi},
  \citenamefont {Gambetta}, \citenamefont {Rebentrost},\ and\ \citenamefont
  {Wilhelm}}]{Motzoi2009}%
  \BibitemOpen
  \bibfield  {author} {\bibinfo {author} {\bibfnamefont {F.}~\bibnamefont
  {Motzoi}}, \bibinfo {author} {\bibfnamefont {J.~M.}\ \bibnamefont
  {Gambetta}}, \bibinfo {author} {\bibfnamefont {P.}~\bibnamefont
  {Rebentrost}},\ and\ \bibinfo {author} {\bibfnamefont {F.~K.}\ \bibnamefont
  {Wilhelm}},\ }\bibfield  {title} {\bibinfo {title} {Simple pulses for
  elimination of leakage in weakly nonlinear qubits},\ }\href
  {https://doi.org/10.1103/PhysRevLett.103.110501} {\bibfield  {journal}
  {\bibinfo  {journal} {Phys. Rev. Lett.}\ }\textbf {\bibinfo {volume} {103}},\
  \bibinfo {eid} {110501} (\bibinfo {year} {2009})}\BibitemShut {NoStop}%
\bibitem [{\citenamefont {Laz\u{a}r}(2023)}]{Lazar2023a}%
  \BibitemOpen
  \bibfield  {author} {\bibinfo {author} {\bibfnamefont {S.}~\bibnamefont
  {Laz\u{a}r}},\ }\emph {\bibinfo {title} {Improving Single-Qubit Gates in
  Superconducting Quantum Devices}},\ \href
  {https://www.research-collection.ethz.ch/entities/publication/1b0d7b07-08f6-4349-9183-5ccb38ec7ad0}
  {Ph.D. thesis},\ \bibinfo  {school} {ETH Zurich} (\bibinfo {year}
  {2023})\BibitemShut {NoStop}%
\bibitem [{\citenamefont {Magesan}\ \emph
  {et~al.}(2012{\natexlab{a}})\citenamefont {Magesan}, \citenamefont
  {Gambetta},\ and\ \citenamefont {Emerson}}]{Magesan2012a}%
  \BibitemOpen
  \bibfield  {author} {\bibinfo {author} {\bibfnamefont {E.}~\bibnamefont
  {Magesan}}, \bibinfo {author} {\bibfnamefont {J.~M.}\ \bibnamefont
  {Gambetta}},\ and\ \bibinfo {author} {\bibfnamefont {J.}~\bibnamefont
  {Emerson}},\ }\bibfield  {title} {\bibinfo {title} {Characterizing quantum
  gates via randomized benchmarking},\ }\href
  {https://doi.org/10.1103/PhysRevA.85.042311} {\bibfield  {journal} {\bibinfo
  {journal} {Phys. Rev. A}\ }\textbf {\bibinfo {volume} {85}},\ \bibinfo
  {pages} {042311} (\bibinfo {year} {2012}{\natexlab{a}})}\BibitemShut
  {NoStop}%
\bibitem [{\citenamefont {Magesan}\ \emph
  {et~al.}(2012{\natexlab{b}})\citenamefont {Magesan}, \citenamefont
  {Gambetta}, \citenamefont {Johnson}, \citenamefont {Ryan}, \citenamefont
  {Chow}, \citenamefont {Merkel}, \citenamefont {da~Silva}, \citenamefont
  {Keefe}, \citenamefont {Rothwell}, \citenamefont {Ohki}, \citenamefont
  {Ketchen},\ and\ \citenamefont {Steffen}}]{Magesan2012}%
  \BibitemOpen
  \bibfield  {author} {\bibinfo {author} {\bibfnamefont {E.}~\bibnamefont
  {Magesan}}, \bibinfo {author} {\bibfnamefont {J.~M.}\ \bibnamefont
  {Gambetta}}, \bibinfo {author} {\bibfnamefont {B.~R.}\ \bibnamefont
  {Johnson}}, \bibinfo {author} {\bibfnamefont {C.~A.}\ \bibnamefont {Ryan}},
  \bibinfo {author} {\bibfnamefont {J.~M.}\ \bibnamefont {Chow}}, \bibinfo
  {author} {\bibfnamefont {S.~T.}\ \bibnamefont {Merkel}}, \bibinfo {author}
  {\bibfnamefont {M.~P.}\ \bibnamefont {da~Silva}}, \bibinfo {author}
  {\bibfnamefont {G.~A.}\ \bibnamefont {Keefe}}, \bibinfo {author}
  {\bibfnamefont {M.~B.}\ \bibnamefont {Rothwell}}, \bibinfo {author}
  {\bibfnamefont {T.~A.}\ \bibnamefont {Ohki}}, \bibinfo {author}
  {\bibfnamefont {M.~B.}\ \bibnamefont {Ketchen}},\ and\ \bibinfo {author}
  {\bibfnamefont {M.}~\bibnamefont {Steffen}},\ }\bibfield  {title} {\bibinfo
  {title} {Efficient measurement of quantum gate error by interleaved
  randomized benchmarking},\ }\href
  {https://doi.org/10.1103/PhysRevLett.109.080505} {\bibfield  {journal}
  {\bibinfo  {journal} {Phys. Rev. Lett.}\ }\textbf {\bibinfo {volume} {109}},\
  \bibinfo {pages} {080505} (\bibinfo {year} {2012}{\natexlab{b}})}\BibitemShut
  {NoStop}%
\bibitem [{\citenamefont {Laz\u{a}r}\ \emph {et~al.}(2023)\citenamefont
  {Laz\u{a}r}, \citenamefont {Ficheux}, \citenamefont {Herrmann}, \citenamefont
  {Remm}, \citenamefont {Lacroix}, \citenamefont {Hellings}, \citenamefont
  {Swiadek}, \citenamefont {Zanuz}, \citenamefont {Norris}, \citenamefont
  {Panah}, \citenamefont {Flasby}, \citenamefont {Kerschbaum}, \citenamefont
  {Besse}, \citenamefont {Eichler},\ and\ \citenamefont
  {Wallraff}}]{Lazar2023}%
  \BibitemOpen
  \bibfield  {author} {\bibinfo {author} {\bibfnamefont {S.}~\bibnamefont
  {Laz\u{a}r}}, \bibinfo {author} {\bibfnamefont {Q.}~\bibnamefont {Ficheux}},
  \bibinfo {author} {\bibfnamefont {J.}~\bibnamefont {Herrmann}}, \bibinfo
  {author} {\bibfnamefont {A.}~\bibnamefont {Remm}}, \bibinfo {author}
  {\bibfnamefont {N.}~\bibnamefont {Lacroix}}, \bibinfo {author} {\bibfnamefont
  {C.}~\bibnamefont {Hellings}}, \bibinfo {author} {\bibfnamefont
  {F.}~\bibnamefont {Swiadek}}, \bibinfo {author} {\bibfnamefont {D.~C.}\
  \bibnamefont {Zanuz}}, \bibinfo {author} {\bibfnamefont {G.~J.}\ \bibnamefont
  {Norris}}, \bibinfo {author} {\bibfnamefont {M.~B.}\ \bibnamefont {Panah}},
  \bibinfo {author} {\bibfnamefont {A.}~\bibnamefont {Flasby}}, \bibinfo
  {author} {\bibfnamefont {M.}~\bibnamefont {Kerschbaum}}, \bibinfo {author}
  {\bibfnamefont {J.}~\bibnamefont {Besse}}, \bibinfo {author} {\bibfnamefont
  {C.}~\bibnamefont {Eichler}},\ and\ \bibinfo {author} {\bibfnamefont
  {A.}~\bibnamefont {Wallraff}},\ }\bibfield  {title} {\bibinfo {title}
  {Calibration of drive nonlinearity for arbitrary-angle single-qubit gates
  using error amplification},\ }\href
  {https://doi.org/10.1103/PhysRevApplied.20.024036} {\bibfield  {journal}
  {\bibinfo  {journal} {Phys. Rev. Applied}\ }\textbf {\bibinfo {volume}
  {20}},\ \bibinfo {pages} {024036} (\bibinfo {year} {2023})}\BibitemShut
  {NoStop}%
\bibitem [{\citenamefont {Magnard}\ \emph {et~al.}(2018)\citenamefont
  {Magnard}, \citenamefont {Kurpiers}, \citenamefont {Royer}, \citenamefont
  {Walter}, \citenamefont {Besse}, \citenamefont {Gasparinetti}, \citenamefont
  {Pechal}, \citenamefont {Heinsoo}, \citenamefont {Storz}, \citenamefont
  {Blais},\ and\ \citenamefont {Wallraff}}]{Magnard2018}%
  \BibitemOpen
  \bibfield  {author} {\bibinfo {author} {\bibfnamefont {P.}~\bibnamefont
  {Magnard}}, \bibinfo {author} {\bibfnamefont {P.}~\bibnamefont {Kurpiers}},
  \bibinfo {author} {\bibfnamefont {B.}~\bibnamefont {Royer}}, \bibinfo
  {author} {\bibfnamefont {T.}~\bibnamefont {Walter}}, \bibinfo {author}
  {\bibfnamefont {J.-C.}\ \bibnamefont {Besse}}, \bibinfo {author}
  {\bibfnamefont {S.}~\bibnamefont {Gasparinetti}}, \bibinfo {author}
  {\bibfnamefont {M.}~\bibnamefont {Pechal}}, \bibinfo {author} {\bibfnamefont
  {J.}~\bibnamefont {Heinsoo}}, \bibinfo {author} {\bibfnamefont
  {S.}~\bibnamefont {Storz}}, \bibinfo {author} {\bibfnamefont
  {A.}~\bibnamefont {Blais}},\ and\ \bibinfo {author} {\bibfnamefont
  {A.}~\bibnamefont {Wallraff}},\ }\bibfield  {title} {\bibinfo {title} {Fast
  and unconditional all-microwave reset of a superconducting qubit},\ }\href
  {https://doi.org/10.1103/PhysRevLett.121.060502} {\bibfield  {journal}
  {\bibinfo  {journal} {Phys. Rev. Lett.}\ }\textbf {\bibinfo {volume} {121}},\
  \bibinfo {pages} {060502} (\bibinfo {year} {2018})}\BibitemShut {NoStop}%
\bibitem [{\citenamefont {Krinner}\ \emph {et~al.}(2020)\citenamefont
  {Krinner}, \citenamefont {Laz\u{a}r}, \citenamefont {Remm}, \citenamefont
  {Andersen}, \citenamefont {Lacroix}, \citenamefont {Norris}, \citenamefont
  {Hellings}, \citenamefont {Gabureac}, \citenamefont {Eichler},\ and\
  \citenamefont {Wallraff}}]{Krinner2020}%
  \BibitemOpen
  \bibfield  {author} {\bibinfo {author} {\bibfnamefont {S.}~\bibnamefont
  {Krinner}}, \bibinfo {author} {\bibfnamefont {S.}~\bibnamefont {Laz\u{a}r}},
  \bibinfo {author} {\bibfnamefont {A.}~\bibnamefont {Remm}}, \bibinfo {author}
  {\bibfnamefont {C.}~\bibnamefont {Andersen}}, \bibinfo {author}
  {\bibfnamefont {N.}~\bibnamefont {Lacroix}}, \bibinfo {author} {\bibfnamefont
  {G.}~\bibnamefont {Norris}}, \bibinfo {author} {\bibfnamefont
  {C.}~\bibnamefont {Hellings}}, \bibinfo {author} {\bibfnamefont
  {M.}~\bibnamefont {Gabureac}}, \bibinfo {author} {\bibfnamefont
  {C.}~\bibnamefont {Eichler}},\ and\ \bibinfo {author} {\bibfnamefont
  {A.}~\bibnamefont {Wallraff}},\ }\bibfield  {title} {\bibinfo {title}
  {Benchmarking coherent errors in controlled-phase gates due to spectator
  qubits},\ }\href {https://doi.org/10.1103/PhysRevApplied.14.024042}
  {\bibfield  {journal} {\bibinfo  {journal} {Phys. Rev. Applied}\ }\textbf
  {\bibinfo {volume} {14}},\ \bibinfo {pages} {024042} (\bibinfo {year}
  {2020})}\BibitemShut {NoStop}%
\bibitem [{\citenamefont {Arnold}\ and\ \citenamefont
  {Sch\"afer}(2022)}]{Arnold2022}%
  \BibitemOpen
  \bibfield  {author} {\bibinfo {author} {\bibfnamefont {J.}~\bibnamefont
  {Arnold}}\ and\ \bibinfo {author} {\bibfnamefont {F.}~\bibnamefont
  {Sch\"afer}},\ }\bibfield  {title} {\bibinfo {title} {Replacing neural
  networks by optimal analytical predictors for the detection of phase
  transitions},\ }\href {https://doi.org/10.1103/PhysRevX.12.031044} {\bibfield
   {journal} {\bibinfo  {journal} {Phys. Rev. X}\ }\textbf {\bibinfo {volume}
  {12}},\ \bibinfo {pages} {031044} (\bibinfo {year} {2022})}\BibitemShut
  {NoStop}%
\end{thebibliography}
\end{document}